\documentclass[preprintnumbers,amsmath,amssymb,twocolumn,showpacs,aps]{revtex4}
\usepackage{amssymb,amsmath}
\usepackage{graphics}
\usepackage{epstopdf}
\usepackage{epsfig}
\usepackage{color}
\usepackage{amsfonts}
\usepackage{amscd}
\usepackage{array}
\usepackage{makeidx}
\usepackage{nomencl}
\usepackage{graphicx}
\usepackage{dcolumn}
\usepackage{dsfont}
\usepackage{bm}
\usepackage{bbm}
\usepackage{amsmath}
\usepackage{amsgen}
\usepackage{amstext}
\usepackage{amsopn}
\usepackage{amscd}
\usepackage{amssymb}
\usepackage{ifthen}
\usepackage{braket}
\usepackage{cases}
\usepackage{amsfonts}
\usepackage{gensymb}
\usepackage{textcomp}
\usepackage{threeparttable}
\usepackage{psfrag}

\begin{document}

\title{Master equation approach to transient quantum transport in nanostructures}
\author{Pei-Yun Yang}
\email{l28001177@phys.ncku.edu.tw}
\affiliation{Department of Physics and Centre for Quantum Information Science, National Cheng Kung University, Tainan 70101, Taiwan}
\author{Wei-Min Zhang}
\email{wzhang@mail.ncku.edu.tw}
\affiliation{Department of Physics and Centre for Quantum Information Science, National Cheng Kung University, Tainan 70101, Taiwan}

\begin{abstract}
In this review article, we present a non-equilibrium quantum transport theory for
transient electron dynamics in nanodevices based on exact master equation derived with
the path integral method in the fermion coherent-state representation.
Applying the exact master equation to nanodevices, we also establish the connection of
the reduced density matrix and the transient quantum transport current with the Keldysh
nonequilibrium Green functions. The theory enables us to study transient quantum
transport in nanostructures with back-reaction effects from the contacts, with
non-Markovian dissipation and decoherence being fully taken into account. In applications,
we utilize the theory to specific quantum transport systems, a variety of quantum decoherence and
quantum transport phenomena involving the non-Markovian memory effect are investigated in both transient
and stationary scenarios at arbitrary initial temperatures of the contacts.

\ \\


\end{abstract}


\date{October 31, 2016}

\maketitle

\section{Introduction}

Generally speaking, a nanostructure refers to any structure with one
or more dimensions measuring in the nanometer ($10^{-9}$m) scale,
which puts the scale of a nanostructure intermediate in size between
a molecule and a bacterium. More specifically, the characteristic
dimension of a nanodevice is smaller than one or more of the
following length scales, the de Broglie wavelength of the electrons
(given by their kinetic energy), mean free path of electrons
(distance between collisions), and phase coherence length of
electrons (distance over which an electron can interfere with
itself). Such devices usually do not follow the Ohmic law because of
the quantum mechanical wave nature of electrons. Studying
nanostructures makes up one of the frontiers of semiconductor
industry due to Moore's Law, which is the observation that the
number of transistors in a dense integrated circuit doubles
approximately every two years. Although the pace of advancement has
slowed down, the current transistor fabrication already runs at $14
nm$, and Intel claim that they will have $10 nm$ technology in
commercial devices in late $2017$. Understanding how electrons
behave over such tiny distant scales is therefore of very obvious
importance to the electronics, communication and computation
industries.

Experimentalists now have access to a huge array of nanostructures
such as quantum heterostructures, quantum wells, superlattices,
nanowires etc. Nanostructures are typically probed either optically
(spectroscopy, photoluminescence, ...) or in electronic transport experiments.
In this review article, we mainly concentrate on the latter. Common
nanodevices for quantum transport include quantum dots
\cite{Kastner1993}, resonant tunneling diodes (RTDs)
\cite{Chang1974}, and two-dimensional electron gases (2DEGs)
\cite{Ando1982}. Quantum dots are the laboratory produced
solid-state structures with nanometer scales, in which the motion of
charge carriers (electrons and holes) is limited in all three
spatial dimensions. The electrons (holes) confined in discrete
quantum states with the properties of quantum dots being similar to
natural atoms. As a result, quantum dots are also called artificial
atoms, and their electronic properties can be modified and
controlled by external fields. Resonant-tunneling diodes (RTDs)
basically consist of two potential barriers and one quantum well
with electrons confined in the small central region. The major
attraction of RTDs is their ultrasensitive response to voltage bias
in going from the high-transmission state to the low-transmission
state. If these devices are able to operate under high bias
(far from equilibrium condition), very high transistor
transconductance and ultra-fast switching are obtainable. In fact,
microwave experimental results indicate the intrinsic speed limit of
RTD to be in the tera-Hz range~\cite{Brown1991}. Two-dimensional
electron gases (2DEGs) means electrons free to move in two
dimensions, but tightly confined in the third, which can then be
ignored. Most 2DEGs are found in transistor-like structures made
from semiconductors. 2DEGs offer a mature system of extremely high
mobility electrons, especially at low temperatures. These enormous
mobilities enable one to explore fundamental physics of quantum
nature, because except for confinement and effective mass, the
electrons do not interact with each other very often, so that they
can travel several micrometers before colliding. As a result, the quantum
coherence of electron wave may play an important role. Indeed, the quantum
Hall effect was first observed in a 2DEG~\cite{Klitzing1980} which
led to two Nobel Prizes, in 1985 and 1998, respectively.

\begin{figure}
\centerline{\scalebox{0.6}{\includegraphics{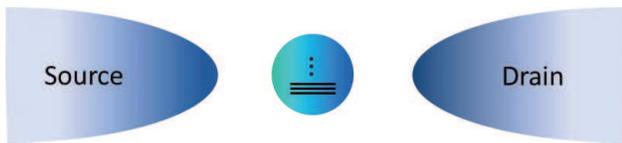}}}
\caption{Theoretical scheme for a quantum transport system}
\label{QT}
\end{figure}

Today, there are many practical applications of nanostructures and
nanomaterials. For example, the Quantum Hall effect now serves as a
standard measurement for resistance. Quantum dots are using in many
modern application areas including quantum dot
lasers in optics, fluorescent tracers in biological and medical
settings, and quantum information processing.
The theory of nanostructures involves a broad range of physical
concepts, from the simple confinement effects to the complex many-body physics,
such as the Kondo and fractional quantum Hall effects. More
traditional condensed matter and quantum many-body theory all have
the role to play in understanding and learning how to control
nanostructures as a practically useful device.
From the theoretical point of view, electrons transport in
nanostructures is described as physical systems consisting of a
nanoscale active region (the device system) attached to two leads
(source and drain), which is presented in Fig.~\ref{QT}.
The quantum transport theory for these physical systems is mainly
based on the following three theoretical approaches.
The Landauer-B\"{u}ttiker
approach~\cite{Imry2002, Buttiker1992}, because of its simplicity,
has often been used to analyze RTDs~\cite{Ohnishi1986} and quantum
wires~\cite{Szafer1989}. In this approach, electrons transport is
simply treated by ballistic transport (pure elastic scattering) near
thermal equilibrium. However, in order
for nanodevices to be functionally operated, it may be subjected to high
source-drain voltages and high-frequency bandwidths, in far from
equilibrium, highly transient and highly nonlinear regimes.
Thus, a more microscopic theory has been
developed for quantum transport in terms of non-equilibrium Green
functions~\cite{Schwinger1961,Kadanoff1962,Chou1985,Rammer1986,Wang2014} for the device system.
Moreover, the device system exchanges the particles, energy and information with the
leads, and is thereby a typical open system. The issues of open quantum
systems, such as dissipation, fluctuation and decoherence inevitably
arise. The third approach, the master equation approach, gets the
advantage by describing the device system in terms of the reduced density matrix.

\begin{figure}
\centerline{\scalebox{0.45}{\includegraphics{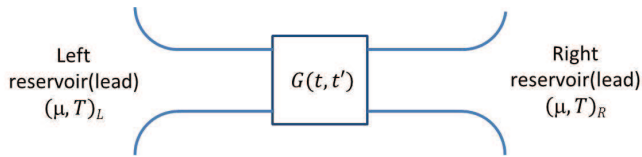}}}
\caption{Theoretical scheme for Keldysh non-equilibrium Green
function technique} \label{scheme2}
\end{figure}

In this review article, we give first a brief description of the
Landauer-B\"{u}ttiker approach~\cite{Imry2002, Buttiker1992}, the
non-equilibrium Green function
technique~\cite{Haug2008,Wingreen1993,Jauho1994}, and the master equation
approach~\cite{Tu2008,Tu2009,Jin2010,Schoeller1994,Gurvitz1996,Jin2008,Li2016,Yan2016}.
The theoretical schemes of these approaches are schematically
presented in Figs.~\ref{scheme1}, \ref{scheme2} and \ref{scheme3}. The main differences
between these three approaches are the ways of characterizing
electron transport flowing through the device system. In the
Landauer-B\"{u}ttiker approach, the device system is depicted as a
potential barrier, and all the information of the device system are
imbedded in the scattering matrix. The actual structure of the
device system is obscure. Comparing to the Landauer-B\"{u}ttiker
approach, the non-equilibrium Green function technique provides a
more microscopic way by describing electrons flowing through the
device system with single-particle non-equilibrium Green functions.
In the master equation approach, the device system is described by
the reduced density matrix, which is the essential quantity for
studying quantum coherent and decoherent phenomena. Better than the
non-equilibrium Green function in which the average over the density
matrix has been done, quantum coherent dynamics is depicted
explicitly by the off-diagonal matrix elements of the reduced density matrix. Then, in the subsequent sections,
we will focus on applications of the master equation approach to
various quantum transport problems in nanostructures.
In particular, using master equation, we investigate transient
current-current correlations and transient noise spectra for a
quantum dot system which contain various time scales associated with
the energy structures of the nanosystem (see Sec.~\ref{Noise}).
The transient quantum transport in nanostructures is also investigated
in the presence of initial correlations (see Sec.~\ref{initial_corr}).
The relation of the phase dependence between the quantum states and the
associated transport current are analyzed in a nanoscale  Aharonov-Bohm (AB) interferometer,
which provides an alternate possibility of quantum tomography in
nanosystems (see Sec.~\ref{AB}). At last, a conclusion is given
in Sec.~IV.

\section{Approaches for studying quantum transport in
semiconductor nanostructures}\label{approach}

\subsection{Landauer-B\"{u}ttiker approach}\label{LB appr}
The Landauer-B\"{u}ttiker formula has been widely utilized to
calculate various transport properties in semiconductor
nanostructures in the steady-state quantum transport
regime~\cite{Datta1995,Imry2002}. It establishes the fundamental
relation between the electron wave functions (scattering amplitudes)
of a quantum system and its conducting properties. In the
Landauer-B\"{u}ttiker formula, the transport current is given in
terms of transmission coefficients, obtained from the
single-particle scattering matrix. This approach is first formulated
by Landauer for the single-channel transport
\cite{Landauer1957,Landauer1970}. Later on, B\"{u}ttiker {\it et.
al.} extended the formula to multi-channel \cite{Buttiker1985} and
multi-terminal cases \cite{Buttiker1986}. The further development of
the Landauer-B\"{u}ttiker approach is the calculation of the current
noise correlations in mesoscopic conductors \cite{Buttiker1992}, the
detailed discussion can be found from the review article by Blanter
{\it et. al.} \cite{Blanter2000}.

\begin{figure}
\centerline{\scalebox{0.45}{\includegraphics{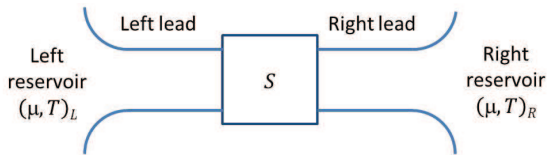}}}
\caption{Theoretical scheme for scattering theory} \label{scheme1}
\end{figure}

A typical system considered in the Landauer-B\"{u}ttiker approach
consists of reservoirs (contacts), quantum leads, and a mesoscopic
sample (scatterer) (see Fig.~\ref{scheme1}). The reservoirs connect
to the mesoscopic sample by quantum leads, and are always in an
equilibrium state in which electrons are always incoherent. However, electron transport
passing through the mesoscopic sample between the reservoirs is
phase coherent. Such coherent transport is described by the electron
wave function scattered in the mesoscopic sample, which
can be characterized by a scattering matrix $\bm{S}$. Therefore, the
key to describe electron quantum transport in Landauer-B\"{u}ttiker
approach is to determine the scattering matrix $\bm{S}$ which relies
crucially on the mesoscopic sample structure.

We start with the single-channel and two-terminal case. Consider an
electron plane wave impinging on a finite potential barrier from
left ($x<0$), and is scattered into the reflected and transmitted
components. Assume that the energy and momentum are conserved in
the scattering process, and the wave function of the electron
incident from the left and right are given respectively,
\begin{align}
\psi_L(x)=& \left\{
\begin{array}{ll}
e^{ikx}+re^{-ikx} & x<0, \\
te^{ikx} & x>0,
\end{array}  \right. \\
\psi_R(x)=& \left\{
\begin{array}{ll}
t'e^{-ikx} & x<0, \\
e^{-ikx}+r'e^{ikx} & x>0,
\end{array}  \right.
\end{align}
where $r$ ($r'$) and $t$ ($t'$) are respectively the complex reflected and
transmitted amplitudes of the wave incoming from the left (right), with
$|r|^2$ ($|r'|^2$) and $|t|^2$ ($|t'|^2$) being the reflected and transmitted
probabilities. These wave functions are the so-called scattering states.
For a general incoming state, $a_Le^{ikx}+a_Re^{-ikx}$, with probability
amplitudes $a_{L,R}$, the total wave function should be
\begin{align}
\Psi(x)=\left\{
\begin{array}{ll}
a_Le^{ikx}+b_Le^{-ikx} & x<0, \\
a_Re^{-ikx}+b_Re^{ikx} & x>0,
\end{array}  \right.
\end{align}
by introducing probability amplitudes $b_{L,R}$ for the outgoing
state such that the incoming and outgoing probability amplitudes are related
to each other by the scattering matrix:
\begin{align}
\left( \begin{array}[c]{cc}b_L\\b_R \end{array} \right)= \left(
\begin{array}[c]{cc}
r & t' \\
t & r'
\end{array} \right)\left( \begin{array}[c]{cc}a_L\\a_R \end{array}
\right)\equiv\bm{S}\left( \begin{array}[c]{cc}a_L\\a_R
\end{array}\right).
\end{align}
The coefficients in the scattering matrix ($r$, $t$, $r'$, and $t'$)
are obtained by solving the Schr\"{o}dinger equation with the
potential that models the mesoscopic sample.

It is straightforward to generalize the formalism to multi-channel
case where there are $N_L$ modes on the left and $N_R$ modes on the
right. The incoming and outgoing amplitudes can be written in
vectors such that
\begin{align}
\bm{a}=\left(
\begin{array}[c]{cc}a_{L1}\\\vdots\\a_{LN_L}\\a_{R1}\\\vdots\\a_{RN_R}
\end{array} \right);~~~~\bm{b}=\left(
\begin{array}[c]{cc}b_{L1}\\\vdots\\b_{LN_L}\\b_{R1}\\\vdots\\b_{RN_R}
\end{array} \right),
\end{align}
and the scattering matrix, leading to $\bm{b}=\bm{S}\bm{a}$,
is in dimension $(N_L+N_R)\times(N_L+N_R)$ and has the following
form,
\begin{align}
\label{scatt_mat}
\bm{S}=\left(
\begin{array}[c]{cc}
\bm{s}_{LL} & \bm{s}_{LR} \\
\bm{s}_{RL} & \bm{s}_{RR}
\end{array} \right)=\left(
\begin{array}[c]{cc}
\bm{r} & \bm{t}' \\
\bm{t} & \bm{r}'
\end{array} \right),
\end{align}
where the matrices $\bm{t}$ ($N_R\times N_L$) and $\bm{r}$
($N_L\times N_L$) describe respectively the transmission and
reflection of electrons incoming from the left with the element
$\bm{t}_{mn}$ and $\bm{r}_{mn}$ characterizing respectively the
electrons transmitted from the left mode $n$ into the right mode $m$
and the electrons reflected from the left mode $n$ into the left
mode $m$. Similarly, the matrices $\bm{r}'$ ($N_R\times N_R$) and
$\bm{t}'$ ($N_L\times N_R$) represent the reflection and
transmission processes for states incoming from the right. The
scattering matrix $\bm{S}$ is unitary due to the flux conservation,
i.e.
\begin{align}
\bm{S}^\dag\bm{S}=\bm{S}\bm{S}^\dag=\mathds{1}.
\end{align}

Consider the Hamiltonian of lead $\alpha$ ($\alpha=L,R$),
\begin{align}
H_\alpha=\frac{p^2_{x_\alpha}}{2m^*}+\frac{p^2_{\perp\alpha}}{2m^*}+U(\bm{r}_{\perp\alpha}),
\end{align}
where $x_\alpha$ and $\bm{r}_{\perp\alpha}$ denote the local
coordinates in the longitudinal and transverse directions,
respectively, and $m^*$ is the effective mass of the electron in the
lead. The motion of electrons in the longitudinal direction is free,
but it is quantized in the transverse direction due to the
confinement potential $U(\bm{r}_{\perp\alpha})$. Then, the
eigenfunctions of the Hamiltonian $H_\alpha$ can be expressed as
\begin{align}
\phi^{\pm}_{\alpha n}(x_\alpha,\bm{r}_{\perp\alpha})=\chi_{\alpha
n}(\bm{r}_{\perp\alpha})e^{\pm ik_{\alpha n}x_{\alpha}},
\end{align}
where the incoming wave $e^{ikx_\alpha}$ and outgoing wave
$e^{-ikx_\alpha}$ characterize the longitudinal motion of elections,
and $\chi_{\alpha n}(\bm{r}_{\perp\alpha})$ satisfies
\begin{align}
\big[~\frac{p^2_{\perp\alpha}}{2m^*}+U(\bm{r}_{\perp\alpha})~\big]\chi_{\alpha
n}(\bm{r}_{\perp\alpha})=\epsilon_{\alpha n}\chi_{\alpha
n}(\bm{r}_{\perp\alpha}),
\end{align}
with each transverse mode contributing a transport channel. As a result, the
dispersion relation of electron is thus given by
\begin{align}
E_{\alpha n}(k_{\alpha n})=\frac{\hbar^2k^2_{\alpha
n}}{2m^*}+\epsilon_{\alpha n}.
\end{align}
In this case, for an electron from mode $m$ of lead $\alpha$
scattering by the mesoscopic sample, the scattering state of the
electron for lead $\alpha$ is
\begin{align}
\label{scatt_state_a} \psi_{\alpha
m}(\alpha)=\sum_n\big\{\delta_{mn}\phi^+_{\alpha
n}+\sqrt{\frac{v_{\alpha m}}{v_{\alpha n}}}S_{\alpha \alpha
nm}\phi^-_{\alpha n}\big\},
\end{align}
where $S_{\alpha \beta nm}$ represents the amplitude of a state
scattered from mode $m$ in lead $\beta$ to mode $n$ in lead
$\alpha$, and the factor $\sqrt{v_{\alpha m}/v_{\alpha n}}$ is
introduced to guarantee the flux conservation, where $v_{\alpha
m}=\hbar k_{\alpha m}/m^*$ is the electron velocity. The
corresponding scattering state for lead $\beta\neq\alpha$ is
\begin{align}
\label{scatt_state_b} \psi_{\alpha
m}(\beta)=\sum_n\sqrt{\frac{v_{\alpha m}}{v_{\beta n}}}S_{\beta
\alpha nm}\phi^-_{\beta n}.
\end{align}
In the second quantization scheme, a general state of the lead-device system is given by an arbitrary
superposition of these scattering states,
\begin{align}
\label{field_opt_k}
\hat{\Psi}(\bm{r},t)\!=\!\frac{1}{\sqrt{2\pi}}\!\!\sum_{\alpha m}\! \int
\!\! dk_{\alpha m}\psi_{\alpha m}(\bm{r})e^{-\frac{i}{\hbar}E_{\alpha m}(k_{\alpha
m})t}\hat{a}_{\alpha m}(k_{\alpha m}),
\end{align}
where $\hat{a}_{\alpha
m}(k_{\alpha m})$ is the annihilation operator satisfying the canonical
anti-commutation relation,
\begin{align}
\big\{\hat{a}_{\alpha m}(k), \hat{a}^\dag_{\beta
n}(k')\big\}=\delta_{\alpha\beta}\delta_{nm}\delta(k-k').
\end{align}
Changing the $k$ space into the energy space, and defining the
incoming operator in the
energy space $\hat{a}_{\alpha m}(E)=\hat{a}_{\alpha m}(k)/[\hbar
v_{\alpha m}(k)]^{1/2}$, one has
\begin{align}
\big\{\hat{a}_{\alpha m}(E), \hat{a}^\dag_{\beta
n}(E')\big\}=\delta_{\alpha\beta}\delta_{nm}\delta(E-E'),
\end{align}
where $\delta(E-E')=1/v_{\alpha m}\delta(k-k')$. The field operator
can be rewritten as
\begin{align}
\label{field_opt_E}
\hat{\Psi}(\bm{r},t)=\sum_{\alpha m}\!\! \int
\!\! \frac{dE_{\alpha m}}{\sqrt{h v_{\alpha m}(E_{\alpha
m})}}\psi_{\alpha m}(\bm{r})e^{-\frac{i}{\hbar}E_{\alpha m}t}\hat{a}_{\alpha
m}(E_{\alpha m}).
\end{align}
With the above solution, the current flowing from contact
$\alpha$ to the mesoscopic sample can be deduced.
The current
operator of lead $\alpha$ is given by
\begin{align}
\hat{I}_{\alpha}(t)=\int
d\bm{r}_{\perp\alpha}\hat{j}(\bm{r}_{\alpha},t),
\end{align}
where the current density operator is expressed as
\begin{align}
\label{current_density}
\hat{j}(\bm{r},t)=\frac{\hbar}{2m^*i}[\hat{\Psi}^\dag\nabla\hat{\Psi}-(\nabla\hat{\Psi}^\dag)\hat{\Psi}].
\end{align}
Substituting the scattering states Eq.~(\ref{scatt_state_a}) and
Eq.~(\ref{scatt_state_b}) into the field operator of
Eq.~(\ref{field_opt_E}) gives the following form,
\begin{align}
\label{field_opt_E_final} \hat{\Psi}(\bm{r}_\alpha,t)=&\sum_{m} \! \int
\!\! \frac{dE_{\alpha m}}{\sqrt{h v_{\alpha m}(E_{\alpha
m})}}e^{-\frac{i}{\hbar}E_{\alpha m}t}\hat{a}_{\alpha m}(E_{\alpha
m})\notag\\&\times\Big\{\sum_n\big[\delta_{mn}\phi^+_{\alpha
n}+\sqrt{\frac{v_{\alpha m}}{v_{\alpha n}}}S_{\alpha \alpha
nm}\phi^-_{\alpha n}\big] \notag \\ &~~~~~~
+\sum_{\beta\neq\alpha}\sum_n\sqrt{\frac{v_{\alpha
m}}{v_{\beta n}}}S_{\beta \alpha nm}\phi^-_{\beta n}\Big\}\notag\\
=&\sum_{m}\!\int\!\! \frac{dE_{\alpha m}}{\sqrt{h v_{\alpha m}(E_{\alpha
m})}}e^{-\frac{i}{\hbar}E_{\alpha m}t} \notag \\ & \times \Big[\phi^+_{\alpha m}\hat{a}_{\alpha
m}(E_{\alpha m})+\phi^-_{\alpha m}\hat{b}_{\alpha m}(E_{\alpha
m})\Big],
\end{align}
where the contribution of the incoming (the first term) and outgoing
(the second term) states are explicitly presented in the above form.
Using the orthogonal properties of
different transverse modes, $\int
d\bm{r}_{\perp\alpha}\chi_{\alpha
m}(\bm{r}_{\perp\alpha})\chi_{\alpha
n}(\bm{r}_{\perp\alpha})=\delta_{mn}$, the current can be reduced to
the following form,
\begin{align}
\hat{I}_{\alpha}(t)=\frac{e}{h}\sum_m \!\int
\!\! & dEdE'\big[\hat{a}^\dag_{\alpha m}(E)\hat{a}_{\alpha
m}(E') \notag \\ & -\hat{b}^\dag_{\alpha m}(E)\hat{b}_{\alpha
m}(E')\big]e^{i(E-E')t/\hbar},
\end{align}
with the approximation $v_{\alpha m}(E)=v_{\alpha m}(E')$ which is
always valid for a slowly-varying function $v(E)$. From the
scattering relation $\bm{b}=\bm{S}\bm{a}$, one can express the
current as
\begin{align}
\label{I_scatt}
\hat{I}_{\alpha}(t)\!=\!\frac{e}{h}\!\! \sum_{\beta\gamma n k} \!\!\int
\!\! dEdE' \hat{a}^\dag_{\beta
n}\!(\!E)A^{nk}_{\beta\gamma}(\!\alpha;\!E,\!E'\!)\hat{a}_{\gamma
k}(\!E'\!)e^{\frac{i}{\hbar}(\!E-E'\!)t}
\end{align}
with the matrix $A$ having the following form
\begin{align}
A^{nk}_{\beta\gamma}(\alpha;E,E')=\delta_{\alpha\beta}
\delta_{\alpha\gamma}\delta_{kn}-\sum_mS^\dag_{\alpha\beta
nm}(E)S_{\alpha\gamma mk}(E').
\end{align}
Because contact $\alpha$ is in equilibrium, the average current at
lead $\alpha$ is then,
\begin{align}
\langle I_{\alpha}\rangle=\frac{e}{h}\sum_{\beta n}\int
dEA^{nn}_{\beta\beta}(\alpha,E,E)f_{\beta}(E),
\end{align}
where $f_{\alpha}(E)=1/(e^{(E-\mu_\alpha)/k_BT_\alpha}+1)$ is the
Fermi-Dirac distribution of contact $\alpha$ at the chemical
potential $\mu_\alpha$ and temperature $T_\alpha$.
Applying with the scattering matrix (\ref{scatt_mat}),
the average current of the left lead becomes
\begin{align}
\langle I_{L}\rangle=\frac{e}{h}\int dE~{\rm
Tr}[\bm{t}^\dag(E)\bm{t}(E)][f_L(E)-f_R(E)].
\end{align}
This gives the famous Landauer-B\"{u}ttiker
formula~\cite{Buttiker1992}. Here, $\bm{t}(E)=\bm{t}'(E)$ coming from
the time-reversal symmetry.

In the steady-state quantum transport regime, the
Landauer-B\"{u}ttiker approach is a powerful method to calculate
various transport properties in semiconductor
nanostructures~\cite{Buttiker1988,Buttiker1990,Samuelsson2006,Rothstein2009}.
However, the scattering theory considers
the reservoirs connecting to the scatterer (the mesoscopic sample)
to be always in equilibrium and electrons in the reservoir are
always incoherent. Thus, the Landauer-B\"{u}ttiker formula becomes
invalid to transient quantum transport. The scattering theory method
could be extended to deal with time-dependent transport phenomena,
through the so-called the Floquet scattering
theory~\cite{Moskalets2004,Wohlman2002,Moskalets2007}, but it is
only applicable to the case of the time-dependent quantum transport
for systems driven by periodic time-dependent external fields.

\subsection{Non-equilibrium Green function technique}\label{NEGF_appr}
Green function techniques are widely used in many-body systems. For
equilibrium systems, zero temperature Green functions and Matsubara
(finite temperature) Green functions are useful tools for
calculating the thermodynamical properties of many-body systems, as
well as the linear responses of systems under small time-dependent
(or not) perturbations \cite{Mahan1990}. However, when systems are
driven out of equilibrium, non-equilibrium Green functions are
utilized \cite{Haug2008,Wingreen1993,Jauho1994}. Non-equilibrium Green
function techniques are initiated by Scwinger \cite{Schwinger1961}
and Kadanoff and Baym \cite{Kadanoff1962}, and popularized by
Keldysh \cite{Keldysh1965}. To deal with non-equilibrium phenomena,
the contour-ordered Green functions which are defined on complex
time contours are introduced such that the equations of motion and
perturbation expansions of contour-ordered Green functions are
formally identical to that of usual equilibrium Green functions.

In this section, the contour-ordered Green functions defined on
Kadanoff-Baym contour which takes into account the initial correlations
and statistical boundary conditions will be discussed.
The real-time non-equilibrium Green functions are deduced from the
contour-ordered Green function by analytic continuation. In
application, one gives a detailed derivation of the
transport current for a mesoscopic system by means of the Keldysh
technique. The resulting transport current is formulating in terms
of the non-equilibrium Green functions of the device system, which
provides a more microscopic picture to the electron transport,
in comparison to the Landauer-B\"{u}ttiker formula, see Fig.~\ref{scheme3}. For a more complete
description of non-equilibrium Green function techniques, we refers
the readers to \cite{Haug2008,Kadanoff1962,Stefanucci2004}.

The contour-ordered Green function of non-equilibrium many-body
theory is defined as,
\begin{align}
\label{con_Green_fn_1} G(\bm{x},\tau;\bm{x}',\tau')& \!=\!-i\langle
T_C[\psi(\bm{x},\tau)\psi^\dag(\bm{x}',\tau')]\rangle \notag \\
& \!=\!-i{\rm
Tr}\big[\rho_{tot}(t_0)T_C\!\big\{\psi(\bm{x},\tau)\psi^\dag(\bm{x}'\!,\!\tau')\!\big\}\big],
\end{align}
where $\psi(\bm{x},\tau)$ and $\psi^\dag(\bm{x}',\tau')$ are the
fermion field operators in the Heisenberg picture with time
variables $\tau$,$\tau'$ (denoted by Greek letters) defined on the
complex contour $C$, and $T_C$ is a contour-ordering operator which
orders the operators according to their time labels on the contour:
\begin{align}
T_C[\psi(\bm{x},\tau)\psi^\dag(\bm{x}',\tau')]
\equiv \left\{
\begin{array}{ll}
\psi(\bm{x},\tau)\psi^\dag(\bm{x}',\tau') &  \tau >_C \tau', \\
-\psi^\dag(\bm{x}',\tau')\psi(\bm{x},\tau) & \tau <_C \tau'.
\end{array} \right.
\end{align}
From the above definition, it is straightforward to rewrite the
contour-ordered Green function as,
\begin{align}
\label{con_Green_fn_2}
G(\bm{x},\tau;\bm{x}',\tau')= & \Theta_C(\tau-\tau')G^>(\bm{x},\tau;\bm{x}',\tau') \notag \\
& +\Theta_C(\tau'-\tau)G^<(\bm{x},\tau;\bm{x}',\tau'),
\end{align}
where $\Theta_C(\tau-\tau')$ is the step function defined on the
contour in a clockwise direction, and $G^>$ and $G^<$ are the greater and lesser Green
functions, respectively. The configuration of complex contour $C$ is
determined by the initial density matrix of the total system
$\rho_{tot}(t_0)$.

\begin{figure}
\centerline{\scalebox{0.5}{\includegraphics{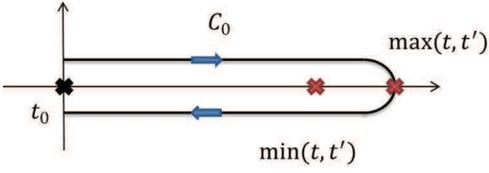}}}
\caption{Closed time path contour $C_0$} \label{contour1}
\end{figure}

The non-equilibrium dynamics considered in the Kadanoff-Baym
formalism is formulated as follows. The physical
system is described by a time-independent Hamiltonian,
\begin{align}
\label{h} h=H_0+H_i,
\end{align}
where $H_0$ represents a free Hamiltonian, and $H_i$ is the
interaction between the particles. The system is initially assumed
at thermal equilibrium, which means the system is in partition-free
scheme~\cite{Cini1980},
\begin{align}
\rho_{tot}(t_0)=\frac{\exp(-\beta h)}{{\rm Tr} [\exp(-\beta
h)]}=\frac{1}{Z}\exp(-\beta h),
\end{align}
where $\beta=1/k_B T$, and the particle energies are measured from
the chemical potential $\mu$. After $t=t_0$, the system is exposed
to external disturbances, e.g. an electric field, a light excitation
pulse, or a coupling to contacts at differing (electro) chemical
potentials that are described by time-dependent Hamiltonian $H'(t)$.
Thus, the total Hamiltonian is,
\begin{align}
H(t)=h+H'(t),
\end{align}
where $H'(t<t_0)=0$. By choosing the time arguments in contour-ordered
Green function (\ref{con_Green_fn_1}) are real time variables $t$
and $t'$, the field operator $\psi(\bm{x},t)$ is then,
\begin{align}
\psi(\bm{x},t)\equiv
\psi(1)=\mathcal{U}(t_0,t)\hat{\psi}_h(1)\mathcal{U}(t,t_0),
\end{align}
where  the shorthand notation $(1)=(\bm{x},t)$ has been used, and
$\mathcal{U}$ is the evolution operator for the time-dependent Hamiltonian
$H'(t)$,
\begin{align}
\mathcal{U}(t,t')=T\big\{\exp\big[-i\int^t_{t'}dt_1\hat{H}'_h(t_1)\big]\big\}.
\end{align}
In the above equations, $\hat{\psi}_h(1)$ and $\hat{H}'_h(t)$ are
operators in the interaction picture with respect to Hamiltonian
$h$,
\begin{subequations}
\begin{align}
\hat{\psi}_h(1)=&~e^{ih(t-t_0)}\psi(\bm{x},t_0)e^{-ih(t-t_0)},\\
\hat{H}'_h(t)=&~e^{ih(t-t_0)}H'(t)e^{-ih(t-t_0)}.
\end{align}
\end{subequations}
The contour-ordered Green function in Kadanoff-Baym formalism is now
written as
\begin{align}
iG(\!1,\!1'\!)\!=&{\rm
Tr}\big[\rho_{tot}(t_0\!)T_C\big\{\mathcal{U}(t_0,\!t\!)\hat{\psi}_h(1)\mathcal{U}(t,t'\!)\hat{\psi}^\dag_h(1'\!)\mathcal{U}(t'\!,\!t_0\!)\!\big\}\big]\notag\\
=&{\rm
Tr}\big[\rho_{tot}(t_0)T_{C_0}\big\{\mathcal{U}_{C_0}(t_0,t_0)\hat{\psi}_h(1)\hat{\psi}^\dag_h(1')\!\big\}\big],
\end{align}
where
$\mathcal{U}_{C_0}(t_0,t_0)=T_{C_0}\{\exp[-i\oint_{C_0}d\tau_1\hat{H}'_h(\tau_1)]\}$
is the evolution operator defined on the close path contour $C_0$ as shown in Fig.~\ref{contour1}.

\begin{figure}
\centerline{\scalebox{0.5}{\includegraphics{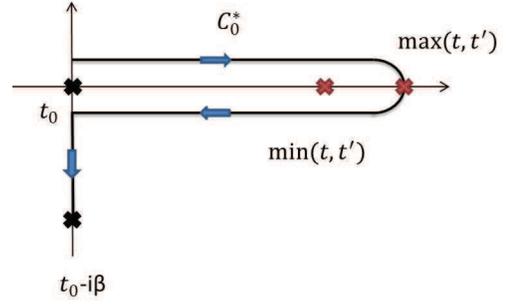}}}
\caption{Kadanoff-Baym contour $C^*_0$} \label{contour2}
\end{figure}

In order to perform the Wick theorem, one needs to further transform
the operators $\hat{\psi}_h(1)$, $\hat{\psi}^\dag_h(1')$ in the
interaction picture with respect to the free Hamiltonian $H_0$:
\begin{align}
\hat{\psi}_h(1)=U(t_0,t)\hat{\psi}(1)U(t,t_0).
\end{align}
Here, $U$ being the evolution operator for the interaction Hamiltonian $H_i$,
\begin{align}
U(t,t')=T\big\{\exp\big[-i\int^t_{t'}dt_1\hat{H}_i(t)\big]\big\},
\end{align}
and $\hat{\psi}(1)$ and $\hat{H}_i(t)$ in the interaction picture
are given by,
\begin{subequations}
\begin{align}
\hat{\psi}(1)=&e^{iH_0(t-t_0)}\psi(\bm{x},t_0)e^{-iH_0(t-t_0)},\\
\hat{H}_i(t)=&e^{iH_0(t-t_0)}H_ie^{-iH_0(t-t_0)}.
\end{align}
\end{subequations}
Furthermore, one can rewrite the factor
$\exp(-\beta h)$ in the initial density matrix,
\begin{align}
\exp(-\beta h)=\exp(-\beta H_0)U(t_0-i\beta,t_0).
\end{align}
Finally, the contour-ordered Green function is reduced to,
\begin{align}
\label{con_Green_fn_Kadanoff} iG(1,1'\!) =&\frac{{\rm
Tr}\big\{\rho_0T_{C^*_0}\big[U_{C^*_0}(t_0\!-\!i\beta,t_0)\mathcal{U}_{C_0}(t_0,t_0)\hat{\psi}(1\!)\hat{\psi}^\dag(1')\big]\!\big\}}{{\rm
Tr}\big\{\rho_0T_{C^*_0}\big[U_{C^*_0}(t_0-i\beta,t_0)\mathcal{U}_{C_0}(t_0,t_0)\big]\big\}},
\end{align}
where $\rho_0=e^{-\beta H_0}/Z_0$ is the equilibrium density matrix
of Hamiltonian $H_0$, and
$U_{C^*_0}(t_0-i\beta,t_0)=T_{C^*_0}\{\exp[-i\int_{C^*_0}d\tau_1\hat{H}_i(\tau_1)]\}$
is the evolution operator defined on contour $C^*_0=C_0\cup[t_0,t_0-i\beta]$ which
is the Kadanoff-Baym contour shown in Fig.~\ref{contour2}.

Eq.~(\ref{con_Green_fn_Kadanoff}) is the exact contour-ordered Green
function in Kadanoff-Baym formalism, which is defined in the
interaction picture with respect to the free Hamiltonian $H_0$, so
that Wick theorem is always applicable. Thus, the perturbative
evaluation of Eq.~(\ref{con_Green_fn_Kadanoff}) could be put in a
form analogous to the usual Feynman diagrammatic technique as in the
equilibrium Green function techniques, which leads to the Keldysh formalism.

On the other hand, the contour-ordered Green function obeys the following Dyson
equation,
\begin{subequations}
\label{Dyson_eq_con_1}
\begin{align}
G(1,1')=G_0(1,1')+\!\! \int \!\! d2\!\! \int \!\! d3G_0(1,2)\Sigma(2,3)G(3,1'),\\
G(1,1')=G_0(1,1')+\!\! \int \!\! d2\!\! \int \!\! d3 G(1,2)\Sigma(2,3)G_0(3,1'),
\end{align}
\end{subequations}
where $G_0(1,1')=-i\langle
T_C[\hat{\psi}(1)\hat{\psi}^\dag(1')]\rangle$ is the unperturbed
Green function, $\Sigma(2,3)$ is the one particle irreducible
self-energy, and the integral sign $\int d2 (3)$ denotes a sum over
all integral variables. The equations can be simply written as,
\begin{subequations}
\label{Dyson_eq_con_2}
\begin{align}
G=G_0+G_0\Sigma G,\\
G=G_0+G\Sigma G_0.
\end{align}
\end{subequations}
The Dyson equation can be regarded as the Schr\"{o}dinger equation
of a particle in the medium subject to the self-energy as the
potential. In the Dyson equation, the single-particle Green function
is entirely determined by the self-energy which contains all the
many-body effects.

The steady-state transport
current in a mesoscopic system is presented by the Keldysh
formalism. Consider a nanostructure consisting of a
quantum device coupled with two leads (the source and drain), which
can be described by the Fano-Anderson  Hamiltonian~\cite{Fano1961,
Anderson1958,Mahan1990},
\begin{align}
H = \sum_{ij}\varepsilon_{ij}a_{i}^{\dag}a_{j} + \sum_{\alpha
k}\epsilon_{\alpha k}c_{\alpha k}^{\dag}c_{\alpha k} \notag\\ +\sum_{i\alpha
k}[V_{i \alpha k}a^{\dag }_ic_{\alpha k} + {\rm H.c.}].
\end{align}
Here $a_{i}^{\dag}$ ($a_{i}$) and $c_{\alpha k}^{\dag}$ ($c_{\alpha
k}$) are creation (annihilation) operators of electrons in the
quantum device and lead $\alpha$, respectively, $\varepsilon_{ij}$
and $\epsilon_{\alpha k}$ are the corresponding energy levels,
$V_{i\alpha k}$ is the tunneling amplitude between the orbital state
$i$ of the device system and the orbital state $k$ of lead $\alpha$.
In the Keldysh approach, the quantum device and the leads are
decoupled in the remote past, and the tunneling between them is
viewed as a perturbation. Then, the time-independent Hamiltonian $h$,
time-dependent Hamiltonian $H'(t)$, and the initial density matrix
$\rho_{tot}(-\infty)$ of the total system are,
\begin{subequations}
\label{Hamiltonian_Keldysh}
\begin{align}
&h = H_0 =
\sum_{ij}\varepsilon_{ij}a_{i}^{\dag}a_{j} +\sum_{\alpha
k}\epsilon_{\alpha k}c_{\alpha k}^{\dag}c_{\alpha k},\\
&H'(t>t_0)= \sum_{i\alpha k}[V_{i \alpha k}a^{\dag }_ic_{\alpha k}
+ {\rm H.c.}],\\
&\rho_{tot}(-\infty)=\rho_L \otimes
\rho(-\infty) \otimes  \rho_R ,
\end{align}
\end{subequations}
where $H_\alpha=\sum_k\epsilon_{\alpha k}c^\dag_{\alpha k}c_{\alpha
k}$ and $N_\alpha=\sum_k c^\dag_{\alpha k}c_{\alpha k}$ are the
Hamiltonian and the total particle number of lead $\alpha$,
respectively. Lead $\alpha$ is initially in thermal equilibrium
$\rho_\alpha = \frac{1}{Z}\big[e^{-\beta_\alpha(H_\alpha-\mu_\alpha N_\alpha)}\big]$ with
inverse temperature $\beta_\alpha$ and chemical potential
$\mu_\alpha$. The device system can be in an arbitrary state
$\rho(-\infty)$, i.e. the total system is in the partitioned
scheme~\cite{Caroli1971a,Caroli1971b}.

In non-equilibrium Green function techniques, the information of the
dissipation and fluctuation dynamics of the device system can be
extracted from the contour-ordered Green function of the device
system $G_{ij}(\tau,\tau')$,
\begin{align}
G_{ij}(\tau,\tau')=&-i\langle T_C[ a_i(\tau)a^\dag_j(\tau')]
\rangle\notag\\
=&\Theta_C(\tau-\tau')G^>_{ij}(\tau,\tau')+\Theta_C(\tau'-\tau)G^<_{ij}(\tau,\tau').
\end{align}
This Green function obeys the following equations of motion,
\begin{subequations}
\label{EOM_con_Green_fn}
\begin{align}
\sum_l & \Big[i\frac{\partial}{\partial\tau}\delta_{il}-\varepsilon_{il}\Big] G_{lj}(\tau,\tau')\notag \\
& =\delta_C(\tau-\tau')\delta_{ij}+\sum_{\alpha
k}V_{i\alpha k}G_{\alpha k,j}(\tau,\tau'),\\
\sum_l & \Big[-i\frac{\partial}{\partial\tau'}\delta_{lj}-\varepsilon_{lj}\Big] G_{il}(\tau,\tau') \notag \\
& =\delta_C(\tau-\tau')\delta_{ij}+\sum_{\alpha
k}G_{i,\alpha k}(\tau,\tau')V^*_{j\alpha k},
\end{align}
\end{subequations}
where the mixed contour-ordered Green functions, $G_{\alpha k,
j}(\tau,\tau')=-i\langle T_C [c_{\alpha
k}(\tau)a^\dag_j(\tau')]\rangle$ and $G_{i,\alpha
k}(\tau,\tau')=-i\langle T_C[a_i(\tau)c^\dag_{\alpha
k}(\tau')]\rangle$, are given as follows 
\begin{subequations}
\label{mix_G}
\begin{align}
 G_{i,\alpha
k}(\tau,\tau')=&\sum_l\int_Cd\tau_1G_{il}(\tau,\tau_1)V_{l\alpha
k}g_{\alpha k}(\tau_1,\tau'),\\
G_{\alpha k,j}(\tau,\tau')=&\sum_l\int_Cd\tau_1g_{\alpha
k}(\tau,\tau_1)V^*_{l\alpha k}G_{lj}(\tau_1,\tau').
\end{align}
\end{subequations}
Inserting Eq.~(\ref{mix_G}a) and (\ref{mix_G}b) into
Eq.~({\ref{EOM_con_Green_fn}}) gives the Dyson equation in the
differential form, i.e. Kadanoff-Byam equation,
\begin{subequations}
\label{Dyson_eq_diff}
\begin{align}
[i\partial_\tau\mathds{1}_D-\bm{\varepsilon}]\bm{G}(\tau,\tau')& =\mathds{1}_D\delta_C(\tau-\tau')\notag \\
& +\int_Cd\tau_1\bm{\Sigma}(\tau,\tau_1)\bm{G}(\tau_1,\tau')\\
\bm{G}(\tau,\tau')[-i\partial_{\tau'}\mathds{1}_D-\bm{\varepsilon}] & =\mathds{1}_D\delta_C(\tau-\tau')\notag \\
& +\int_Cd\tau_1\bm{G}(\tau,\tau_1)\bm{\Sigma}(\tau_1,\tau'),
\end{align}
\end{subequations}
with self energy,
\begin{align}
\Sigma_{ij}(\tau,\tau')=\sum_\alpha\Sigma_{\alpha
ij}(\tau,\tau')=\sum_{\alpha k}V_{i\alpha k}g_{\alpha
k}(\tau,\tau')V^*_{j\alpha k}.
\end{align}
Here, $\mathds{1}_D$ is an identity matrix in the dimension of the
device system. On the other hand, the equation of
unperturbed contour-ordered Green function of the device system is,
\begin{subequations}
\begin{align}
&[i\partial_\tau\mathds{1}_D-\bm{\varepsilon}]\bm{G}_0(\tau,\tau')=\mathds{1}_D\delta_C(\tau-\tau')\\
&\bm{G}_0(\tau,\tau')[-i\partial_{\tau'}\mathds{1}_D-\bm{\varepsilon}]=\mathds{1}_D\delta_C(\tau-\tau').
\end{align}
\end{subequations}
Consequently, the Dyson equation (\ref{Dyson_eq_diff}) can be rewritten
in the following form,
\begin{subequations}
\label{Dyson_eq_con_Keldysh}
\begin{align}
&\bm{G}^{-1}_0\bm{G}=\mathds{1}+\bm{\Sigma}\bm{G},\\
&\bm{G}\bm{G}^{-1}_0=\mathds{1}+\bm{G}\bm{\Sigma}.
\end{align}
\end{subequations}
Here, the matrix product means a product of all the internal
variables (energy level and time). Equation
(\ref{Dyson_eq_con_Keldysh}) reproduces the integral form of Dyson
equation (\ref{Dyson_eq_con_2}).

Using the Dyson equation (\ref{Dyson_eq_con_2}) and the Langreth
theorem \cite{Langreth1976}, one has
\begin{subequations}
\label{real_time_Green_fn_Keldysh}
\begin{align}
&\bm{G}^{R,A}=\bm{G}^{R,A}_0+\bm{G}^{R,A}_0\bm{\Sigma}^{R,A}\bm{G}^{R,A},\\
&\bm{G}^{\gtrless}=\bm{G}^{\gtrless}_0+\bm{G}^{R}_0\bm{\Sigma}^R\bm{G}^{\gtrless}+\bm{G}^{R}_0\bm{\Sigma}^{\gtrless}\bm{G}^{A}+\bm{G}^{\gtrless}_0\bm{\Sigma}^{A}\bm{G}^{A}.
\end{align}
\end{subequations}
One can further iterates Eq.~(\ref{real_time_Green_fn_Keldysh}b)
respect to $\bm{G}^{\gtrless}$ and obtains,
\begin{align}
\bm{G}^{\gtrless}=&(1\!+\!\bm{G}^R_0\bm{\Sigma}^R)\bm{G}^{\gtrless}_0(1\!+\!\bm{\Sigma}^A\bm{G}^A)\notag \\
& \!+\!(\bm{G}^{R}_0\!+\!\bm{G}^R_0\bm{\Sigma}^R\bm{G}^{R}_0)\bm{\Sigma}^{\gtrless}\bm{G}^{A}
\!+\!\bm{G}^R_0\bm{\Sigma}^{R}\bm{G}^{R}_0\bm{\Sigma}^R\bm{G}^{\gtrless}.
\end{align}
After iterates infinite orders, one can get,
\begin{align}
\bm{G}^{\gtrless}=(1+\bm{G}^R\bm{\Sigma}^R)\bm{G}^{\gtrless}_0(1+\bm{\Sigma}^A\bm{G}^A)+\bm{G}^{R}\bm{\Sigma}^{\gtrless}\bm{G}^{A}.
\end{align}
In the Keldysh technique, the first term is neglected because it
usually vanishes at steady-state limit. Then,
\begin{align}
\label{G^<_Keldysh}
\bm{G}^{\gtrless}=\bm{G}^{R}\bm{\Sigma}^{\gtrless}\bm{G}^{A}.
\end{align}
Eq.~(\ref{real_time_Green_fn_Keldysh}a) and Eq.~(\ref{G^<_Keldysh})
are the final results of real time non-equilibrium Green functions
in the Keldysh formalism which all the transport properties are
determined by.

The transport current from lead $\alpha$ to the device system is
defined as,
\begin{align}\label{current_NEGF}
I_\alpha(t)& \equiv -e\frac{d}{dt}\langle N_\alpha(t)
\rangle =-\frac{ie}{\hbar}\langle [ H, N_\alpha ]\rangle\notag \\
& =\frac{2e}{\hbar}{\rm
Re}\sum_{ik}V^*_{i\alpha k}G^<_{i,\alpha k}(t,t),
\end{align}
where the mixed lesser Green function, $G^<_{i,\alpha
k}(t,t')=i\langle c^\dag_{\alpha k}(t')a_i(t)\rangle$, can be
obtained by applying the Langreth theorem to the mixed
contour-ordered Green function (\ref{mix_G}a),
\begin{align}
G^<_{i,\alpha
k}(t,t')=\sum_j\int^{\infty}_{-\infty} & \!\!\! dt_1\big[G^R_{ij}(t,t_1)V_{j\alpha
k}g^<_{\alpha k}(t_1,t') \notag \\
& +G^<_{ij}(t,t_1)V_{j\alpha k}g^A_{\alpha
k}(t_1,t')\big].
\end{align}
In the steady-state limit, all the Green functions usually depend on
the differences of time arguments, i.e. $G(t,t')=G(t-t')$ because of
time translation symmetry. Thus Green function
$G^<_{i,\alpha k}(t,t)$ in the frequency domain can be expressed as,
\begin{align}\label{mix_Green}
G^<_{i,\alpha k}(t,t)=\sum_j\int
\frac{d\omega}{2\pi} & \big[G^R_{ij}(\omega)V_{j\alpha k}g^<_{\alpha
k}(\omega) \notag \\
& +G^<_{ij}(\omega)V_{j\alpha k}g^A_{\alpha k}(\omega)\big],
\end{align}
Combinding Eq.~(\ref{current_NEGF}) and (\ref{mix_Green}), the
steady-state transport current reduces to,
\begin{align}
I_\alpha\!=\! \frac{ie}{\hbar} \! \int \!\! \frac{d\omega}{2\pi}{\rm
Tr}\bm{\Gamma}_\alpha(\omega)\big\{f_\alpha(\omega)[\bm{G}^R(\omega)
\!-\!\bm{G}^A(\omega)]\!+\!\bm{G}^<(\omega)\big\},
\end{align}
where $\Gamma_{\alpha ij}(\omega)=2\pi\sum_k V_{i\alpha
k}V^*_{j\alpha k}\delta(\omega-\epsilon_{\alpha k})$ is a
level-width function, and we have used the results of the
free-particle Green functions given \cite{Haug2008}. Now,
the transport current is fully determined by Green functions of the
device system. Besides, for the non-interacting device system, one
has,
\begin{align}
\label{G^R-G^A}
\bm{G}^R(\omega)-\bm{G}^A(\omega)& =\bm{G}^>(\omega)-\bm{G}^<(\omega) \notag \\
& =-i\bm{G}^R(\omega)\sum_\alpha
\bm{\Gamma}_\alpha(\omega)\bm{G}^A(\omega).
\end{align}
Then, the steady-state transport current
becomes,
\begin{align}
I_\alpha=\frac{e}{\hbar}\sum_\beta\int\frac{d\omega}{2\pi}T_{\alpha\beta}(\omega)[f_\alpha(\omega)-f_{\beta}(\omega)],
\end{align}
where $T_{\alpha\beta}={\rm
Tr}\big[\bm{\Gamma}_\alpha(\omega)\bm{G}^R(\omega)\bm{\Gamma}_\beta(\omega)\bm{G}^A(\omega)\big]$
is the transmission coefficient. This expression of the steady-state
transport current reproduces the Landauer-B\"{u}ttiker formula with
the transmission probability being derived microscopically. The
non-equilibrium Green function technique based on Keldysh
formalism~\cite{Schwinger1961,Keldysh1965} has been used extensively
to investigate the steady-state quantum transport in mesoscopic
systems~\cite{Wingreen1993,Haug2008,Jauho1994}.

Wingreen {\it et al.}~extended Keldysh's non-equilibrium Green
function technique to time-dependent quantum transport under
time-dependent external bias and gate voltages~\cite{Wingreen1993,Jauho1994}.
Explicitly, the parameters in Hamiltonian
(\ref{Hamiltonian_Keldysh}), controlled by the external bias and
gate voltages, become time dependent,
\begin{subequations}
\begin{align}
&\varepsilon_{ij}\rightarrow\varepsilon_{ij}(t),\\
&\epsilon_{\alpha k}\rightarrow\epsilon_{\alpha
k}(t)=\epsilon_{\alpha k}+\Delta_{\alpha k}(t),\\
&V_{i\alpha k}\rightarrow V_{i\alpha k}(t)
\end{align}
\end{subequations}
Then, the time-dependent transport current becomes,
\begin{align}
\label{I_NEGF_time_dep}
I_\alpha(t)=-\frac{2e}{\hbar}\!\!\int^t_{-\infty}& \!\!\!d\tau \!\!\!\int\frac{d\omega}{2\pi}{\rm
ImTr}\big\{e^{-i\omega(\tau-t)}\bm{\Gamma}_{\alpha}(\omega,\tau,t)\notag\\&\times\big[f_\alpha(\omega)\bm{G}^R(t,\tau)+\bm{G}^<(t,\tau)\big]\big\},
\end{align}
where the level-width function is also time dependent,
\begin{align}
\Gamma_{\alpha ij}&(\omega,t_1,t_2) \notag\\
=&2\pi\!\sum_k \!V_{i\alpha
k}(t_1) e^{\!-i\!\int^{t_1}_{t_2}ds\Delta_{\alpha
k}(s)} V^*_{j\alpha k}(t_2) \delta(\omega-\epsilon_{\alpha k}).
\end{align}
In particular, the Green functions in time domain are given by,
\begin{subequations}
\label{Green_fn_time_dep}
\begin{align}
&\bm{G}^R(t,t'\!)\!\!=\!\bm{G}^R_0(t,t')\!+\!\!\!\int\!\! dt_1\!\!\!\int\!\!\! dt_2
\bm{G}^R_0(\!t,t_1\!)\bm{\Sigma}^R(\!t_1,t_2\!)\bm{G}^R(\!t_2,t'\!),\\
&\bm{G}^<(t,t')\!\!=\!\!\!\int\!\! dt_1\!\!\!\int\!\!\! dt_2
\bm{G}^R(t,t_1)\bm{\Sigma}^<(t_1,t_2)\bm{G}^A(t_2,t').
\end{align}
\end{subequations}
with self-energy defined as,
\begin{subequations}
\begin{align}
&\Sigma^R_{ij}(t_1,t_2)\!=\!-i\Theta(t_1\!-\!t_2)\!\!\sum_{\alpha}\!\!
\int\!\!\frac{d\omega}{2\pi}\Gamma_{\alpha ij
}(\omega,t_1,t_2)e^{-i\omega(t_1-t_2)},\\
&\Sigma^<_{ij}(t_1,t_2)=i\sum_\alpha
\int\frac{d\omega}{2\pi}f_\alpha(\omega)\bm{\Gamma}_{\alpha
}(\omega,t_1,t_2)e^{-i\omega(t_1-t_2)}
\end{align}
\end{subequations}
This gives a general formalism for time-dependent current through
the device system valid for non-linear response, where electron
energies can be varied time-dependently by external gate voltages.
However, in the Keldysh formalism, non-equilibrium Green functions
are defined with the initial time $t_0 \rightarrow -\infty$, where
the initial correlations are hardly taken into account. This limits
the Keldysh technique to be useful mostly in the non-equilibrium
steady-state regime.

\subsection{Master equation approach}\label{Master_Eq_appr}
\begin{figure}
\centerline{\scalebox{0.45}{\includegraphics{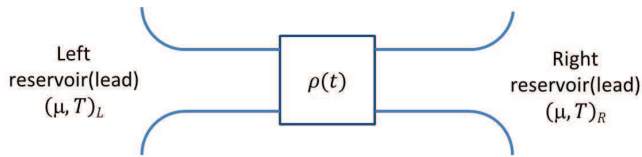}}}
\caption{Theoretical scheme for master equation approach}
\label{scheme3}
\end{figure}

The master equation approach concerns the dynamic properties of the
device system in terms of the time evolution of the reduced density
matrix $\rho(t)={\rm Tr_E}[\rho_{tot}(t)]$, where $\rm{Tr_E}$ is the
trace over all the environmental (leads) degrees of freedom. The
dissipation and fluctuation dynamics of the device system induced by
the reservoirs (leads) are fully manifested in the master equation.
The transient transport properties can be naturally addressed within
the framework of the master equation. Compared to the
non-equilibrium Green function technique, the master equation
approach manifests the state information of the device system, see Fig.~\ref{scheme3},
which is a key element in studying quantum phenomena.

In principle, the master equation for quantum transport can be
solved in terms of the real-time diagrammatic expansion approach up
to all orders~\cite{Schoeller1994}. However, most of the master
equations used in nanostructures are obtained by the perturbation
theory up to the second order of the system-lead couplings, which is
mainly applicable in the sequential tunneling regime~\cite{Li2005}.
A recent development of master equations in quantum transport
systems is the hierarchical expansion of the equations of motion for
the reduced density matrix~\cite{Jin2008,Yan2016}, which provides a
systematical and also very useful numerical calculation scheme for
quantum transport.

A few years ago, we derived an exact master equation for
non-interacting nanodevices~\cite{Tu2008,Tu2009,Jin2010}, using the
Feynman-Vernon influence functional approach~\cite{Feynman1963} in
the fermion coherent-state representation~\cite{Zhang1990}. The
obtained exact master equation not only describes the quantum state
dynamics of the device system but also takes into account all the
transient electronic transport properties. The transient transport
current is obtained directly from the exact master
equation~\cite{Jin2010}, which turns out to be expressed precisely
with the non-equilibrium Green functions of the device
system~\cite{Haug2008,Wingreen1993,Jauho1994}. This new theory has also been
used to study quantum transport (including the transient transport)
for various nanostructures recently~\cite{Tu2008,Tu2009,Jin2010,Tu2011,Lin2011,Tu2012a,Tu2012b,Jin2013,Tu2014,Yang2014,Yang2015,Tu2016,Liu2016}. In
the following, an introduction of this
exact master equation approach~\cite{Tu2008,Tu2009,Jin2010} is given,
and the transient transport current derived using the exact master
equation is explicitly presented.

\begin{figure}
\centerline{\scalebox{0.6}{\includegraphics{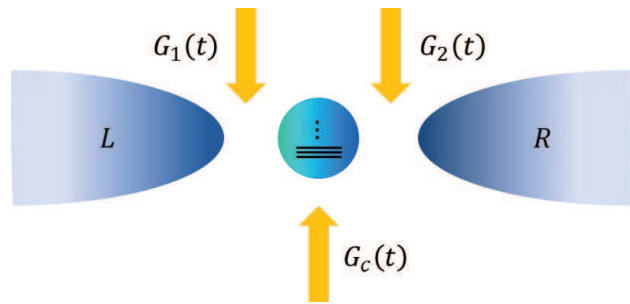}}} \caption{A
schematic plot of a nanoscale quantum device in which the bias
voltage is applied to the source and the drain electrode leads
labeled $L$ and $R$, and other gates labeled $G_c$, $G_1$, $G_2$
control the energy levels of the central region as well as the
couplings between the central region and the leads.} \label{QT_ME}
\end{figure}
We begin with a nanostructure consisting of a quantum device coupled
with two leads (the source and the drain), described by a time-dependent
Fano-Anderson Hamiltonian~\cite{Fano1961, Anderson1958,Mahan1990},
\begin{align}
H(t)=&H_S(t)+H_E(t)+H_{SE}(t),\notag
\end{align}
with
\begin{align}\label{Hamiltonian_Mas}
&H_S(t)=\sum_{ij}\bm{\varepsilon}_{ij}(t) a^\dag_i a_j,\notag\\
&H_E(t)=\sum_{\alpha k}\epsilon_{\alpha k}(t) c_{\alpha k}^\dag
c_{\alpha k},\notag\\
&H_{SE}(t)=\sum_{i \alpha k} \big[V_{i \alpha k}(t)a^\dag_i
c_{\alpha k}+V^*_{i \alpha k}(t)c_{\alpha k}^\dag a_i \big],
\end{align}
where $a_{i}^{\dag}$ ($a_{i}$) and $c_{\alpha k}^{\dag}$ ($c_{\alpha
k}$) are creation (annihilation) operators of electrons in the
device system and lead $\alpha$, respectively;
$\bm{\varepsilon}_{ij}(t)$ and $\epsilon_{\alpha k}(t)$ are the
corresponding energy levels, and $V_{i\alpha k}(t)$ is the tunneling
amplitude between the orbital state $i$ in the device system and the
orbital state $k$ in lead $\alpha$. These time-dependent parameters
in Eq.~(\ref{Hamiltonian_Mas}) can be manipulated by external bias
and gate voltages in experiments (see Fig.~\ref{QT_ME}).

The density matrix of the total system follows the unitary evolution,
\begin{align}
\rho_{tot}(t)=\mathbf{U}(t,t_0)\rho_{tot}(t_0)\mathbf{U}^\dag(t,t_0),
\end{align}
with the evolution operator
$\mathbf{U}(t,t_0)=T\exp\{-i\int^t_{t_0}H(\tau)d\tau\}$, where $T$ is the
time-ordering operator. Here we assume, as usual, that
the device system is uncorrelated with the reservoirs
(leads) before the tunneling couplings are turned
on~\cite{Leggett1987}: $\rho_{tot}(t_0)=\rho(t_0) \otimes
\rho_E(t_0)$, in which the system can be in an arbitrary state
$\rho(t_0)$, and the reservoirs are initially in equilibrium,
$\rho_E(t_0)=\frac{1}{Z}e^{-\sum_\alpha \beta_\alpha
(H_\alpha-\mu_\alpha N_\alpha)}$, where
$\beta_\alpha=1/(k_BT_\alpha)$ is the inverse temperature of lead
$\alpha$ at initial time $t_0$, and $N_\alpha=\sum_k c^\dag_{\alpha
k}c_{\alpha k}$ is the total particle number for lead $\alpha$. In
other words, the system is in the so-called partitioned
scheme~\cite{Caroli1971a,Caroli1971b} as in the Keldysh framework. After $t_0$,
the device system and the leads evolve into dynamically
non-equilibrium states. These dynamically non-equilibrium processes
are fully taken into account when we completely and exactly
integrated over all the dynamical degrees of freedom of leads
through the Feynman-Vernon influence functional. Here we do not need
to specify or assume the lead distribution function after the
initial time, since the quantum evolution operator of the total
system (the dot, the leads and the coupling between them) in the
Feynman-Vernon influence functional theory has automatically taken
into account all possible states of the leads.

The non-equilibrium electron dynamics of an open system are
determined by the reduced density matrix: $\rho(t)={\rm
Tr_E}[\rho_{tot}(t)]$. In the fermion coherent-state
representation~\cite{Zhang1990}, the reduced density matrix at an
arbitrary later time $t$ is expressed as,
\begin{align}\label{red_dens_cohe}
\langle \bm{\xi}_f\!|\rho(t)\!|\bm{\xi}'_f\rangle\!=\!\!\!\int\!\!\!
d\mu(\bm{\xi}_0\!)d\mu(\!\bm{\xi}'_0\!)\langle\bm{\xi}_0|\rho(t_0)|\bm{\xi}'_0\!\rangle
\mathcal{J}\!(\bar{\bm{\xi}}_f\!,\!\bm{\xi}'_f\!,\!t|\bm{\xi}_0,\!\bar{\bm{\xi}}'_0,\!t_0),
\end{align}
with $\bm{\xi} = (\xi_1, \xi_2, . . .)$ and $\bar{\bm{\xi}}=
(\xi^*_1, \xi^*_2, . . .)$ being the Grassmann variables and their
complex conjugate defined through the fermion coherent states:
$a_i|\bm{\xi}\rangle =\xi_i|\bm{\xi}\rangle $and $\langle \bm{\xi}|
a^\dag_i =\langle \bm{\xi}|\xi^*_i $. As these coherent states obey
the completeness relation, $\int
d\mu(\bm{\xi})|\bm{\xi}\rangle\langle\bm{\xi}|=\mathds{1}$, where
the integration measure is defined by
$d\mu(\bm{\xi})=\prod_ie^{-\xi^*_i\xi_i}d\xi^*_id\xi_i$. The
propagating function in equation (\ref{red_dens_cohe}) is given in
terms of Grassmann variable path integrals,
\begin{align}
\mathcal{J}\!(\bar{\bm{\xi}}_f\!,\!\bm{\xi}'_f\!,\!t|\bm{\xi}_0,\!\bar{\bm{\xi}}'_0,\!t_0\!)
\!&=\!\!\!\int\!\!\mathcal{D}[\bar{\bm{\xi}}\bm{\xi};\!\bar{\bm{\xi}}'\!\bm{\xi}']e^{i(\!S_c[\bar{\bm{\xi}},\bm{\xi}]-S^*_c[\bar{\bm{\xi}}'\!,\bm{\xi}'])}
\!\mathcal{F}[\bar{\bm{\xi}}\bm{\xi};\!\bar{\bm{\xi}}'\!\bm{\xi}']\notag\\
&=\!\!\!\int\!\!\mathcal{D}[\bar{\bm{\xi}}\bm{\xi};\bar{\bm{\xi}}'\bm{\xi}']e^{iS_{eff}[\bar{\bm{\xi}}\bm{\xi};\bar{\bm{\xi}}'\bm{\xi}']},
\end{align}
where $S_c[\bar{\bm{\xi}},\bm{\xi}]$ and
$S^*_c[\bar{\bm{\xi}}'\bm{\xi}']$ are respectively the forward and
backward actions of the device system in the fermion coherent-state
representation. The influence functional
$\mathcal{F}[\bar{\bm{\xi}}\bm{\xi};\bar{\bm{\xi}}'\bm{\xi}']$ takes fully into account the
back-action effects of the environments (leads) to the device
system, it modifies the original action of the device system into an
effective one, which dramatically changes the dynamics of the device
system. After integrating out all the environmental degrees of
freedom, the influence functional has the following form,
\begin{align}
\mathcal{F}[\bar{\bm{\xi}}&\bm{\xi};\bar{\bm{\xi}}'\bm{\xi}']\notag\\=&\!\exp\!\Big\{\!\!\!-\!\!\!\sum_{\alpha
}\!\!\int^t_{t_0}\!\!\!d\tau\!\!\!\int^\tau_{t_0}\!\!\!\!d\tau'\big[\bar{\bm{\xi}}(\!\tau\!)\bm{g}_{\alpha
}(\!\tau,\!\tau'\!)\bm{\xi}(\!\tau'\!)\!+\!\bar{\bm{\xi}}'(\!\tau'\!)\bm{g}_{\alpha
}(\!\tau'\!,\!\tau\!)\bm{\xi}'(\!\tau\!)\big]\notag\\&~~~~~-\!\sum_{\alpha
}\!\int^t_{t_0}\!\!d\tau\!\!\int^t_{t_0}\!\!d\tau'\big\{\bar{\bm{\xi}}'(\tau)\bm{g}_{\alpha
}(\tau,\tau')\bm{\xi}(\tau')\notag\\&~~~~~~~~~~~-[\bar{\bm{\xi}}(\tau)+\bar{\bm{\xi}}'(\tau)]\widetilde{\bm{g}}_{\alpha
}(\tau,\tau')[\bm{\xi}(\tau')+\bm{\xi}'(\tau')]\big\}\Big\}.
\end{align}
In the above equation, the time non-local integral kernels,
$\bm{g}_{\alpha}(\tau,\tau')$ and
$\widetilde{\bm{g}}_{\alpha}(\tau,\tau')$ characterize all the
memory effects between the device system and lead $\alpha$,
\begin{subequations}
\label{integral_kernel}
\begin{align}
\bm{g}_{\alpha ij}(\tau,\tau')\!=\!&\sum_k\! V_{i\alpha
k}(\tau)V^*_{j\alpha
k}(\tau')e^{-i\!\int^\tau_{\tau'}d\tau_1\epsilon_{\alpha
k}(\tau_1)}\!,\\
\widetilde{\bm{g}}_{\alpha ij}(\tau,\tau')\!=\!&\sum_k\! V_{i\alpha
k}(\tau)V^*_{j\alpha k}(\tau')\!f_\alpha(\epsilon_{\alpha
k})e^{-i\!\int^\tau_{\tau'}d\tau_1\epsilon_{\alpha
k}(\tau_1)}\!,
\end{align}
\end{subequations}
where $f_\alpha(\epsilon_{\alpha
k})=1/(e^{\beta_\alpha(\epsilon_{\alpha k}-\mu_\alpha)}+1)$ is the
Fermi-Dirac distribution function of lead $\alpha$ at initial time
$t_0$.

After integrating over all the forward paths $\bar{\bm{\xi}}(\tau)$,
$\bm{\xi}(\tau)$ and the backward paths $\bar{\bm{\xi}}'(\tau)$,
$\bm{\xi}'(\tau)$ in the Grassmann space bounded by
$\bar{\bm{\xi}}(t)=\bar{\bm{\xi}}_f$, $\bm{\xi}(t_0)=\bm{\xi}_0$,
$\bar{\bm{\xi}}'(t_0)=\bar{\bm{\xi}}'_0$, and
$\bm{\xi}'(t)=\bm{\xi}'_f$, and by introducing a transformation,
\begin{subequations}
\label{stationary_path}
\begin{align}
&\bm{\xi}(\tau)=\bm{u}(\tau,t_0)\bm{\xi}(t_0)+\bm{v}(\tau,t)[\bm{\xi}(t)+\bm{\xi}'(t_0)],\\
&\bm{\xi}(\tau)+\bm{\xi}'(\tau)=\bm{u}^\dag(t,\tau)[\bm{\xi}(t)+\bm{\xi}'(t)],
\end{align}
\end{subequations}
the propagating function becomes,
\begin{align}
\label{propagator}
\mathcal{J}(&\bar{\bm{\xi}}_f,\bm{\xi}'_f,t|\bm{\xi}_0,\bar{\bm{\xi}}'_0,t_0)=\!\frac{1}{{\rm
det}[\bm{w}(t)]}\!\exp\!\Big\{\!\bar{\bm{\xi}}_f\bm{J}_1(t)\bm{\xi}_0\!\notag\\&+\!\bar{\bm{\xi}}_f
\bm{J}_2(t)\bm{\xi}'_f\!+\!\bar{\bm{\xi}}'_0\bm{J}_3(t)\bm{\xi}_0\!+\!\bar{\bm{\xi}}'_0\bm{J}^\dag_1(t)\bm{\xi}'_f\Big\},
\end{align}
where the time-dependent coefficients are given explicitly as,
\begin{align}\label{J}
\bm{J}_1(t)=&\bm{w}(t)\bm{u}(t,t_0), ~~~\bm{J}_2(t)=\bm{w}(t)-\mathds{1}, \notag\\
\bm{J}_3(t)=&\bm{u}^\dag(t,t_0)\bm{w}(t)\bm{u}(t,t_0)-\mathds{1},
\end{align}
with $\bm{w}(t)=[\mathds{1}-\bm{v}(t,t)]^{-1}$. As one can see, the
propagating function is determined by the two Green functions
$\bm{u}(t,t_0)$ and $\bm{v}(t,t)$, which are $N_S \times N_S$ matrix
with $N_S$ being the total number of single-particle energy levels
in the device system. They satisfies the following
integro-differential equations,
\begin{subequations}
\label{Dyson_eq_uv}
\begin{align}
&\frac{d}{d\tau}\bm{u}(\tau,t_0) +
i\bm{\varepsilon}(\tau)\bm{u}(\tau,t_0)\notag   \label{ut} \\&~~~~~~~~~~~+
\sum_{\alpha} \int_{t_0}^{\tau}d\tau'\bm{g}_{\alpha}(\tau,\tau')\bm{u}(\tau',t_0) = 0,\\
&\frac{d}{d\tau}\bm{v}(\tau,t) +
i\bm{\varepsilon}(\tau)\bm{v}(\tau,t) +
\sum_{\alpha}\int_{t_0}^{\tau}d\tau'\bm{g}_{\alpha}(\tau,\tau')\bm{v}(\tau',t) \notag \\
& ~~~~~~~~~~~=
\sum_{\alpha}\int_{t_0}^td\tau'\bm{\widetilde{g}}_{\alpha}(\tau,\tau')\bm{u}^{\dag}(t,\tau'),
\end{align}
\end{subequations}
subject to the boundary conditions $\bm{u}(t_0,t_0)=\mathds{1}$ and
$ \bm{v}(t_0,t)=0$ with $t_0\leq\tau\leq t$. Actually,
$\bm{u}(\tau,t_0)$ and $\bm{v}(\tau,t)$ are related to the
non-equilibrium Green functions of the device system as we will show
later.

Taking the time derivative of the reduced density matrix
(\ref{red_dens_cohe}) with the solution of the propagating function
(\ref{propagator}), together with the fermion
creation and annihilation operator properties in the fermion coherent-state
representation, one can obtain the final form of the exact master
equation,
\begin{align}
\label{Master Equation} {d\rho(t)\over dt} =&
\!-i\big[H'_S(t),\rho(t)\big]\!\!
+\!\!\sum_{ij}\!\big\{ \bm{\gamma}_{ij}(t)\big[2a_j\rho(t) a_i^{\dag}
 \notag\\
&-a_i^{\dag}a_j\rho(t) -\rho(t) a_i^{\dag}a_j\big]
+\widetilde{\bm{\gamma}}_{ij}(t)\big[ a_i^{\dag}\rho(t)a_j
 \notag\\
&-a_j\rho(t)a_i^{\dag} -\rho(t) a_ja_i^{\dag} +a_i^{\dag}a_j\rho(t) \big] \big\}  \notag\\
=&-i\big[H_S(t),\rho(t)\big]+\sum_{\alpha}\big[\mathcal{L}^{+}_{\alpha}(t)
+\mathcal{L}^{-}_{\alpha}(t)\big]\rho(t) .
\end{align}
The first term describes the unitary evolution of electrons in the
device system, where the renormalization effect, after integrating
out all the lead degrees of freedom, has been fully taken into
account. The resultant renormalized Hamiltonian is $H'_S(t) =
\sum_{ij}\varepsilon'_{ij}(t)a_i^{\dag}a_j$, with
$\bm{\varepsilon}'_{ij}(t)$ being the corresponding renormalized
energy matrix of the device system, including the energy shift of
each level and the lead-induced couplings between different levels.
The remaining terms give the non-unitary dissipation and fluctuation
processes induced by back-actions of electrons from the leads, and
are described by the dissipation and fluctuation coefficients
$\bm{\gamma}(t)$ and $\bm{\widetilde{\gamma}} (t)$, respectively. On
the other hand, the current superoperators of lead $\alpha$,
$\mathcal{L}^{+}_{\alpha}(t)$ and $\mathcal{L}^{-}_{\alpha}(t)$,
determine the transport current flowing from lead $\alpha$ into the
device system:
\begin{align}
\label{current_Mas} I_\alpha(t)=&-e\Big\langle
\frac{dN_\alpha(t)}{dt}\Big\rangle
\notag\\=&\frac{e}{\hbar}\rm{Tr}\big[\mathcal{L}^{+}_{\alpha}(t)\rho(t)\big]=
-\frac{e}{\hbar}\rm{Tr}\big[\mathcal{L}^{-}_{\alpha}(t)\rho(t)\big],
\end{align}
where $N_\alpha(t)=\sum_k c^\dag_{\alpha k}(t)c_{\alpha k}(t)$ is
the total particle number of lead $\alpha$.

All the time-dependent coefficients in Eq.~(\ref{Master Equation})
are found to be,
\begin{subequations}
\label{ecoff_Mas}
\begin{align}
\bm{\varepsilon}'_{ij}(t)= &
\frac{i}{2}\big[\dot{\bm{u}}(t,t_0)\bm{u}^{-1}(t,t_0) - {\rm
H.c.}\big]_{ij}
\notag\\&=\bm{\varepsilon}_{ij}(t)-\frac{i}{2}\sum_\alpha[\bm{\kappa}_\alpha(t)-\bm{\kappa}^\dag_\alpha(t)]_{ij},\\
\bm{\gamma}_{ij}(t)= &
-\frac{1}{2}\big[\dot{\bm{u}}(t,t_0)\bm{u}^{-1}(t,t_0) + {\rm
H.c.}\big]_{ij}
\notag\\&=\frac{1}{2}\sum_\alpha[\bm{\kappa}_\alpha(t)+\bm{\kappa}^\dag_\alpha(t)]_{ij},\\
\widetilde{\bm{\gamma}}_{ij}(t)=&\, \, \frac{d}{dt}\bm{v}_{ij}(t,t)
-[\dot{\bm{u}}(t,t_0)\bm{u}^{-1}(t,t_0)\bm{v}(t,t)+ {\rm
H.c.}]_{ij}\notag\\&=-\sum_\alpha[\bm{\lambda}_\alpha(t)+\bm{\lambda}^\dag_\alpha(t)]_{ij},
\end{align}
\end{subequations}
The current superoperators of lead $\alpha$,
$\mathcal{L}^{+}_{\alpha}(t)$ and $\mathcal{L}^{-}_{\alpha}(t)$, are
also explicitly given by
\begin{subequations}
\label{current_super}
\begin{align}
\!\mathcal{L}^{+}_{\alpha}(t)\rho(t)=&-\!\!\sum_{ij}\big\{\bm{\lambda}_{\alpha
ij}(t)\big[a^\dag_ia_j\rho(t)+a^\dag_i\rho(t)a_j\big]\notag\\&+\bm{\kappa}_{\alpha
ij}(t)a^\dag_ia_j\rho(t)+\rm H.c.\big\},\\
\!\mathcal{L}^{-}_{\alpha}(t)\rho(t)=&\sum_{ij}\big\{\bm{\lambda}_{\alpha
ij}(t)\big[a_j\rho(t)a^\dag_i+\rho(t)a_ja^\dag_i\big]\notag\\&+\bm{\kappa}_{\alpha
ij}(t)a_j\rho(t)a^\dag_i+\rm H.c.\big\}.
\end{align}
\end{subequations}
The functions $\bm{\kappa}_{\alpha}(t)$ and
$\bm{\lambda}_{\alpha}(t)$ in Eq.~(\ref{ecoff_Mas}) and
Eq.~(\ref{current_super}) are solved from Eq.~(\ref{Dyson_eq_uv}),
\begin{subequations}
\label{kappalambda}
\begin{align}
\bm{\kappa}_\alpha(t)=&\int^t_{t_0}d\tau\bm{g}_\alpha(t,\tau)
\bm{u}(\tau,t_0)[\bm{u}(t,t_0)]^{-1},
\\\bm{\lambda}_\alpha(t)=&\int^t_{t_0}d\tau[\bm{g}_\alpha(t,\tau)
\bm{v}(\tau,t)-\widetilde{\bm{g}}_\alpha(t,\tau)\bm{u}^\dag(t,\tau)]
\notag\\&-\bm{\kappa}_\alpha(t)\bm{v}(t,t).
\end{align}
\end{subequations}

The master equation (\ref{Master Equation}) takes a convolution-less
form, so the non-Markovian dynamics are fully encoded in the
time-dependent coefficients (\ref{ecoff_Mas}). These coefficients
determined by the functions $\bm{u}(t,t_0)$, and $\bm{v}(\tau,t)$
are governed by integro-differential equations (\ref{Dyson_eq_uv}),
where the integral kernels (\ref{integral_kernel}) manifest the
non-Markovian memory effects. The master equation is derived exactly
so that the positivity, hermiticity of the trace of the reduced
density matrix are guaranteed. It is also worth mentioning that the
master equation (\ref{Master Equation}) is valid for various
nano-devices coupled to various surroundings through particle
tunnelings, even when initial-correlations are presented as long as
the electron-electron interaction can be ignored (including the initial correlation
effect is given in Sec.~\ref{initial_corr}).

From Eqs.~(\ref{Master Equation}-\ref{current_Mas}), the transient
transport current is given explicitly as follows:
\begin{align}
\label{I_trans} I_\alpha(t)
=&-\frac{e}{\hbar}\rm{Tr}[\bm{\lambda}_\alpha(t)+\bm{\kappa}_\alpha(t)\bm{\rho}^{(1)}(t)+H.c.]\notag\\
=&\!-\!\frac{2e}{\hbar}{\rm Re}\!\!\!\int_{t_0}^t \!\!\!d\tau {\rm
Tr}\big[\bm{g}_\alpha(t,\!\tau)\bm{\rho}^{(1)}\!(\tau, t)
\!-\!\bm{\widetilde{g}}_\alpha(t,\tau)\bm{u}^{\dag}(t,\tau)\big] .
\end{align}
In Eq.~(\ref{I_trans}), the single-particle correlation function of
the device system $\bm{\rho}^{(1)}(\tau,t)$ is determined by
\begin{align}
\label{rho(1)} \bm{\rho}^{(1)}_{ij}(\tau,t)
=\big[\bm{u}(\tau,t_0)\bm{\rho}^{(1)}(t_0)\bm{u}^{\dag}(t,t_0)+
\bm{v}(\tau,t)\big]_{ij} ,
\end{align}
where $\bm{\rho}^{(1)}_{ij}(t_0)={\rm Tr_S}[a^\dag_ja_i\rho(t_0)]$,
is the initial single-particle density matrix. The transient
transport current obtained from the master equation actually has
exactly the same formula as the one used in the non-equilibrium Green
function technique~\cite{Wingreen1993}, except for the first term of
the single-particle correlation function (\ref{rho(1)}) that is
originated from the initial occupation $\bm{\rho}^{(1)}_{ij}(t_0)$
in the device system, which was missing in Ref.~\cite{Wingreen1993}.

To be explicitly, here we present the relation between
$\bm{u}(\tau,\tau')$ and $\bm{v}(\tau,t)$ and the non-equilibrium
Green functions. As one see, both the master equation (\ref{Master
Equation}) and the transient current (\ref{I_trans}) are completely
determined by the Green functions $\bm{u}(\tau,\tau')$ and
$\bm{v}(\tau,t)$, which are introduced in
Eq.~(\ref{stationary_path}). The equations (\ref{stationary_path})
show that $\bm{u}(\tau, t_0)$ describes the electron forward
propagation from time $t_0$ to time $\tau$, $\bm{u}^\dag(t,\tau)$
describes the electron backward propagation from time $t$ to time
$\tau$, and $\bm{v}(\tau,t)$ describes the electron correlation
between the forward and backward paths. These Green functions
satisfy the integro-differential equations (\ref{Dyson_eq_uv}).
Solving inhomogeneous equation (\ref{Dyson_eq_uv}b) with initial
condition $\bm{v}(t_0,t)=0$, we obtain
\begin{align}
\label{v_tau_t} \bm{v}(\tau,t) =\sum_\alpha\int_{t_0}^{\tau} d\tau_1
\int_{t_0}^t d\tau_2
\bm{u}(\tau,\tau_1)\widetilde{\bm{g}}_\alpha(\tau_1,\tau_2)
\bm{u}^{\dag}(t,\tau_2),
\end{align}
where $\bm{u}(\tau,\tau')$ is determined by Eq.~(\ref{ut}).

It is easy to infer that
\begin{align}
\bm{u}_{ij}(\tau,\tau')=&\langle
\{a_i(\tau),a^\dag_j(\tau')\}\rangle\notag\\=&i[\bm{G}^R(\tau,\tau')-\bm{G}^A(\tau,\tau')]_{ij},
\end{align}
which is the spectral function in non-equilibrium Green function
techniques, with
\begin{align}
\bm{g}_{\alpha
ij}(\tau,\tau')=&i[\bm{\Sigma}^R_{\alpha}(\tau,\tau')-\bm{\Sigma}^A_{\alpha}(\tau,\tau')]_{ij}
\notag\\=&\int\frac{d\omega}{2\pi}\Gamma_{\alpha
ij}(\omega,\tau,\tau')e^{-i\omega(\tau-\tau')}.
\end{align}
As a result, matrix function $\bm{v}(\tau,t)$ (\ref{v_tau_t}) can be
written in terms of non-equilibrium Green functions,
\begin{align}
\label{v_tau_t_Green}
\bm{v}(\tau,t)=-i\!\!\int^\tau_{t_0}\!\!d\tau_1\!\!\int^t_{t_0}\!\!d\tau_2\bm{G}^R(\tau,\tau_1)\bm{\Sigma}^<(\tau_1,\tau_2)\bm{G}^A(\tau_2,t),
\end{align}
where
\begin{align}
\widetilde{\bm{g}}_{\alpha}(\tau,\tau')=&-i\bm{\Sigma}^<_\alpha(\tau,\tau')
\notag\\=&\int\frac{d\omega}{2\pi}f_\alpha(\omega)\bm{\Gamma}_\alpha(\omega,t_1,t_2)e^{-i\omega(\tau-\tau')}.
\end{align}
Comparing Eq.~(\ref{v_tau_t_Green}) with
Eq.~(\ref{Green_fn_time_dep}b), one can see that $\bm{v}(\tau,t)$
exactly has the same form as the lesser Green function in the
Keldysh formalism. However, when one considers transient electron
dynamics, the general solution of the lesser Green function is
related to the single-particle correlation function in the master
equation approach,
\begin{align}
\label{lesser_Green_fn_general}
\bm{G}^<(t,t')\!=&i\bm{\rho}^{(1)}(t,t'\!)\!=i\big[\bm{u}(\tau,t_0)\bm{\rho}^{(1)}\!(t_0)\bm{u}^{\dag}(t,t_0)\!+\!
\bm{v}(\tau,t)\big]\notag\\=&\bm{G}^R(t,t_0)\bm{G}^<(t_0,t_0)\bm{G}^A(t_0,t')\notag\\&+\!\!\int^\infty_{t_0}\!\!\!d\tau\!\!\int^\infty_{t_0}\!\!\!d\tau'\bm{G}^R(t,\tau)\bm{\Sigma}^<(\tau,\tau')\bm{G}^A(\tau',t').
\end{align}
The first term depends on the initial occupation of the device
system. According to the above results, one can
express the transient transport current in terms of non-equilibrium
Green functions:
\begin{align}
\label{I_trans_Green_fn} I_\alpha(t)\! =&\!-\!\frac{2e}{\hbar}{\rm
Re}\!\!\int_{t_0}^t\!\! d\tau {\rm
Tr}\big[\bm{g}_\alpha(t,\tau)\bm{\rho}^{(1)}\!(\tau, t)
\!-\!\bm{\widetilde{g}}_\alpha(t,\tau)\bm{u}^{\dag}(t,\tau)\big]\notag\\
=&\!-\!\frac{2e}{\hbar}{\rm Re}\!\!\int_{t_0}^t \!\!d\tau {\rm
Tr}\!\big[\bm{\Sigma}^R_\alpha(t,\!\tau\!)\bm{G}^<(\tau,\! t)
\!+\!\bm{\Sigma}^<_\alpha(t,\!\tau\!)\bm{G}^A(\tau,\!t)\big].
\end{align}
It is easy to check the consistency between
Eq.~(\ref{I_trans_Green_fn}) and Eq.~(\ref{I_NEGF_time_dep}),
Thus, we have proved
that the transient transport current obtained from the master
equation has exactly the same formula as the one using the
non-equilibrium Green function technique~\cite{Wingreen1993}, except
for the term that is originated from the initial occupation
$\bm{\rho}^{(1)}_{ij}(t_0)$ in the device system. This also
indicates further that the Keldysh's non-equilibrium Green function
technique is mostly valid in the steady-state limit.

\section{Application of Master equation approach}\label{application}
From the above discussion, the master equation approach provides
a more essential way to study the quantum transport problem. In the master equation approach,
the device system is described by the reduced density matrix which
contains full information of quantum coherence and decoherence, as well as
the non-Markovian memory effects induced by the environment.
That makes the master equation approach valid in both the transient dynamics and steady-state limit phenomena.
In the following contents, we discuss different quantum transport problems in nanostructures using the master equation approach.
\subsection{Transient current-current correlations and noise
spectra}\label{Noise} Noise spectra provide the information of
temporal correlations between individual electron transport events.
It has been shown that noise spectra can be a powerful tool to
reveal different possible mechanisms which are not accessible to the
mean current measurement. Examples include electron
kinetics~\cite{Landauer1998}, quantum statistics of charge
carriers~\cite{Beenakker2003}, correlations of electronic wave
functions~\cite{Gramespacher1998}, and effective quasiparticles
charges~\cite{Saminadayar1997, Quirion2003}. Noise spectra can also
be used to reconstruct quantum states via a series of measurements
known as quantum state tomography~\cite{Samuelsson2006}.
Conventionally, evaluations of noise spectra are largely limited to
the rather low frequency ($\hbar\omega\ll k_BT$), where the noise
spectrum is symmetric at zero bias~\cite{Blanter2000}.
However, experimental measurements of high frequency noise
spectra~\cite{Schoelkopf1997,Deblock2003, Onac2006,Zakka2007}
inspired the exploration of the frequency-resolved noise spectrum
both in symmetric~\cite{Aguado2004,Lambert2007,Wu2010} and
asymmetric form~\cite{Engel2004,Wohlman2007,
Rothstein2009,Orth2012}. The asymmetric noise spectrum, which is
directly proportional to the emission-absorption spectrum of the
system~\cite{Gavish2000,Aguado2000}, has been demonstrated
experimentally~\cite{Deblock2003,Onac2006,Zakka2007,
Billangeon2009}. In recent years, the higher order
current-correlations in a non-equilibrium steady state are also
explored both in experimental and theoretical studies
~\cite{Ubbelohde2012, Zazunov2007}.

The above investigations were mainly focused on the steady-state
transport regime. Owing to the theoretic development on quantum
transient transport dynamics~\cite{Haug2008}, the transient current
fluctuations (correlations at equal time) and noises in the time
domain are a subject of considerable interest. Recently, the
transient current fluctuations of a two-probe transport junction in
response to the sharply turning off the bias voltage were analyzed
by Feng {\it et al.}~\cite{Feng2008}. The transient evolution of
finite-frequency current noises after abruptly switching on the
tunneling coupling in the resonant level model and the Majorana
resonant level model were studied by Joho {\it et
al.}~\cite{Joho2012}. In this section, we shall investigate the
transient current-current correlations of a biased quantum dot
system in the nonlinear transient transport regime~\cite{Yang2014}. Using the exact
master equation~\cite{Tu2008,Jin2010}, a
general formalism for transient current-current correlations and
transient noise spectra are presented for non-interacting
nanostructures with arbitrary spectral density.
This formalism unveils how the electron correlations change in the
system when the system evolves far away from the equilibrium to the
steady state. Besides, various time-scales in the system when it
reaches the steady state can also be obtained. These time-scales are
important for understanding the role of quantum coherence and
non-Markovian behaviors in quantum transport dynamics. It is also
essential for one to reconstruct quantum states of electrons in
nanostructures \cite{Samuelsson2006} for further applications in nanotechnology, such as
the controlling of quantum information processing and quantum
metrology on quantum states, etc.

The current-current auto-correlation
($\alpha = \alpha'$) and cross-correlation ($\alpha \neq \alpha'$)
functions are defined as follows,
\begin{align}
\label{currentcorrelationdef} S_{\alpha \alpha'}(t+\tau, t)\equiv
\langle\delta I_{\alpha}(t+\tau)\delta I_{\alpha'}(t) \rangle,
\end{align}
where $\delta I_\alpha(t)\equiv I_\alpha(t)-\langle
I_{\alpha}(t)\rangle$ is the fluctuation of the current in the lead
$\alpha$ at time $t$. $I_\alpha$(t) is the current operator of
electrons flowing from the lead $\alpha$ into the central dot. It is
determined by
\begin{align}
\label{currentoperator}
I_{\alpha}(t) =& -e\frac{d}{dt}N_\alpha(t)=i\frac{e}{\hbar}[N_\alpha(t),H(t)]\notag\\
= &-i\frac{e}{\hbar}\!\sum_{ik}[V_{i\alpha k}(t)a^{\dag}_i(t)c_{\alpha
k}(t)\!-\!V^{*}_{i\alpha k}(t)c^{\dag}_{\alpha k}(t)a_i(t)],
\end{align}
where $e$ is the electron charge, $N_\alpha(t)=\sum_k c_{\alpha
k}^{\dag}(t)c_{\alpha k}(t)$ is the particle number operator of the
lead $\alpha$. The angle brackets in
Eq.~(\ref{currentcorrelationdef}) take the mean value of the
operator over the whole system, which is defined as $\langle
O(t)\rangle = {\rm Tr}[O(t)\rho_{\rm tot}(t_0)]$. Here $\rho_{\rm
tot}(t_0)$ is the initial state of the total system. Current-current
correlations measure the correlations between currents flowing in
different time. Explicitly,
\begin{widetext}
\begin{align}
\label{currentcorrelation} &S_{\alpha\alpha'}(t+\tau,t) =
\frac{e^{2}}{\hbar^{2}}\sum_{ijkk'}\notag\\&\times\Big\{-V_{i\alpha
k}(t)V_{j\alpha'k'}(t)[\langle a_{i}^{\dag}(t+\tau)c_{\alpha
k}(t+\tau) a_{j}^{\dag}(t)c_{\alpha' k'}(t)\rangle - \langle
a_{i}^{\dag}(t+\tau)c_{\alpha k}
(t+\tau)\rangle\langle a_{j}^{\dag}(t)c_{\alpha' k'}(t)\rangle]\notag\\
&- V_{i\alpha k}^{*}(t)V_{j\alpha'k'}^{*}(t)[\langle c_{\alpha
k}^{\dag}(t+\tau)a_{i}
(t+\tau)c_{\alpha'k'}^{\dag}(t)a_{j}(t)\rangle - \langle c_{\alpha
k}^{\dag}(t+\tau)
a_{i}(t+\tau)\rangle\langle c_{\alpha'k'}^{\dag}(t)a_{j}(t)\rangle]\notag\\
&+ V_{i\alpha k}(t)V_{j\alpha'k'}^{*}(t)[\langle
a_{i}^{\dag}(t+\tau)c_{\alpha k}(t+\tau)
c_{\alpha'k'}^{\dag}(t)a_{j}(t)\rangle - \langle
a_{i}^{\dag}(t+\tau)c_{\alpha k}
(t+\tau)\rangle\langle c_{\alpha'k'}^{\dag}(t)a_{j}(t)\rangle]\notag\\
&+ V_{i\alpha k}^{*}(t)V_{j\alpha'k'}(t)[\langle c_{\alpha
k}^{\dag}(t+\tau)a_{i}(t+\tau)a_{j}^{\dag}(t)c_{\alpha'k'}(t)\rangle
- \langle c_{\alpha k}^{\dag}(t+\tau)a_{i}(t+\tau)\rangle\langle
a_{j}^{\dag}(t)c_{\alpha'k'}(t)\rangle]\Big\} .
\end{align}
\end{widetext}
Current-current correlations are in general complex and physical
observables are related to its real or imaginary parts,
\begin{align}
S_{\alpha \alpha'}(t+\tau, t)& = S'_{\alpha \alpha'}(t+\tau, t) +
iS''_{\alpha \alpha'}(t+\tau, t),
\end{align}
where
\begin{subequations}
\begin{align}
S'_{\alpha \alpha'}(t+\tau, t)& = \frac{1}{2}\langle\{\delta
I_{\alpha}(t+\tau) , \delta I_{\alpha'}(t)\}\rangle \\
S''_{\alpha \alpha'}(t+\tau, t)& = \frac{1}{2i}\langle[\delta
I_{\alpha}(t+\tau) , \delta I_{\alpha'}(t)]\rangle
\end{align}
\end{subequations}
are directly proportional to the fluctuation function and the
response function, respectively, in the linear response
theory~\cite{Zwanzig2001,Mazenko2006}. On the other hand, we may
introduce the total current-current correlation defined by
\begin{align}
S(t+\tau,t)\equiv\langle\delta I(t+\tau)\delta I(t) \rangle,
\label{totcorrelation}
\end{align}
where the total current operator $I(t)$ is given by
\begin{align}
I(t)=aI_L(t)-bI_R(t),
\end{align}
and the coefficients satisfying $a + b = 1$, which are associated
with the symmetry of the transport setup (e.g., junction
capacitances). Then Eq.~(\ref{totcorrelation}) can be written as
\begin{align}
S(t+\tau,t)=&a^2S_{LL}(t+\tau,t)+b^2S_{RR}(
t+\tau,t)\notag\\&-ab\big[S_{LR}(t+\tau,t)+S_{RL}(t+\tau,t)\big].
\end{align}
Taking different values of $a$ and $b$ can also give other
current-current correlations, such as the auto-correlation ($a=1,
b=0$ or $a=0, b=1$), etc. Taking Fourier transform of the total current-current
correlation with $\tau$,  an
asymmetric noise spectrum of the electronic transport at time $t$ is
obtained, denoted as $S(t,\omega)\equiv\int^{\infty}_{-\infty}d\tau e^{-i\omega\tau}\langle\delta I(t+\tau)\delta I(t)\rangle$.
The asymmetric noise spectra is proportional to the emission-absorption spectrum of the system,
so $S(t,\omega)$ can be viewed as the probability of a quantum energy
$\hbar\omega$ being transferred from the system to a measurement apparatus.

Now, we shall calculate  these correlation functions in terms of the
exact master equation represented in Sec.~\ref{Master_Eq_appr} and
the extended quantum Langevin equation for the dot operators \cite{Yang2014}. The
later can be derived formally from the Heisenberg equation of motion
\begin{align}
\label{quantumLangevineq} \frac{d}{dt}a_i(t) =& -i\sum_j
\varepsilon_{ij}(t) a_j(t) -\sum_{\alpha j}\int_{t_0}^t d\tau
g_{\alpha ij}(t,\tau)a_j(\tau)\notag\\& -i\sum_{\alpha k} V_{i\alpha k}(t)
c_{\alpha k}(t_0)e^{-i\int^t_{t_0}d\tau\epsilon_{\alpha k}(\tau)}.
\end{align}
In the above quantum Langevin equation, the first term is determined
by the evolution of the dot system itself, the second term is the
dissipation risen from the coupling to the leads, and the last term
is the fluctuation induced by the environment (the leads), and
$c_{\alpha k}(t_0)$ is the electron annihilation operator of the
lead $\alpha$ at initial time $t_0$. The time non-local correlation
function $g_{\alpha ij}(t,\tau)$ in Eq.~(\ref{quantumLangevineq}) is
also given by Eq.~(\ref{integral_kernel}a), which characterizes
back-actions between the dot system and the leads. Because the
quantum Langevin equation~(\ref{quantumLangevineq}) is linear in
$a_i$, its general solution can be written as
\begin{align}
\label{at} a_i(t)=\sum_j u_{ij}(t,t_0)a_j(t_0)+F_i(t),
\end{align}
where $u_{ij}(t,t_0)$ is the same non-equilibrium Green's function
of Eq.~(\ref{Dyson_eq_uv}a) that determines the energy level
renormalization and electron dissipations of the dot system, as
described by the master equation. The noise operator $F_i(t)$ obeys
the following equation,
\begin{align}
\frac{d}{dt}F_i(t) &+i \sum_{j}\epsilon_{ij}(t) F_j(t) +\sum_{\alpha
j}\int_{t_0}^t d\tau g_{\alpha ij}(t,\tau) F_j(\tau) \notag\\& =-i
\sum_{\alpha k}V_{i\alpha k}(t) c_{\alpha k}(t_0)
 e^{-i\int^t_{t_0}d\tau\epsilon_{\alpha k}(\tau)}  \label{foe}
\end{align}
with the initial condition $F_i(t_0)=0$. Since the system and the
leads are initially decoupled to each other, and the leads are
initially in equilibrium, it can be shown that the solution of
Eq.~(\ref{foe}) gives
\begin{align}
\langle F_j^{\dag}\!(t) F_i(\tau)\rangle\! = & v_{ij}(\tau,t)\notag\\ =&\!
\sum_{\alpha}\!\!\int_{t_0}^{\tau}\!\!\!dt_1\!\!\!\int_{t_0}^t\!\!\!dt_2 \big[\bm
u(\tau,t_1)\widetilde{\bm g}_{\alpha}\!(t_1,t_2) \bm
u^{\dag}\!(t,t_2)\big]_{ij}, \label{v}
\end{align}
which is indeed the solution of Eq.~(\ref{Dyson_eq_uv}b). Thus the
connection of the solution of the quantum Langevin equation to the
dissipation and fluctuation dynamics in the master equation is
explicitly established.

Furthermore, the time-dependent operator $c_{\alpha k}(t)$ of the
lead $\alpha$ can also be obtained from its equation of motion:
\begin{align}
\label{ct} c_{\alpha k}(t) =& c_{\alpha
k}(t_0)e^{-i\int^t_{t_0}d\tau\epsilon_{\alpha k}(\tau)}\notag\\& -i\sum_i
\int_{t_0}^t d\tau V^*_{i\alpha k}(\tau)
a_i(\tau)e^{-i\int^t_{\tau}d\tau_1\epsilon_{\alpha k}(\tau_1)}.
\end{align}
Using the solutions of Eq.~(\ref{at}) and (\ref{ct}), we can
calculate explicitly and exactly the current-current correlation
function (\ref{currentcorrelation}). The explicit expression is
still very complicated so we consider the situation that the dot has
no initial occupation. Then, the four terms in
Eq.~(\ref{currentcorrelation}), denoted simply as
$S^{(1)}$, $S^{(2)}$, $S^{(3)}$ and $S^{(4)}$, are given by
\begin{widetext}
\begin{subequations}
\label{cccf}
\begin{align}
\label{1st} S_{\alpha \alpha'}^{(1)}(t+\tau,
t)=-\frac{e^2}{\hbar^2}{\rm
Tr}\Big\{&\big[\int_{t_0}^{t+\tau}ds\bm{g}_{\alpha}(t+\tau,s)
\overline{\bm{v}}(s,t)-\int_{t_0}^{t}ds\bm{\widetilde{\bar{g}}}_{\alpha}(t+\tau,s)
\bm{u}^{\dag}(t,s)\big]
\notag\\
\times&\big[\int_{t_0}^{t+\tau}ds'\bm{\widetilde{g}}_{\alpha'}(t,s')
\bm{u}^{\dag}(t+\tau,s')-\int_{t_0}^{t}ds'\bm{g}_{\alpha'}(t,s')
\bm{v}(s',t+\tau)\big] \Big\},
\end{align}
\begin{align}
\label{2nd} S_{\alpha \alpha'}^{(2)}(t+\tau,
t)=-\frac{e^2}{\hbar^2}{\rm Tr}\Big\{&\big[\int_{t_0}^{t+\tau}ds
\bm{v}(t,s)\bm{g}_{\alpha}(s,t+\tau)-\int_{t_0}^{t}ds\bm{u}(t,s)
\bm{\widetilde{g}}_{\alpha}(s,t+\tau)\big]
\notag\\
\times&\big[\int_{t_0}^{t+\tau}ds'\bm{u}(t+\tau,s')
\bm{\widetilde{\bar{g}}}_{\alpha'}(s',t)-\int_{t_0}^{t}ds'
\overline{\bm{v}}(t+\tau,s')\bm{g}_{\alpha'}(s',t)\big]\Big\},
\end{align}
\begin{align}
\label{3rd} S_{\alpha \alpha'}^{(3)}(t+\tau,
t)=&+\frac{e^2}{\hbar^2}{\rm
Tr}\Big\{\big[\bm{\widetilde{\bar{g}}}_{\alpha}(t+\tau,t)
\delta_{\alpha\alpha'}+\int_{t_0}^{t+\tau}ds\int_{t_0}^{t}ds'
\bm{g}_{\alpha}(t+\tau,s)\overline{\bm{v}}(s,s')\bm{g}_{\alpha'}(s',t)
\notag\\
&-\!\!\int_{t_0}^{t+\tau}\!\!\!\!ds\!\!\int_{t_0}^{s}\!\!ds'\bm{g}_{\alpha}(t+\tau,s)
\bm{u}(s,s')\bm{\widetilde{\bar{g}}}_{\alpha'}(s',t)-\!\!\int_{t_0}^{t}\!\!ds\!\!
\int_{t_0}^{s}ds'\bm{\widetilde{\bar{g}}}_{\alpha}(t+\tau,s')
\bm{u}^{\dag}(s,s')\bm{g}_{\alpha'}(s,t)\big]\bm{v}(t,t+\tau)\Big\},
\end{align}
\begin{align}
\label{4th} S_{\alpha \alpha'}^{(4)}(t+\tau,
t)=&+\frac{e^2}{\hbar^2}{\rm
Tr}\Big\{\overline{\bm{v}}(t+\tau,t)\big[\bm{\widetilde{g}}_{\alpha}(t,t+\tau)
\delta_{\alpha\alpha'}+\int_{t_0}^{t+\tau}ds\int_{t_0}^{t}ds'
\bm{g}_{\alpha'}(t,s')\bm{v}(s',s)\bm{g}_{\alpha}(s,t+\tau)
\notag\\
&-\int_{t_0}^{t+\tau}ds\int_{t_0}^{s}ds'\bm{\widetilde{g}}_{\alpha'}(t,s')
\bm{u}^{\dag}(s,s')\bm{g}_{\alpha}(s,t+\tau)-\int_{t_0}^{t}ds\int_{t_0}^{s}ds'
\bm{g}_{\alpha'}(t,s)\bm{u}(s,s')\bm{\widetilde{g}}_{\alpha}(s',t+\tau)\big]\Big\}.
\end{align}
\end{subequations}
\end{widetext}
Here, $\overline{v}_{ij}(\tau,t)=\langle
a_i(\tau)a_j^{\dag}(t)\rangle$ is related to the greater Green's
function in non-equilibrium Green functions approach. Its general
solution is given by
\begin{align}
\overline{\bm{v}}(\tau,t)=
\theta(\tau-t)\bm{u}(\tau,t)+\theta(t-\tau)\bm{u}^{\dag}(t,\tau)-\bm{v}(\tau,t).
\end{align}
The function
$\bm{\widetilde{\bar{g}}}_{\alpha}(\tau,\tau')=\int\frac{d\omega'}
{2\pi}\bm{\Gamma}_{\alpha}(\omega',\tau,\tau')[1-f_{\alpha}(\omega')]e^{-i\omega'(\tau-\tau')}$
is a self-energy correlation of electron holes. As one can see, the
transient current-current correlations have been expressed
explicitly in terms of non-equilibrium Green's functions $\bm
u(\tau,\tau')$ and $\bm v(\tau,t)$ that determine the dissipation
and fluctuation coefficients in the exact master equation
(\ref{Master Equation}).

As an example, we consider the transient current-current correlations of
a single-level quantum dot coupled to the source and the drain, where the noise spectra have been
recently investigated~\cite{Engel2004,Rothstein2009} in the wide band limit (WBL). The
Hamiltonian is expressed as
\begin{align}
H=&\varepsilon a^\dag a +\sum_{\alpha k}\epsilon_{\alpha
k}c^\dag_{\alpha k}c_{\alpha k}\notag\\&+\sum_{i \alpha k}(V_{i\alpha
k}a^\dag c_{\alpha k}+V^*_{i\alpha k}c^\dag_{\alpha k}a).
\end{align}
For the sake of generality, we assume that the electronic structure of the
leads has a Lorentzian line shape
\cite{Maciejko2006,Jin2008,Jin2010,Tu2008}.
\begin{align}
\Gamma_{\alpha}(\omega)=\frac{\Gamma_{\alpha}W_{\alpha}^2}{(\omega-\mu_{\alpha})^2+W_{\alpha}^2},
\end{align}
where $W_{\alpha}$ is the band width and $\Gamma_{\alpha}$ is the
coupling strength to the lead $\alpha$. The current-current
correlations describe how the correlations persist until they are
averaged out through the coupling with the surroundings. Thus, by
fixing the observing time $t$, one can see how the correlations vary via
the time difference $\tau$ of measurements. Hereafter, the initial time
is set $t_0=0$.
\begin{figure}
\centering
\includegraphics[width =5.0 cm]{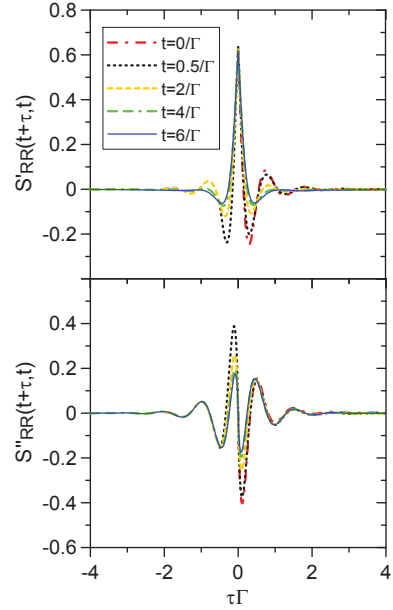}
\caption{Auto-correlation function $S_{RR}$ 
in terms of their real and imaginary parts 
(in units of $e^2\Gamma^2/\hbar^2$) in a single-level nanostructure
for different $t$ as a function of $\tau$. Where
$\varepsilon=\Gamma$, with $\Gamma_L=\Gamma_R=0.5\Gamma$,
$W_L=W_R=5\Gamma$, $eV=10\Gamma$, at $k_{B}T=0.5\Gamma$ for both two
leads \cite{Yang2014}.} \label{correlaitons_difft}
\end{figure}

Figure \ref{correlaitons_difft} plots the auto-correlation
function of the right lead for several different $t$. This allows
one to monitor the transient processes until the system reaches its
steady state, at which these correlations come to only depend on the
time difference $\tau$ between the measurements. As one can see,
both the real and imaginary parts of correlation functions approach
the steady-state values at $t \simeq 5/\Gamma$. The real part of the
auto-correlation has a maximal value at $\tau=0$ (namely when it is
measured in the same time), this gives the current fluctuation,
 $\langle I^2(t)\rangle -\langle I(t) \rangle^2$, and
this current fluctuation is independent of the
observing time $t$ (less transient).  In fact,  the current
fluctuation, $\langle I^2(t)\rangle -\langle I(t) \rangle^2$, is
mainly contributed from $S^{(3)}$ and $S^{(4)}$ in
Eq.~(\ref{cccf}). From the expression of
Eq.~(\ref{currentcorrelation}), one can see that $S^{(3)}$ describes
the current correlation between an electron tunneling from the dot
to the leads at time $t$ and another electron tunneling from the
leads to the dot at time $t+\tau$, and $S^{(4)}$ is given by the
opposite processes. These processes have the maximum contribution to
the current correlation at $\tau=0$.  While, $S^{(1)}$ and $S^{(2)}$
describe the correlations of electron tunnelings in the same
direction (namely both tunnelings from the leads to the dot or from
the dot to the leads), and has a minimum contribution at $\tau=0$,
due to the Pauli exclusion principle. When the time difference
$\tau$ gets larger, the auto-correlation decays rather faster, and
it reaches to zero after $\tau > 2/\Gamma$, i.e.~the correlation
vanishes.
With the observing time goes on, the real part of auto-correlation
becomes more and more symmetric, and the imaginary part gets more
antisymmetric. Eventually they become fully symmetric and
antisymmetric functions of $\tau$, respectively, in the steady-state
limit, as one expected. It is also found that the cross-correlation is
rather small (about of one order of magnitude smaller in comparison
with the auto-correlation) that it is not presented in
Fig.~\ref{correlaitons_difft}.

\begin{figure}
\centering
\includegraphics[width =9.5 cm]{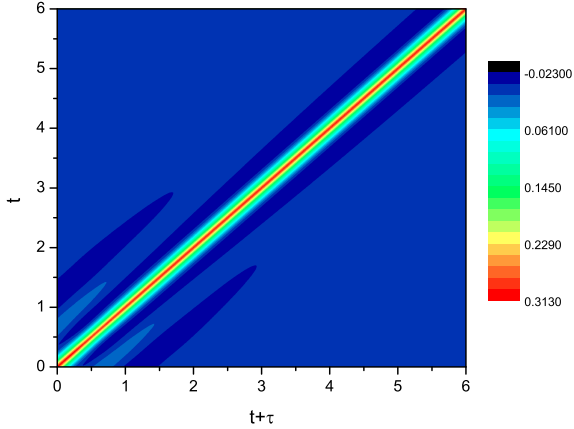}
\caption{The contour plot of the real part of the total
current-current correlation, $S'(t+\tau,t)$ (in units of
$e^2\Gamma^2/\hbar^2$), in the single-level nanostructure in the
two-time plane (scaled by $\Gamma$). Here the parameter
$\varepsilon=\Gamma$, with $\Gamma_L=\Gamma_R=0.5\Gamma$,
$W_L=W_R=5\Gamma$, $eV=10\Gamma$, at $k_{B}T=0.5\Gamma$ for both two
leads \cite{Yang2014}.} \label{contour_plot}
\end{figure}

To have a more general picture how the system reaches the steady
state, here a contour plot of the real part of the
total-correlation in the 2-D time domain is presented in Fig~\ref{contour_plot}.
As one can see, it is symmetric in the diagonal line ($\tau=0$), as
a consequence of the identity: $S_{\alpha \alpha'}(t+\tau,
t)=S^{*}_{\alpha' \alpha}(t, t+\tau)$.  The contour-plot clearly
shows an oscillating profile of the correlation in the region
$t<3/\Gamma$. The oscillation quickly decays for the time period
$3/\Gamma < t < 5/\Gamma$. The correlation reaches a steady-state
value after $t  \simeq 5/\Gamma$. The imaginary part has much the
same behavior, except that it has an antisymmetric profile in terms
of $t$ and $t+\tau$. This gives the whole picture of the transient
current-current correlations.

\begin{figure}
\centering
\includegraphics[width =5.0 cm]{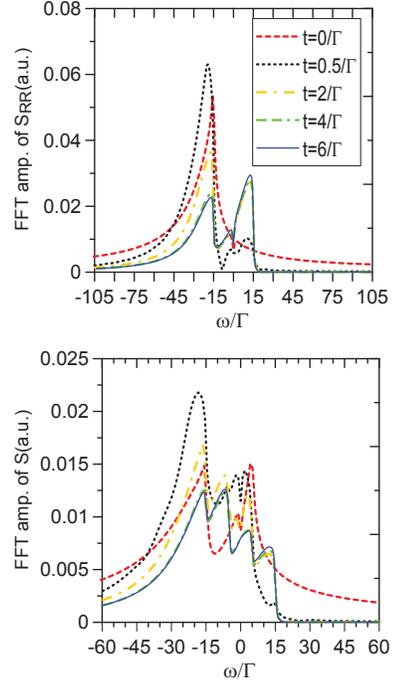}
\caption{The FFT amplitude of the auto-correlation $S_{RR}$ and the
total correlation $S$ in the single-level nanostructure as a
function of $\omega$ (in units of $\Gamma$). Where
$\varepsilon=5\Gamma$, $\Gamma_L=\Gamma_R=0.5\Gamma$,
$W_L=W_R=15\Gamma$, $eV=20\Gamma$, $k_{B}T=0.1\Gamma$ for both two
leads \cite{Yang2014}.  } \label{Amp}
\end{figure}

To see the energy structure in electron transports through the
transient current-current correlations, one can use the fast Fourier
transform (FFT) to convert the correlation functions from the time
domain ($\tau$) into the frequency domain for different observing
time $t$. The result gives the standard definition of the transient
noise spectra. Figure~\ref{Amp} plots the FFT amplitude of the
auto-correlation $S_{RR}(t+\tau, t)$ and the total-correlation
$S(t+\tau, t)$.
From Fig.~\ref{Amp}, one can analyze the electron transport
properties through the noise spectra not only just in the steady
state, but also in the entire transient regime. To make the energy
structures manifest in the transient noise spectra, one can let the
initial temperature approach zero ($k_BT=0.1\Gamma$). The right-lead
auto-correlation shows only one single peak at
$\omega_-=-\omega_R=-|\mu_R- \varepsilon|$ in the beginning. This is
because the dot is initially empty so that electron tunnelings from
the Fermi surface of the right lead to the dot have a maximum
probability. This peak corresponds to the energy absorption of the
electron tunnelings. On the other hand, we also observed that the
tunneling process for $\omega>eV$ can happen in the transient
regime, which is forbidden in the steady state near zero
temperature~\cite{Rothstein2009,Yang2014}. As the time $t$ varies,
the second peak shows up. This comes from backward electron tunnelings (i.e. emission processes)
from dot to the right lead, with the peak edge locating at the
resonance frequency $\omega_+=\omega_R=|\mu_R-\varepsilon|$. Note
that with a finite bandwidth spectral density, the spectrum decays
when the frequency passes over the resonant frequencies, which is
different from the WBL where the spectrum is flat~\cite{Rothstein2009,Yang2014}.
The noise spectrum still has a dip at zero frequency in both the transient and
steady-state regimes.  Furthermore, as one see it needs more time to
reach steady state when electrons transit from the leads to the dot,
due to the difference of the degrees of freedom between the dot and
the leads. Specifically, since there are infinity energy levels in
the lead but only one level in the dot system, electrons transiting
from the dot to the lead has much smaller probability to return back
to the dot, in comparison of the electrons transiting from lead into
the dot, as a dissipation effect. Thus, it takes longer time to
reach steady state for electrons tunneling from the leads to the
central dot. This effect will be reduced if we take a small band
width.

The FFT amplitude of the total-correlation has the same properties
as the right auto-correlation, with two more peaks coming from the
left auto-correlation functions as effects of the emission and
absorption processes between the left lead and the central dot.  By
calculating the individual contribution of the four terms in the
auto-correlation expression (Eq.~(\ref{1st})-(\ref{4th})),  it shows
that $S^{(3)}$ and $S^{(4)}$ dominate the noise of the current
correlations for an electron tunneling from the dot to the lead and
another electron tunneling from the lead to the dot.  The
contributions from $S^{(1)}$ and $S^{(2)}$ are much smaller because
they describe the correlations of electron tunnelings in the same
direction (namely both tunnelings from the leads to the dot or from
the dot to the leads), and mostly contribute to the noise around
zero frequency, due to the Pauli exclusion principle.

\subsection{Master equation approach to transient quantum transport in nanostructures
incorporating initial correlations}\label{initial_corr}

Quantum transport incorporating initial correlations in
nanostructures is a long-standing problem in mesoscopic
physics~\cite{Haug2008}. In the past two decades, investigations of
quantum transport have been mainly focused on steady-state
phenomena~\cite{Datta1995,Blanter2000,Imry2002}, where initial
correlations are not essential due to memory loss. Recent
experimental developments allow one to measure transient quantum
transport in different nano and quantum devices~\cite{Lu2003,
Bylander2005,Gustavsson2008}. In the transient transport regime,
initial correlations could induce different transport effects. In
this section, using the exact master equation
approach~\cite{Tu2008,Jin2010},one can address the transient
quantum transport incorporating initial correlations.

Transient quantum transport was first proposed by
Cini~\cite{Cini1980},
 under the so-called partition-free scheme. In this
scheme, the whole system (the device system plus the leads together)
is in thermal equilibrium up to time $t=0$, and then one applies an
external bias to let electrons flow. Thus, the device system and the
leads are initially correlated. Stefanucci {\it et.
al.}~\cite{Stefanucci2004,Stefanucci2007} adopted non-equilibrium Green
functions with the Kadanoff-Baym formalism~\cite{Kadanoff1962} to
investigate transient quantum transport with the partition-free
scheme. They obtained an
analytic transient transport current in the wide-band limit. In
these works~\cite{Stefanucci2004,Stefanucci2007}, the transport solution is
given in terms of the non-equilibrium Green functions of the total
system, rather than the Green functions for the device part in the
nanostructures~\cite{Wingreen1993}.

In fact, earlier investigations of the time-dependent electron
transport in solid-state physics had largely used the Kubo formula
in the linear response regime~\cite{Kubo1965,Mahan1990} and the
semiclassical Boltzmann equation~\cite{Cer1975,Smith1989}. For
nanostructural devices, which have an extremely short length scale
($\sim nm$) and an extremely fast time scale ($\sim ps$ to $fs$),
the semiclassical Boltzmann equation is most likely inapplicable and
the nonlinear response effect must be taken into
account~\cite{Haug2008}.
An alternative approach to investigate transient quantum transport
is the master equation approach developed particularly for
nanostructures~\cite{Schoeller1994,Jin2008,Tu2008,Jin2010,Gurvitz1996}
which we have given a complete description in
Sec.~\ref{Master_Eq_appr}.
However, the exact master equation given in
Sec.~\ref{Master_Eq_appr} is derived in the partitioned scheme in
which the system and the leads is initially uncorrelated, the same
situation considered in the non-equilibrium Green function technique
in Keldysh formalism in Sec.~\ref{NEGF_appr}. Realistically, it is
possible and often unavoidable in experiments that the device system
and the leads are initially correlated. Therefore, the transient
transport theory based on the master equation that takes the effect
of initial correlations into account becomes necessary.

In this subsection, we present the exact master equation
including the effect of initial correlations for non-interacting
nanostructures through the extended quantum Langevin
equation~\cite{Yang2015}. It is found that the initial correlations only
affect the fluctuation dynamics of the device system, while the
dissipation dynamics remains the same as in the case of initially
uncorrelated systems. The transient transport current in the
presence of initial system-lead and lead-lead correlations is also
obtained directly from the exact master equation. Both the
partitioned and the partition-free schemes studied in previous works
~\cite{Wingreen1993,Stefanucci2004,Stefanucci2007} are naturally reproduced in
this theory. Taking an experimentally realizable nano-fabrication
system, a single-level quantum dot coupled to two one-dimensional
tight-binding leads, as a specific example, the
initial correlation effects in the transient transport current as
well as in the density matrix of the device system are discussed in details.

Consider a nanostructure consisting of a quantum
device coupled with two leads (the source and the drain), described
by a Fano-Anderson Hamiltonian~(\ref{Hamiltonian_Mas}).
Because the system and the leads are coupled through electron
tunnelings, and the electron-electron interactions in the device are
ignored, the total Hamiltonian has a bilinear form of the
electron creation and annihilation operators, the master equation
describing the time evolution of the reduced density matrix of the
device system, $\rho(t)= \rm{Tr}_E[\rho_{tot}(t)]$, can have the
following general bilinear
form~\cite{Tu2008,Tu2012a,Jin2010,Yang2014} as shown in
Sec.~\ref{Master_Eq_appr}:
\begin{align}
\label{Master Equation_ini_corr} {d\rho(t)\over dt} &=
-i\big[H'_S(t),\rho(t)\big]+\sum_{ij}\big\{
\bm{\gamma}_{ij}(t)\big[2a_j\rho(t) a_i^{\dag} \notag \\
& -a_i^{\dag}a_j\rho(t) -\rho(t) a_i^{\dag}a_j\big]
 +\widetilde{\bm{\gamma}}_{ij}(t)\big[ a_i^{\dag}\rho(t)a_j \notag \\
& -a_j\rho(t)a_i^{\dag}+a_i^{\dag}a_j\rho(t) -\rho(t)
a_ja_i^{\dag}\big] \big\}\notag\\
=&-i\big[H_S(t),\rho(t)\big]+\sum_{\alpha}\big[\mathcal{L}^{+}_{\alpha}(t)
+\mathcal{L}^{-}_{\alpha}(t)\big]\rho(t) .
\end{align}
Here the renormalized Hamiltonian
$H'_S(t)=\sum_{ij}\bm{\varepsilon}'_{ij}(t)a_i^{\dag}a_j$, and the
coefficient $\bm{\varepsilon}'_{ij}(t)$ is the corresponding
renormalized energy matrix of the device system, including the
energy shift of each level and the lead-induced couplings between
different levels. The time-dependent dissipation coefficients
$\bm{\gamma}_{ij}(t)$ and the fluctuation coefficients
$\widetilde{\bm{\gamma}}_{ij}(t)$ take into account all the
back-action effects between the device system and the reservoirs.
The current superoperators of lead $\alpha$,
$\mathcal{L}^{+}_{\alpha}(t)$ and $\mathcal{L}^{-}_{\alpha}(t)$,
determining the transport current from lead $\alpha$ to the device
system is given by Eq.~(\ref{current_Mas})~\cite{Jin2010}.

When the device system and the leads are initially correlated, i.e.
$\rho_{tot}(t_0)\neq\rho(t_0) \otimes \rho_E(t_0)$, it would be
challenging to use the Feynman-Vernon influence functional approach
to derive the master equation. Alternately, one can use the extended quantum
Langevin equation (\ref{quantumLangevineq}) to determine the time-dependent
coefficients in the master equation when the initial system-lead correlations are
presented \cite{Yang2015}. Since the quantum Langevin equation is derived
exactly from the Heisenberg equation of motion,
it is valid for an arbitrary initial state of the device system and the leads.

To determine the time-dependent coefficients in the master equation
(\ref{Master Equation_ini_corr}), one can compute the equation of motion
of the single-particle density matrix of the device system,
$\bm{\rho}^{(1)}_{ij}(t)=\langle a_j^\dag(t) a_i(t) \rangle = {\rm
Tr}[a_j^\dag a_i \rho(t)]$ from the master equation (\ref{Master
Equation_ini_corr}). The result is given by
\begin{align}
\frac{d}{dt}\bm{\rho}^{(1)}_{ij}(t)=&\big\{\bm{\rho}^{(1)}(t)
\big[i\bm{\varepsilon}'(t)-\bm{\gamma}(t) \big]\big\}_{ij}
\notag\\&-\big\{\big[i\bm{\varepsilon}'(t) + \bm{\gamma}(t) \big]
\bm{\rho}^{(1)}(t)\big\}_{ij} + \widetilde{\bm{\gamma}}_{ij}(t) .
\label{aa1}
\end{align}
It is interesting to see that the homogenous master equation of motion
generates an inhomogeneous equation of motion for the single particle density
matrix. The inhomogeneous term in Eq.~(\ref{aa1}) is indeed induced
by various initial system-lead and lead-lead correlations, which will be shown
next.

On the other hand, Eq.~(\ref{aa1}) can also be derived from the
exact solution of the quantum Langevin equation, Eq.~(\ref{at}).
Explicitly, the single-particle correlation function of the device
system calculated from the solution of Eq.~(\ref{at}) is given by
\begin{align}
\bm{\rho}^{(1)}_{ij}(\tau,t)&= \langle a^\dag_j(t) a_i(\tau)\rangle
\notag\\&=\big[\bm{u}(\tau,t_0)\bm{\rho}^{(1)}(t_0)\bm{u}^{\dag}(t,t_0)+\bm{v}(\tau,t)\big]_{ij}
 ,   \label{rho1}
\end{align}
which indeed has exactly the same form as Eq.~(\ref{rho(1)}) for the
initially partitioned state, and
\begin{align}
\bm{v}_{ij}(\tau, t)\!=&\!\sum_{\alpha}\!\! \int_{t_0}^{\tau} \!\!d\tau_1
\!\!\int_{t_0}^{t}\!\!\! d\tau_2 [\bm{u}(\tau, \tau_1)
\widetilde{\bm{g}}_{\alpha}(\tau_1, \tau_2)
\bm{u}^{\dag}(t,\tau_2)]_{ij} ,    \label{vt1}
\end{align}
which also has the same form as Eq.~(\ref{v_tau_t}), but the time
non-local integral kernel, $\widetilde{\bm{g}}_{\alpha}(\tau,
\tau')$ of Eq.~(\ref{integral_kernel}b), is now modified by
the additional initial system-lead correlations as
\begin{align}
\widetilde{\bm{g}}_{\alpha ij}(\tau,
\tau')=\widetilde{\bm{g}}^{se}_{\alpha
ij}(\tau,\tau')+\widetilde{\bm{g}}^{ee}_{\alpha ij}(\tau ,\tau'),
\label{wtg}
\end{align}
where
\begin{widetext}
\begin{subequations}
\label{corr}
\begin{align}
\widetilde{\bm{g}}^{se}_{\alpha
ij}(\tau,\tau')=&-2i\sum_{k}\!\!\Big[ V_{i\alpha k}(\tau)
e^{-i\int_{t_0}^{\tau} \epsilon_{\alpha k}(\tau_1)d\tau_1}\langle
a_{j}^\dag(t_0)c_{\alpha k}(t_0)\rangle\delta(\tau'-t_0) \notag\\
&~~~~~~~~~~~~~ -V^*_{j\alpha k}(\tau') e^{i\int_{t_0}^{\tau'} \epsilon_{\alpha
k}(\tau_1)d\tau_1}\langle
c^\dag_{\alpha k}(t_0)a_{i}(t_0)\rangle\delta(\tau-t_0)\Big], \\
\widetilde{\bm{g}}^{ee}_{\alpha ij}(\tau
,\tau')=&\sum_{\alpha'}\sum_{kk'} V_{i\alpha
k}(\tau)e^{-i\int_{t_0}^{\tau}\epsilon_{\alpha k}(\tau_1)d\tau_1}
V^*_{j\alpha' k'}(\tau')e^{i\int_{t_0}^{\tau'}\epsilon_{\alpha'
k'}(\tau_1)d\tau_1}\langle c_{\alpha' k'}^\dag(t_0)c_{\alpha
k}(t_0)\rangle.
\end{align}
\end{subequations}
\end{widetext}
As one can see, $\widetilde{\bm{g}}^{se}_{\alpha}(\tau ,\tau')$ is
proportional to all the initial electron correlations between the
system and the leads, and
$\widetilde{\bm{g}}^{ee}_{\alpha }(\tau,\tau')$ is associated with
the initial electron correlations in the leads. Physically, the electron correlation Green
function $\bm{v}(\tau, t)$ characterizes all possible electron
fluctuation processes due to the initial system-lead correlations
and initial lead-lead correlations, both are induced by the
inhomogeneity of the quantum Langevin equation (\ref{quantumLangevineq}).  Also,
Eq.~(\ref{rho1}) indeed gives the exact solution of the lesser Green
function incorporating initial correlations.
Thus, through
the extended quantum Langevin equation, we obtain the most general
solution for the single-particle correlation function
$\bm{\rho}^{(1)}(\tau,t)$ (the lesser Green function)  and the
electron correlation Green function $\bm{v}(\tau, t)$ in the
Keldysh nonequilibrium Green function technique.

With the above general solution Eq.~(\ref{rho1}), it is found that
\begin{align}
&\frac{d}{dt} \bm{\rho}_{ij}^{(1)}(t)\!=\!
[\dot{\bm{u}}(t,t_0)\bm{u}^{-1}(t,t_0)\bm{\rho}^{(1)}(t)+{\rm H.c.}]_{ij}
\notag\\&~~~~~~~~~~~~~~-\![\dot{\bm{u}}(t,t_0)\bm{u}^{-1}\!(t,t_0)\bm{v}(t,t)\!+\!{\rm
H.c.}]_{ij} \!+\!\frac{d}{dt}\bm{v}_{ij}(t,t)\label{aa2} .
\end{align}
The last two terms in the above equation are inhomogeneous and
proportional to the electron correlation Green function
$\bm{v}(\tau, t)$ and, therefore, are purely induced by various
initial system-lead and lead-lead correlations through the integral
kernel $\widetilde{\bm{g}}_{\alpha}(\tau,\tau')$. Now, by comparing
Eq.~(\ref{aa1}) with Eq.~(\ref{aa2}), the
time-dependent renormalized energy $\bm{\varepsilon}'_{ij}(t)$,
dissipation, and fluctuation coefficients $\bm{\gamma}_{ij}(t)$ and
$\widetilde{\bm{\gamma}}_{ij}(t)$ in the master equation
incorporating initial correlations are uniquely determined as follows,
\begin{subequations}
\begin{align}
\bm{\varepsilon}'_{ij}(t)&=
\frac{i}{2}\big[\dot{\bm{u}}(t,t_0)\bm{u}^{-1}(t,t_0) - {\rm
H.c.}\big]_{ij}\notag\\&=\bm{\varepsilon}_{ij}(t)-\frac{i}{2}
\sum_\alpha[\bm{\kappa}_\alpha(t)-\bm{\kappa}^\dag_\alpha(t)]_{ij}, \notag \\
\bm{\gamma}_{ij}(t)&=
-\frac{1}{2}\big[\dot{\bm{u}}(t,t_0)\bm{u}^{-1}(t,t_0) + {\rm
H.c.}\big]_{ij}\notag\\&=\frac{1}{2}\sum_\alpha[\bm{\kappa}_\alpha(t)
+\bm{\kappa}^\dag_\alpha(t)]_{ij}, \notag \\
\widetilde{\bm{\gamma}}_{ij}(t)&= \frac{d}{dt}\bm{v}_{ij}(t,t)
-[\dot{\bm{u}}(t,t_0)\bm{u}^{-1}(t,t_0)\bm{v}(t,t)+ \rm H.c.]_{ij}
\notag\\&=-\sum_\alpha[\bm{\lambda}_\alpha(t)+\bm{\lambda}^\dag_\alpha(t)]_{ij}.
\notag
\end{align}
\end{subequations}
From the above results, one can see that the renormalized energy
$\bm{\varepsilon}'_{ij}(t)$ and the dissipation coefficients
$\bm{\gamma}_{ij}(t)$  are independent of the initial correlations
and are identical to the results given in Eqs.~(\ref{ecoff_Mas}a) and
(\ref{ecoff_Mas}b) for the decoupled initial state. The fluctuation
coefficients $\widetilde{\bm{\gamma}}_{ij}(t)$ also have the same
form of Eq.~(\ref{ecoff_Mas}c) as for the initially partitioned state,
but the electron correlation Green function $\bm{v}(t,t)$
takes into account both the initial system-lead and the initial
lead-lead correlations through Eqs.~(\ref{wtg}-\ref{corr}). In
other words, initial correlations only contribute to the
fluctuation-related dynamics of the device system, and the
expressions of all the time-dependent coefficients in the master
equation (\ref{Master Equation_ini_corr}) remain the same.
Correspondingly, the current superoperators in the master equation
(\ref{Master Equation_ini_corr}) incorporating initial system-lead
correlations, $\mathcal{L}^{+}_{\alpha}(t)$ and
$\mathcal{L}^{-}_{\alpha}(t)$, are still given by the same form of
Eq.~(\ref{current_super}) as in the initially partitioned state. As
a result, the transient transport current $I_\alpha(t)$
incorporating with the initial system-lead correlations is still
given by the same equation (\ref{I_trans})
\begin{align}
I_\alpha(t)\!
=\!- 2e{\rm Re}\!\!\int_{t_0}^t \!\! d\tau {\rm Tr}&
[\bm{g}_\alpha(t,\tau)\!\bm{\rho}^{(1)}(\tau, t)
\! -\!\widetilde{\bm{g}}_\alpha(t,\tau)\bm{u}^{\dag}(t,\tau)] .
\end{align}
Thus, the transient quantum transport incorporating initial
correlations is fully expressed in terms of the standard
non-equilibrium Green functions of the device system. The initially
uncorrelated case (the partitioned scheme) in
Sec.~\ref{Master_Eq_appr} is a special case in which the initial
system-lead correlations vanish so that
$\widetilde{\bm{g}}^{se}_\alpha(t,\tau)=0$, and then the time
non-local integral kernel  $\widetilde{\bm{g}}_\alpha(t,\tau)$ is
simply reduced to Eq.~(\ref{integral_kernel}b).

In conclusion, the exact master equation (\ref{Master
Equation_ini_corr}) describes the non-Markovian dynamics and
transient quantum transport of nano-device systems coupled to leads
involving various initial system-lead and lead-lead correlations. In
fact, the exact master equation with or without the initial
system-lead correlations is given by the same formula, except for
the time non-local integral kernel
$\widetilde{\bm{g}}_\alpha(t,\tau)$, which is determined by
Eq.~(\ref{integral_kernel}b) for the initially uncorrelated
states between the system and the leads, but it must be modified by
Eqs.~(\ref{wtg}-\ref{corr}) for the initially correlated states.

In the literature~\cite{Breuer2002}, it is claimed that in the master equation formally
derived through the Nakajima-Zwanzig (NZ) operator projective
technique~\cite{Nakajima1958, Zwanzig1960}, the initial
system-lead correlations would induce an inhomogeneous term in the
master equation. However, the so-called inhomogeneous term in the
NZ master equation is a misunderstanding in \cite{Breuer2002}.
In a recent work \cite{Yang2016}, we show explicitly that the so-called initial system-lead
correlations induced inhomogeneous term in the NZ master equation is indeed a homogeneous
term  both in terms of projected Hilbert subspaces in the original NZ master equation formalism and
in the master equation in terms of the reduced density matrix after taking trace over the environment states.
The result must be similar to  Eq.~(\ref{Master Equation_ini_corr}) for Fano-Anderson model where
the initial system-lead correlations are embedded in the fluctuation
coefficients, as given explicitly in this section.

It should be pointed out that if the leads are made by
superconductors, there may be initial pairing correlations. Then,
the master equation (\ref{Master Equation_ini_corr}) may need to be
modified. Further investigation of this problem is in
progress~\cite{Lai}. Nevertheless, the master equation (\ref{Master
Equation_ini_corr}) is sufficient for the description of transient
quantum transport in nanostructures with the initial correlations
given in Eq.~(\ref{corr}). In fact, because the total Hamiltonian
has a bilinear form of the electron creation and annihilation
operators, [see Eq.~(\ref{Hamiltonian_Mas})], all other correlation
functions can be fully determined by the two basic nonequilibrium
Green's functions, $\bm{u}(t,t_0)$ and $\bm{v}(\tau, t)$. The
non-Markovain memory effects, including the initial-state
dependence, which are fully embedded in the time-dependent
dissipation and fluctuation coefficients in the exact master
equation (\ref{Master Equation_ini_corr}), are consistently
determined by these two basic nonequilibrium Green functions.

\begin{figure}
\centerline{\scalebox{0.4}{\includegraphics{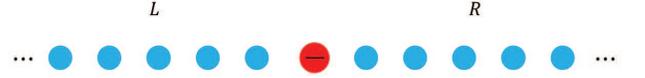}}}
\caption{A schematic plot of a single-level quantum dot coupled to
two one-dimensional tight-binding leads} \label{TB}
\end{figure}

To be specific, we consider an experimentally
realizable nano-fabrication system, a single-level quantum dot
coupled to the source and the drain, which are modeled by two
one-dimensional tight-binding leads (see Fig.~\ref{TB}). The
Hamiltonian of the whole system is given by
\begin{align}
H(t)=&\, \, \varepsilon_ca^\dag a -\sum_\alpha (\lambda_{\alpha 1}
a^{\dag}c_{\alpha 1}+ \lambda^*_{\alpha 1} c_{\alpha
1}^{\dag}a)\notag\\&+\sum_{\alpha}\sum_{n=1}^{\mathcal{N}}
[\epsilon_{\alpha }+U_\alpha(t)] c_{\alpha n}^\dag c_{\alpha n}
\nonumber\\& -\sum_{\alpha}\sum_{n=1}^{\mathcal{N}-1}
(\lambda_{\alpha}c_{\alpha n}^{\dag}c_{\alpha
n+1}+\lambda^{*}_{\alpha}c_{\alpha n+1}^{\dag}c_{\alpha
n}),\label{H2}
\end{align}
where $a$ ($a^\dag$) is the annihilation (creation) operator of the
single-level dot with the energy level $\varepsilon_c$, and
$c_{\alpha n}$ ($c_{\alpha n}^\dag$) the annihilation (creation)
operator of lead $\alpha$ at site $n$. All the sites in lead
$\alpha$ have an equal on-site energy $\epsilon_{\alpha}$.
$U_\alpha(t)$ is the time-dependent bias voltage applied on lead
$\alpha$ to shift the on-site energy. The second term in
Eq.~(\ref{H2}) describes the coupling between the quantum dot and
the first site of the lead $\alpha$ with the coupling strength
$\lambda_{\alpha 1}$. The last term characterizes the electron
tunneling between two consecutive sites in lead $\alpha$ with
tunneling amplitude $\lambda_{\alpha}$, and $\mathcal{N}$ is the
total number of sites on each lead.

\begin{figure}
\centerline{\scalebox{0.38}{\includegraphics{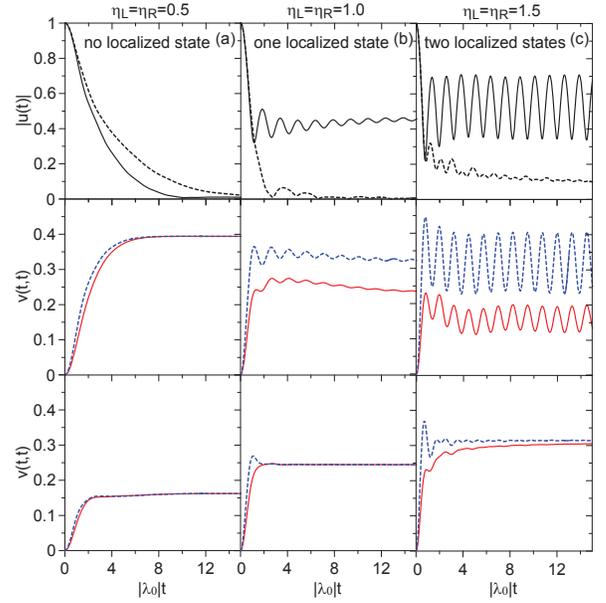}}} \caption{The
absolute value of time-dependent propagating Green function $|u(t)|$
and the electron correlation Green function $v(t,t)$ at different
coupling ratios (a) $\eta_{L}=\eta_{R}=0.5$, (b)
$\eta_{L}=\eta_{R}=1.0$, (c) $\eta_{L}=\eta_{R}=1.5$ with zero bias
$eV_{SD}=\mu_L-\mu_R=U_L-U_R=0$ and a finite bias
$eV_{SD}=\mu_L-\mu_R=U_L-U_R=3|\lambda_0|$. The energy level of the
quantum dot $\varepsilon_c=3|\lambda_0|$, the band center of the two
leads $\epsilon_0=2.5|\lambda_0|$, and
$k_BT_L=k_BT_R=3|\lambda_0|=k_BT$. For the unbiased case,
$\mu_L=\mu_R=2.5|\lambda_0|$, and $U_L=U_R=1.5|\lambda_0|$. For the
biased case, $\mu_L=4|\lambda_0|$, $\mu_R=|\lambda_0|$,
$U_L=3|\lambda_0|$, and $U_R=0$. In the graph of $|u(t)|$, solid
curves denote the unbiased case, and dash curves denote the biased
case. The value $v(t,t)$ in partition-free (blue dash line) and
partitioned (red solid line) schemes for the unbiased (the second
row) and biased (the third row) cases is presented \cite{Yang2015}.}\label{u}
\end{figure}

In the $k$-space, the Hamiltonian (\ref{H2}) becomes
\begin{align}
H(t)=&\varepsilon_ca^\dag a+\sum_{\alpha k}\epsilon_{\alpha
k}(t)c_{\alpha k}^\dag c_{\alpha k}\notag\\&+\sum_{\alpha k}[V_{\alpha
k}a^{\dag}c_{\alpha k}+V^{*}_{\alpha k}c^{\dag}_{\alpha k}a],
\label{H3}
\end{align}
where
$\epsilon_{\alpha k}(t)=\epsilon_{\alpha k} +U_\alpha(t)$,
$\epsilon_{\alpha k}=\epsilon_\alpha-2|\lambda_\alpha|\cos k$, and
$V_{\alpha k}=-\sqrt{\frac{2}{\mathcal{N}}}\lambda_{\alpha
1}e^{-i\phi}\sin k$.
Hamiltonian (\ref{H3}) has the same structure as Hamiltonian
(\ref{Hamiltonian_Mas}). The time non-local integral kernel
$g_\alpha(t,\tau)$ is given by
\begin{align}
g_\alpha(\tau,\tau')=\sum_k |V_{\alpha k}|^2
e^{-i\int^\tau_{\tau'}\epsilon_{\alpha k}(\tau_1)d\tau_1}. \label{g}
\end{align}
Because both the left and the right leads are modeled by the same
tight-binding model, namely, $\epsilon_L=\epsilon_R=\epsilon_0$ and
$\lambda_L=\lambda_R=\lambda_0$. When the site number $\mathcal{N}
\rightarrow \infty$, without applying bias [$U_\alpha(t)=0$], the general
solution of the Green function $u(t,t_0)$ is \cite{Zhang2012},
\begin{align}
u(t,t_0)=\int \frac{d\epsilon}{2\pi}{\cal D}(\epsilon)
e^{-i\epsilon(t-t_0)},
\end{align}
with
\begin{align}
{\cal D}(\epsilon) =&
2\pi\sum_{j\pm}Z_j\delta(\epsilon-\epsilon_j)\Theta(\eta^2-\eta^2_\pm)
\notag\\&+  \frac{\Gamma( \epsilon)}{\left[ \epsilon-\varepsilon_c -
\eta^2(\epsilon-\epsilon_0)/2 \right] ^{2} +  \Gamma^{2} (
\epsilon)/4}, \label{spectral}
\end{align}
where $\eta^2=\eta^2_L+\eta^2_R$ and $\eta_\alpha$ is the coupling
ratio $|\lambda_{\alpha 1}|/|\lambda_0|$ of lead $\alpha$. The
spectral density $\Gamma(\epsilon)=
\Gamma_L(\epsilon)+\Gamma_R(\epsilon)$ with
\begin{align}
\Gamma_\alpha(\epsilon) 
& =  \left\{\begin{array}{ll}
\eta^2_\alpha\sqrt{4|\lambda_0|^2-(\epsilon-\epsilon_0)^2}
& \mbox{if } |\epsilon-\epsilon_0|\leq2|\lambda_0|, \\
\\ 0 & \mbox{otherwise .} \end{array} \right.
\end{align}
In the solution (\ref{spectral}), the first term characterizes the
localized state~\cite{Mahan1990} with energy $\epsilon_j$ lying
outside the energy band when the total coupling ratio
$\eta^2\geq\eta^2_\pm$, where
$\eta^2_\pm=2\mp\frac{\Delta}{|\lambda_0|}$ is the critical coupling
ratio. Localized states are also referred to as dressed bound
states. Since the
energy bands of the two leads overlap, there are at most two
localized states. The amplitude and the frequency of the localized
state are given by~\cite{Yin}
\begin{subequations}
\label{amplitude}
\begin{align}
Z_\pm&=\frac{1}{2}\frac{(\eta^2-2)\sqrt{4(\eta^2-1)|\lambda_0|^2+\Delta^2}
\pm\eta^2\Delta}{(\eta^2-1)\sqrt{4(\eta^2-1)|\lambda_0|^2+\Delta^2}},\\
\epsilon_\pm&=\epsilon_0+\frac{(\eta^2-2)\Delta}{2(\eta^2-1)}
\pm\frac{\eta^2\sqrt{4(\eta^2-1)|\lambda_0|^2+\Delta^2}}{2(\eta^2-1)},
\end{align}
\end{subequations}
where $\Delta=\varepsilon_c-\epsilon_0$. When a finite bias is
applied, the above result should be modified accordingly, see
Fig.~\ref{u}, and the discussion given over there.

As a result, the effect of initial correlations will be maintained
in the steady-state limit through the localized states, the first
term in the solution of Eq.~(\ref{spectral}). This manifests a
long-time non-Markovian memory effect. The second term in
Eq.~(\ref{spectral}) is the contribution from the continuous energy
spectra, which causes electron dissipation (damping) in the dot
system. Once the solution of  $u(t,t_0)$ is given, the electron
correlation Green function $v(\tau,t)$ can be easily calculated with
the following general relation:
\begin{align}
v(\tau,t) =\sum_\alpha \int_{0}^{\tau}  d\tau_1  \int_{0}^t d\tau_2
u(\tau,\tau_1)\widetilde{g}_\alpha(\tau_1,\tau_2) u^{*}(t,\tau_2) .
\label{v0}
\end{align}
Thus, by solving the Green function $u(t,t_0)$ and the
correlation Green function $v(\tau,t)$, the density matrix and the transient
transport current can be fully determined,
\begin{subequations}
\label{ecpc}
\begin{align}
&\rho^{(1)}(t)= |u(t,t_0)|^2\rho^{(1)}(t_0) +v(t,t) =n(t),   \\
&I_\alpha(t)= -2e{\rm Re} \!\!\int_{t_0}^t \!\!d\tau
[g_\alpha(t,\tau)\rho^{(1)}(\tau, t)
-\widetilde{g}_\alpha(t,\tau)u^{*}(t,\tau)].
\end{align}
\end{subequations}

Consider two different initial states as examples. One is the partition-free
scheme, in which the whole system is in equilibrium before the
external bias is switched on. The other is the partitioned scheme in
which the initial state of the dot system is uncorrelated with the
leads before the tunneling couplings are turned on, the dot can be
in any arbitrary initial state $\rho(t_0)$ and the leads are
initially at separated equilibrium state. Both of these schemes can
be realized through different experimental setups. By comparing the
transient transport dynamics for these two initial schemes, one will
see in what circumstances the initial correlations will affect
quantum transport in the transient regime as well as in the
steady-state limit.

In the partition-free scheme, the whole system is in equilibrium
before the external bias voltage $U_\alpha(t)$ is switched on. The
applied bias voltage is set to be uniform on each lead such that
$U_\alpha(t)=U_\alpha\Theta(t-t_0)$, so $H(t\leq t_0)\equiv H$ is
time-independent. The initial density matrix of the whole system is
given by $\rho_{tot}(t_0)=\frac{1}{Z}e^{-\beta(H-\mu N)}$, where $H$
and $N$ are respectively the total Hamiltonian and the total
particle number operator at initial time $t_0$. The whole system is
initially at the temperature $\beta=1/k_BT$ with the chemical
potential $\mu$. When $t>t_0$, a uniform bias voltage is applied
to each lead, the whole system then suddenly change into a
non-equilibrium state. In this case, the calculations of initial
correlations, $\langle a^\dag(t_0)c_{\alpha k}(t_0)\rangle$ and
$\langle c^\dag_{\alpha'k'}(t_0)c_{\alpha k}(t_0)\rangle$, and the
corresponding time non-local integral kernel,
$\widetilde{g}_\alpha(\tau,\tau')=\widetilde{g}^{se}_\alpha(\tau,\tau')
+\widetilde{g}^{ee}_\alpha(\tau,\tau')$, are very complicated,
see the detailed calculations given in Appendix B of Ref. \cite{Yang2015}.

For the partitioned scheme, the dot and the leads are initially
uncorrelated, and the leads are initially in equilibrium state
$\rho_E(t_0)=\frac{1}{Z}e^{-\sum_\alpha\beta_\alpha(
H_\alpha-\mu_\alpha N_\alpha)}$. After $t_0$ one can turn on the
tunneling couplings between the dot and the leads to let the system
evolve~\cite{Giblin2012}. In comparison with the partition-free
scheme, each energy level in lead $\alpha$ shifts by
$U_\alpha$ to preserve the charge neutrality,
i.e.,~$\epsilon_{\alpha k}\rightarrow \epsilon_{\alpha k}+U_\alpha$.
Also, $\beta_L=\beta_R=\beta$ is taken~\cite{Stefanucci2004}. The
initial-state differences between the partition-free and the
partitioned schemes can be demonstrated simply in an initial empty dot in the
partitioned scheme. In this case, the non-local time system-lead
correlation function vanishes,
$\widetilde{g}^{se}_\alpha(\tau,\tau')=0$; the only non-vanishing
initial correlation for the partitioned scheme is given by the
initial Fermi distribution of the leads: $\langle
c^\dag_{\alpha'k'}(t_0)c_{\alpha k}(t_0)\rangle =\delta_{\alpha
k,\alpha'k'}f_\alpha(\epsilon_{\alpha k})$, which leads to the time
non-local integral kernel
$\widetilde{g}_\alpha(\tau,\tau')= \!\! \int \!\!
\frac{d\epsilon}{2\pi}\Gamma_\alpha(\epsilon)f_\alpha(\epsilon+
U_\alpha)e^{-i(\epsilon+U_\alpha)(\tau-\tau')}$.

The dissipation and localized state dynamics of the electron in the dot
system, given by the time evolution of the Green
function $u(t,t_0)$ is shown in Fig.~\ref{u}. The dissipation dynamics
is independent of the initial correlations, so the results of
$|u(t)|\equiv|u(t,t_0=0)|$ shown in Fig.~\ref{u} are the same for
both the partition-free and the partitioned schemes. Without
applying a bias, in the weak coupling regime: $\eta^2=\eta^2_L +
\eta^2_R <2-\frac{\Delta}{|\lambda_0|}$, no localized state occurs
so the propagating Green function monotonically decays to zero. In
the intermediate coupling regime: $2-\frac{\Delta}{|\lambda_0|}\leq
\eta^2<2+\frac{\Delta}{|\lambda_0|}$, one localized state occurs
[see the detailed discussion following Eq.~(\ref{spectral})].
Correspondingly, $|u(t)|$ decays very fast in the beginning and then
gradually approaches to a non-zero constant value in the
steady-state limit, as shown in Fig.~\ref{u}(b). This non-zero
steady-state value is the contribution of the localized state. In
the strong coupling regime:
$\eta^2\geq2+\frac{\Delta}{|\lambda_0|}$, two localized states occur
simultaneously. One can find that $|u(t)|$ will oscillate in time
forever. The oscillation frequency is the energy difference between
the two localized states energies, as shown in Fig.~\ref{u}(c).

When a finite bias is applied, $|u(t)|$ decays slowely in comparison
with the unbiased case in the weak coupling regime, where no
localized state occurs. In the intermediate coupling regime, it is
different from the unbiased case that $|u(t)|$ continuously decays
and eventually approaches to zero, see the dashed curve in
Fig.~\ref{u}(b). This implies that the localized state is suppressed
by the applied bias. This suppression comes from the fact that the
localized states always lie in the band gaps not far away from the
band edges~\cite{Lo2015}. The applied bias enlarges the band energy
regime, which could exclude the occurrence of the localized state
when the dot-lead coupling strength is not strong enough. The
localized state will reappear if one increases the coupling strength.
Therefore, in the strong coupling regime, the dissipation dynamics
is changed accordingly, in comparison with the unbiased case, where
one of the two localized states is suppressed by the applied bias,
as shown in Fig.~\ref{u}(c). As a result, the long-time oscillation
behavior seen in the unbiased case does not occur. Only in the very
strong coupling regime, the long-time oscillation induced by two
localized states could happen, but this may go beyond  the
physically feasible regime that we are interested in. In summary,
for the same dot-lead coupling strength, the applied bias suppresses
the effect of one localized state. As a result, $|u(t)|$ still
decays to zero in the intermediate coupling regime, and eventually approaches
to a constant value in the strong coupling regime.

The above different dissipation dynamics with or without a finite
bias will significantly affect the electron correlation Green
function $v(t,t)$ which characterizes all the system-lead and
lead-lead initial correlation effects through the time non-local
integral kernel $\widetilde{g}_\alpha(\tau,\tau')$, see
Eqs.~(\ref{wtg}-\ref{corr}). The numerical results are shown in the
second row (without bias) and the third row (with a finite bias) in
Fig.~\ref{u}. In the weak-coupling regime $\eta^2
<2-\frac{\Delta}{|\lambda_0|}$, as we can see that in both the
unbiased or biased cases, electron correlation Green function
$v(t,t)$ are not significantly different for different initial
states.  In particular, $v(t,t)$ becomes independent of initial
states in the steady state limit. In the intermediate coupling
regime $2-\frac{\Delta}{|\lambda_0|}\leq \eta^2
<2+\frac{\Delta}{|\lambda_0|}$,
$v(t,t)$ is quite different for the partitioned and partition-free
schemes in the transient regime, and also approach to different
steady-state values for the unbiased case. This shows that the
initial correlation effects can be manifested though the localized
state in the dot. However, when a finite bias is applied, this
significant initial correlation effect disappears. This is because a
finite bias suppresses the effect of the localized state, as discussed
in the solution of $u(t,t_0)$. In the strong coupling
regime, $\eta^2\geq2+\frac{\Delta} {|\lambda_0|}$,
the initial correlations effects are more significant. For zero
bias, the two localized states generate a strong oscillation in the
steady-state solution of $v(t,t)$. The oscillating frequency is just
the energy difference of the two localized states. When a bias is
applied, one localized state is suppressed so that the oscillation
cannot occur in the steady state, as shown in Fig.~\ref{u}.

\begin{figure}
\centerline{\scalebox{0.38}{\includegraphics{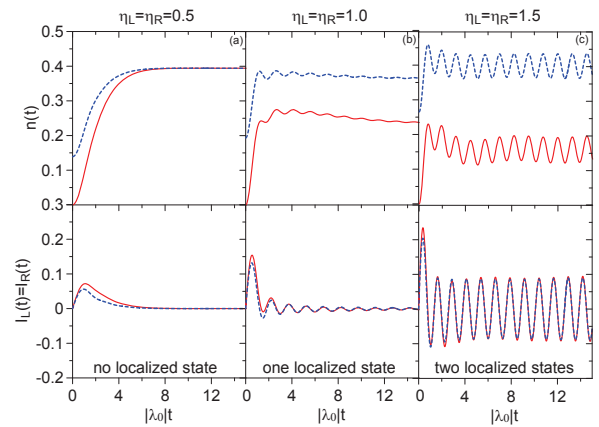}}} \caption{The
transient electron occupation of the dot and the transient transport
current for the unbiased case at different coupling ratios (a)
$\eta_{L}=\eta_{R}=0.5$, (b) $\eta_{L}=\eta_{R}=1.0$, (c)
$\eta_{L}=\eta_{R}=1.5$ for the partition-free (blue-dash line) and
partitioned (red solid line) schemes. The energy level of the
quantum dot $\varepsilon_c=3|\lambda_0|$, the band center of the two
leads $\epsilon_0=2.5|\lambda_0|$. For the partitioned scheme the
leads are prepared at $\mu_L=\mu_R=2.5|\lambda_0|$, and
$k_BT_L=k_BT_R=3|\lambda_0|$. For the partition-free scheme, the
system is initially at equilibrium with $\mu=|\lambda_0|$ and
$k_BT=3|\lambda_0|$. The applied bias voltage
$U_L=U_R=1.5|\lambda_0|$ after $t_0=0$ \cite{Yang2015}.}\label{unbiased_case}
\end{figure}

Figure~\ref{unbiased_case} shows the electron occupation in the dot
and the transient transport current $I_L(t)=I_R(t)$ in the unbiased
case for the partitioned and partition-free schemes. The
partition-free system is initially at equilibrium so that the dot
contains electrons, while the dot is initially empty in the
partitioned scheme. One can see that the effect of the initial
correlations vanish in the steady-state limit when the coupling ratio
$\eta^2<2-\frac{\Delta}{|\lambda_0|}$, where the dot does not have
localized state. This is because after $u(t,t_0)$ decays to zero,
the steady-state electron occupation is purely determined by
$v(t,t)$, which is the same for the partition-free and partitioned
schemes, as shown in Fig.~\ref{u}. This is an evidence of the dot
system reaching equilibrium with the leads so that the steady-state
electron occupation inside the dot must be independent of the
initial states.

However, in the coupling regime $2-\frac{\Delta}{|\lambda_0|}\leq
\eta^2 <2+\frac{\Delta}{|\lambda_0|}$, the localized state play a
significant role in manifesting the initial correlation effects. The
different electron occupation in the dot for the partitioned and
partition-free schemes is very similar to the behavior of $v(t,t)$,
see Fig.~\ref{u} and Fig.~\ref{unbiased_case}, except for a slightly
difference due to the initial occupation, caused by the first term
in Eq.~(\ref{ecpc}). Thus, the electron occupation in the dot
depends significantly on initial states. Physically, this result
implies the breakdown of the equilibrium hypothesis of statistical
mechanics, namely after reached the steady state, the system does
not approach equilibrium with its environment, and the particle
distribution depends on the initial states. This result with
localized states agrees indeed with the fact Anderson pointed out in
Anderson localization~\cite{Anderson1958}, namely, the system cannot
approach equilibrium when localization occurs. In the strong
coupling regime, $\eta^2\geq2+\frac{\Delta}{|\lambda_0|}$, two
localized states occur, which generates a strong oscillation in the
density matrix with the oscillating frequency being the energy
difference of the two localized states. This oscillation is
maintained in the steady state, where the initial-state dependence
becomes more significant, as shown in Fig.~\ref{unbiased_case}(c).

The corresponding transient transport current for the partitioned
and partition-free schemes approaches to the same value in a every
short time scale regardless whether the localized states exist or
not. This is because at zero bias, the steady-state transport
average current must approach to zero.  The transport current will
oscillate slightly around the zero value in
Fig.~\ref{unbiased_case}(b) because one localized state occurs which
causes the oscillation of electrons in the dot in the transient regime. When two localized
states occur, electrons in the dot oscillate between the two
localized states, so that the corresponding transport current
follows the same oscillation. In the meantime, the
initial-correlation dependence in the transport current is not as
significant as in the electron occupation in both the transient
regime and the steady-state limit. In fact, the initial correlation
effects even can be ignored for the transport current in the
steady-state limit, as shown in Fig.~\ref{unbiased_case}. The
current only oscillates around zero value because of the zero bias.

\begin{figure}
\centerline{\scalebox{0.38}{\includegraphics{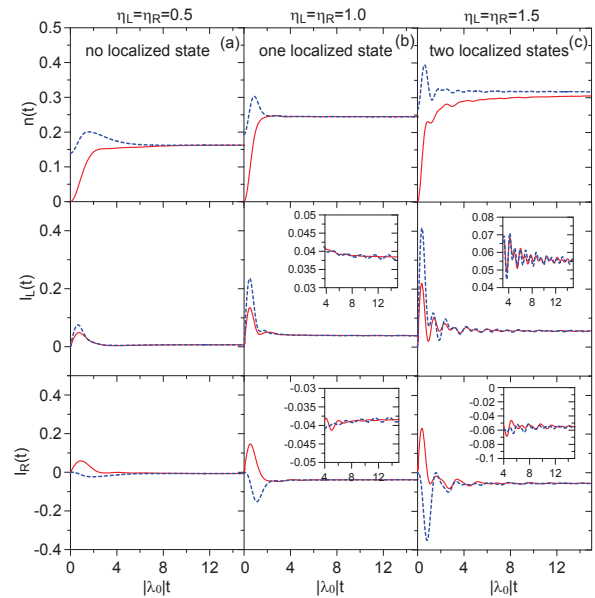}}} \caption{The
transient electron occupation of the dot and the transient transport
current for the biased case at different coupling ratios (a)
$\eta_{L}=\eta_{R}=0.5$, (b) $\eta_{L}=\eta_{R}=1.0$, (c)
$\eta_{L}=\eta_{R}=1.5$ for the partition-free (blue-dash line) and
partitioned (red solid line) schemes. The energy level of the
quantum dot $\varepsilon_c=3|\lambda_0|$, the band center of the two
leads $\epsilon_0=2.5|\lambda_0|$. For the partitioned scheme the
leads are prepared at $\mu_L=4|\lambda_0|$, $\mu_R=|\lambda_0|$, and
$k_BT_L=k_BT_R=3|\lambda_0|$. For the partition-free scheme, the
system is initially at equilibrium with $\mu=|\lambda_0|$ and
$k_BT=3|\lambda_0|$. The applied bias voltage $U_L=3|\lambda_0|$ and
$U_R=0$ after $t_0=0$ \cite{Yang2015}.}\label{biased_case}
\end{figure}

The time evolution of the electron occupation in the dot and the
transient transport current for both the partitioned and the
partition-free schemes for the biased case are shown in
Fig.~\ref{biased_case}. Compare Fig.~\ref{unbiased_case} with
Fig.~\ref{biased_case}, one can find that the applied bias restrains
most of the oscillation behavior in the electron occupation as well
as in the transport current, except for the very beginning of the
transient regime. Also, regardless of the existence of localized
states, the electron occupation in the dot and also the transport
current all approach to steady-state values other than zero due to
the non-zero bias. In other words, the localized state has a less
effect on the electron occupation and the transport current when a
bias is applied. This is because the applied bias suppresses one of
the localized states. However, the remaining localized state will
result in a slightly different steady-state values for
partition-free and partitioned schemes for the electron occupation
in the dot.
The corresponding transient current flow through the left and right
leads are quite different for these two schemes when a bias voltage
is applied. In particular, the transient transport current in the
right lead is positive in the beginning for the partitioned scheme
because the dot is initially empty, and it approaches to a negative
steady-state value in both schemes. But the steady-state current is
almost independent of the initial correlations as shown in the inset
graphs in Fig.~\ref{biased_case}. These results show that the
initial correlation effects have a significant effects in the
transient regime for both the electron occupation in the dot and the
transport currents between the dot and the leads when a finite bias
is applied. In the steady-state limit, it is expected that the
initial correlation effects are less important in electron transport
currents in comparison with the electron occupation in the dot.

In fact, the quantum
transport in the presence of localized states was previously
studied~\cite{boundstate1,boundstate2,boundstate3,
Dhar2006,Stefanucci2007}. In particular, Dhar and Sen considered a wire
connected to reservoirs that is modeled by a tight-binding
noninteracting Hamiltonian in the partitioned
scheme~\cite{Dhar2006}, and they gave the steady-state solution of
the density matrix and the current. Their results show that the
memory effects induced by the localized states can be observed such
that the density matrix of the system is initial-state dependent.
Stefanucci used the Kadanoff-Baym formalism to formally study the
localized state effects in the quantum transport in the
partition-free scheme~\cite{Stefanucci2007}. He found that the biased
system with localized states does not evolve toward a stationary
state. The results here using the master equation approach agree with these results obtained in
Refs.~\cite{Stefanucci2007} and \cite{Dhar2006}. The initial-state
dependance of the density matrix in the partitioned scheme is indeed
obvious in the master equation formalism, as given in Eq.~(\ref{ecpc}). In fact, all
these results with the existence of the localized states are fully
determined by the solution of the nonequilibrium Green functions of
the device system.

\subsection{Quantum coherence of the molecular states and their corresponding currents in nanoscale Aharonov-Bohm interferometers}\label{AB}
Quantum coherence of electrons in nanostructures is expected to
manage quantum computation and quantum information.
It is essential to prepare and read out the state of the qubit in quantum information processing.
There have been many experiments and theoretical analyses on quantum coherence manipulation
of electron states in DQDs which are thought to be a promising charge qubit
\cite{Loss1998,Hayashi2003,Elzerman2003,Petta2004,Gorman2005,Johnson2005,Petta2005,Petersson2010,Maune2012,Fricke2013,Shi2014,Fujisawa2006,Hanson2007}.
The techniques to reconstruct quantum states from series of measurements about the
system are known as quantum state tomography \cite{Kim2014,Samuelsson2006,Wu2006,Foletti2009,Medford2013}.
Quantum state tomography is resource demanding and it aims at very detailed description of coherence of quantum states.
On the other hand, transport measurement utilizing quantum interference
has revealed the main coherent properties of traveling electrons.
How the latter can be associated with the coherence of
local quantum states in the DQDs is interesting to investigate.

Quantum coherence has been detected through the Aharonov-Bohm (AB) interference \cite{Aharonov1959}.
Double quantum dots embedded in AB geometry were achieved in Refs.~\cite{Holleitner2001,Hatano2004,Sigrist2004}
The AB phase coherence of electrons through each dot would induce oscillating current as a function of the magnetic flux,
which is simply called the AB oscillation in the literature.
The results show that the AB phase coherence can be easily manipulated in these devices.
In Coulomb blockade and cotunneling regimes, it is predicted theoretically that currents through
spin-singlet and triplet states carry AB phases with a half of period difference \cite{Loss2000}.
For one-electron states, the half-period difference of AB oscillation is also anticipated
in transport currents through the bonding and antibonding state channels \cite{Kang2004,Kubo2006},
demonstrated in electron conductance.
In particular, it has been revealed \cite{Kang2004} that there are two resonances,
the Breit-Wigner resonance and the Fano resonance,
in the electron conductance that are associated to the
bonding and antibonding states and the interference between them.
it has also been found \cite{Kubo2006} that the Fano resonance can be suppressed as the indirect coupling strength decreases,
and the remaining Breit-Wigner resonance contains two peaks associated with the bonding and the antibonding states, respectively.
Motivated with these theoretical investigations, the transport currents passing through
the bonding and antibonding state channels has been detected experimentally \cite{Hatano2011}.
The half-period difference of AB oscillation in electron current through
the bonding and antibonding state channels, respectively,
is thought to be resulted from the parity of the wave functions of the bonding and antibonding states,
which is a property of the device geometry.
In Ref.~\cite{Hatano2011}, two different energy configurations are used,
which are succeeded by two different gate voltage settings.
Under the assumption that the transport currents flowing
through the bonding state channel in different energy configurations are almost the same,
the transport currents under these two configurations are measured.
The measured currents are used to determine the transport currents flowing
through the bonding and/or antibonding state channels in one of the configurations.
In this subsection, the validity of this assumption is justified
using the theoretical framework of the quantum transport theory based on master equation approach \cite{Tu2008,Jin2010,Tu2012a,Yang2014}.
Then, the relations between the probabilities of
the bonding and antibonding states and the transport currents flowing through the corresponding channels is investigated \cite{Liu2016}.
The results provide useful information for experimental reconstruction of quantum states
of the promising charge qubit in terms of two physical dot states through measurements of transport current.

The nanoscale AB interferometer consists of two coupled single-level QDs coupled to two leads,
its Hamiltonian is given by
\begin{equation}
H=H_{DQD}+H_{B}+H_{T},
\end{equation}
where $H_{DQD}$ is Hamiltonian of DQDs.
\begin{equation}
H_{DQD}=\sum_{i=1}^{2}\epsilon_{ij}d_{i}^{\dagger}d_{j},
\end{equation}
and $d_{i}\;(d_{i}^{\dagger})$ is annihilation (creation) operator
in $i$th QD, $\epsilon_{ii}$ is the energy level of $i$th QD and $\epsilon_{ij}$
with $i\neq j$ is the tunneling matrix element between the DQDs.
The Hamiltonian of the two leads is given by $H_{B}$:
\begin{equation}
H_{B}=\sum_{\alpha=L,R}\sum_{k}\varepsilon_{\alpha k}c_{\alpha k}^{\dagger}c_{\alpha k},
\end{equation}
where the label $\alpha$ denotes the left or right lead, and $c_{\alpha k}\;(c_{\alpha k}^{\dagger})$
is the annihilation (creation) operator of the $k$th level in lead $\alpha$.
The Hamiltonian $H_{T}$ describes the tunnelings between the QDs and the leads.
\begin{equation}
H_{T}=\sum_{\alpha=L,R}\sum_{i=1}^{2}\sum_{k}(V_{i\alpha k}d_{i}^{\dagger}c_{\alpha k}+h.c).
\end{equation}
By threading a magnetic flux $\Phi$ to the above system, the tunneling matrix elements would carry a AB phase, $V_{i\alpha k}=\bar{V}_{i\alpha k}e^{i\phi_{i\alpha}}$,
$\phi_{i\alpha}$ is the AB phase that electrons carry during the tunneling from $\alpha$ lead to $i$th dot, and $\bar{V}_{i\alpha k}$ is the
real tunneling amplitude.
The AB phase will also affect on $H_{DQD}$, i.e. for $i\neq j$, $\epsilon_{ij}=\bar{\epsilon}_{ij}e^{i\phi_{ij}}$
where $\bar{\epsilon}_{ij}=-t_{c}$ is a real amplitude and $\phi_{ij}$ is AB phase from $j$th dot to $i$th dot.
The relation of the AB phases with the magnetic flux  $\Phi$ is given by $\phi_{1L}-\phi_{1R}+\phi_{2R}-\phi_{2L}=2\pi \Phi/\Phi_{0}=\varphi$,
where $\Phi_0$ is the flux quanta. We also set $\phi_{12}=0$ according to Refs.~\cite{Kang2004,Kubo2006,Hatano2011}

The physics of a coupled double quantum dots system is better to understand in  the molecular basis than the computational basis.
By denoting the antibonding state (AS) and the bonding state (BS) with the signs $+$ and $-$ respectively, the Hamiltonian of the DQDs becomes,
\begin{equation}
H_{DQD}=\sum_{\nu=\pm}\epsilon_{\nu}d_{\nu}^{\dagger}d_{\nu},
\end{equation}
where $\epsilon_{\pm}$ is the corresponding energy level, and $d_{\pm}$ ($d_{\pm}^{\dagger}$) is the corresponding annihilation (creation) operator, which are given by:
\begin{subequations}
\begin{align}
& ~~ \epsilon_{\pm} =  ~ \frac{1}{2}\Big[\left(\epsilon_{11}+\epsilon_{22}\right)\pm\sqrt{\left(\epsilon_{11}-\epsilon_{22}\right)^{2}+4t_{c}^{2}}\Big],
\\
& \begin{pmatrix}d_{+}\\
d_{-}
\end{pmatrix} =  \begin{pmatrix} \cos\frac{\theta}{2}  & -\sin\frac{\theta}{2} \\
\sin\frac{\theta}{2} & \cos\frac{\theta}{2}
\end{pmatrix}\begin{pmatrix}d_{1}\\
d_{2}
\end{pmatrix}=\boldsymbol{S}\begin{pmatrix}d_{1}\\
d_{2}
\end{pmatrix},
\end{align}
\end{subequations}
and $\tan\theta=2t_{c}/(\epsilon_{11}-\epsilon_{22})$.
The reduced density matrix of the DQDs can be solved from the exact master equation.
By denoting the empty state with $|0\rangle$, the states AS and BS with $\left|\nu\right>:=\left|\pm\right>$,
and doubly occupied state by $|d\rangle$,
the reduced density matrix elements in molecular basis are expressed as follows,
\begin{subequations}
\begin{align}
&\rho_{00}(t) = \frac{1}{{\rm det}\bm{w}(t)}\Big\{ \rho_{00}(t_{0})+\rho_{dd}(t_{0}){\rm det}\left[\boldsymbol{J}_3(t)\right]
\notag\\&~~~~~~~~~~- \!\!\!\!\!\sum_{\nu,\nu'=: \pm}\!\!\! \rho_{\nu\nu'}(t_{0})
J_{3\nu\nu'} (t) \Big\}, \\
&\rho_{++}(t) = 1\!-\!\rho_{00}(t)\!-\!\rho^{(1)}_{--}(t),~\rho_{+-}(t) = \rho^{(1)}_{+-}(t),\\
& \rho_{--}(t) = 1\!-\!\rho_{00}(t)\!-\!\rho^{(1)}_{++}(t),~ \rho_{-+}(t) = \rho_{+-}^{*}(t),\\
& \rho_{dd}(t) = 1-\rho_{00}(t)-\rho_{++}(t)-\rho_{--}(t),
\end{align}
\label{eq rABt}
\end{subequations}
and the other off-diagonal density matrix elements between the different states are all zero.
Here, $\bm{w}(t)$ and $\bm{J}_3(t)$ are defined in Eq.~(\ref{J}).

The experiment in Ref.~\cite{Hatano2011} is given under the following conditions.
The energy of each dot is the same, $\epsilon_{11}=\epsilon_{22}=\epsilon_{0}$,
and the spectral density of lead $\alpha$ is energy independent, $\boldsymbol{\Gamma}_{\alpha}(\varepsilon)=\boldsymbol{\Gamma}_{\alpha}$ (wide band limit) with the level-width of the left lead $\Gamma_{L11}=\Gamma_{L22}=\Gamma_{L}$ and the right lead $\Gamma_{R11}=\Gamma_{R22}=\Gamma_{R}$. Also the indirect interdot couplings of the left lead $\Gamma_{L12}=a_{L}\Gamma_{L}e^{i\frac{\varphi}{2}}$ and the right lead $\Gamma_{R12}=a_{R}\Gamma_{R}e^{-i\frac{\varphi}{2}}$, where the indirect coupling parameter $a_{L,R}$ was originally introduced in Ref.~\cite{Kubo2006} in order to characterize the strength of the indirect coupling between two quantum dots via leads.
In the molecular basis, the energies of the bonding and antibonding states are $\epsilon_{\pm}=\epsilon_{0} \pm |t_{c}|$.
With the above conditions, the annihilation operators of the bonding and antibonding states become,
\begin{equation}
\begin{pmatrix} d_+ \\ d_- \end{pmatrix} = \frac{1}{\sqrt{2}}\begin{pmatrix} 1 & -1 \\ 1 & 1\end{pmatrix}\begin{pmatrix} d_1 \\ d_2 \end{pmatrix}.
\end{equation}
The tunneling Hamiltonian between the molecular states and the leads is reduced to,
\begin{equation}
H_T = \sum_{\alpha=L,R}\sum_{\nu=\pm}\sum_{k}(V_{\nu \alpha k} d_{\nu}^{\dagger} c_{\alpha k}+{\rm H.c.}),
\end{equation}
with the tunneling matrix elements,
\begin{equation}
\begin{pmatrix} V_{+ \alpha k} \\ V_{- \alpha k} \end{pmatrix}
= \frac{1}{\sqrt{2}}
\begin{pmatrix} 1 & -1 \\ 1 & 1 \end{pmatrix}
\begin{pmatrix} V_{1 \alpha k} \\ V_{2 \alpha k} \end{pmatrix}.
\end{equation}
The level-width matrix $\boldsymbol{\Gamma}_{\alpha}$ is given by
\begin{equation}
\begin{pmatrix}\Gamma_{++} & \Gamma_{+-} \\ \Gamma_{-+} & \Gamma_{--} \end{pmatrix}_{L,R}
=\Gamma_{L,R}\left(\boldsymbol{I}-\vec{\alpha}_{L,R}\cdot\vec{\boldsymbol{\sigma}}\right),
\end{equation}
where $\vec{\alpha}_{L,R}\!=\!(\alpha_{L,R}^{x},\alpha_{L,R}^{y},\alpha_{L,R}^{z})\!=\!a_{L,R}(0,\pm\sin\frac{\varphi}{2},\cos\frac{\varphi}{2})$ and $\vec{\boldsymbol{\sigma}}$ are the Pauli matrices.
Then the Green function $\bm{u}(t,t_0)$ has a simple solution,
\begin{eqnarray}
\boldsymbol{u}(t,t_{0})\!\! & = & \!\!
\begin{pmatrix}
u_{++}(t,t_{0}) & u_{+-}(t,t_{0}) \\
u_{-+}(t,t_{0}) & u_{--}(t,t_{0})
\end{pmatrix}\nonumber \\
\!\! & = & \! \exp\!\Big[\Big(\!\!-i\boldsymbol{\epsilon}-\frac{1}{2}\boldsymbol{\Gamma}_{L}-\frac{1}{2}\boldsymbol{\Gamma}_{R} \Big)(t-t_{0})\Big]\!,
\end{eqnarray}
where $\boldsymbol{\epsilon}=\begin{pmatrix} \epsilon_{+}& \!\! 0\\0& \epsilon_{-}\end{pmatrix}$.
The retarded Green function in energy domain has a simple form,
\begin{align}
\mathbf{G}^{R}(\varepsilon) & = -i\int_{0}^{\infty}e^{i\varepsilon t}\boldsymbol{u}(t)dt, \nonumber \\
 & = \Big(  \varepsilon \boldsymbol{I}-  \boldsymbol{\epsilon}+ \frac{i}{2} \boldsymbol{\Gamma} \Big)^{-1},
\label{eq Gr}
\end{align}
with $\boldsymbol{\Gamma}=\boldsymbol{\Gamma}_{L}+\boldsymbol{\Gamma}_{R}$.
The Green function $\boldsymbol{v}$ in the steady-state limit is
\begin{equation}
\boldsymbol{v}=\int_{-\infty}^{\infty}\frac{d\varepsilon}{2\pi}\sum_{\alpha}f_{\alpha}(\varepsilon)
\mathbf{G}^{R}(\varepsilon)\boldsymbol{\Gamma}_{\alpha}\mathbf{G}^{A}(\varepsilon),  \label{sv}
\end{equation}
and $\mathbf{G}^{a}(\varepsilon)=[\mathbf{G}^{r}(\varepsilon)]^\dag$.

As one can see, in the steady-state limit, $\boldsymbol{u}(t \rightarrow \infty) = 0$ so that the
single-particle reduced density matrix (\ref{rho(1)}) of the DQDs is reduced to
\begin{align}
\rho_{\nu\nu'}^{(1)}(t \rightarrow \infty)= [\boldsymbol{v}]_{\nu\nu'}.
\label{ssrdm}
\end{align}
where $\boldsymbol{v}$ is given by Eq.~(\ref{sv}).  Following the experiment \cite{Hatano2011},
the initial DQDs is empty so that $\rho_{00}(t_{0})=1$ and other initial density matrix elements of
the DQDs all equal to zero. Then, Eq.~(\ref{eq rABt}) in the steady-state limit can be simplified to
\begin{subequations}
\label{rdm0}
\begin{align}
&\rho_{00}= det[\boldsymbol{I}-\boldsymbol{v}], \\
&\rho_{++} = 1\!-\!\rho_{00}\!-\!\boldsymbol{v}_{--},~~~\rho_{+-} = \boldsymbol{v}_{+-},\\
& \rho_{--} = 1\!-\!\rho_{00}\!-\!\boldsymbol{v}_{++},~~~ \rho_{-+} = \boldsymbol{v}_{+-}^{*},\\
& \rho_{dd} = det[\boldsymbol{v}],
\end{align}
\end{subequations}
Thus, the reduced density matrix elements of the DQDs are fully determined by the Green function solution Eq.~(\ref{sv})
through the solution Eq.~(\ref{eq Gr}).

The steady-state electron current of Eq.~(\ref{I_trans}) in the wide band limit can be reduced to:
\begin{equation}
I_{\alpha} \! = -2e\textrm{ReTr}\Big\{ \frac{1}{2} \boldsymbol{\Gamma}_{\alpha} \boldsymbol{v} - i\! \int_{-\infty}^{\infty}\frac{d\varepsilon}{2\pi} f_{\alpha}(\varepsilon) \boldsymbol{\Gamma}_{\alpha} \mathbf{G}^{r}(\varepsilon) \Big\}.
\label{eq steady I_alpha}
\end{equation}
Carrying out explicitly the real part of Eq.~(\ref{eq steady I_alpha}), the transport current in the steady-state limit obeys the generalized Landauer-B\"{u}ttiker formula,
\begin{equation}
I=\frac{e}{2\pi}\int d\varepsilon\left[f_{L}(\varepsilon)-f_{R}(\varepsilon)\right]T(\varepsilon),
\end{equation}
where the electron transmission is
\begin{equation}
T(\varepsilon)=\textrm{Tr}\left[\mathbf{G}^{A}(\varepsilon)\boldsymbol{\Gamma}_{R}\mathbf{G}^{R}(\varepsilon)\boldsymbol{\Gamma}_{L}\right],
\end{equation}
According to the analyses in Ref.~\cite{Hatano2011}, the total transport current can be divided into components flowing through the bonding and antibonding state channels, plus the interference between them:
\begin{equation}
I=I_{+}+I_{-}+I_{+-}.
\end{equation}
These current components are explicitly given by:
\begin{equation}
I_{\pm}=\frac{e}{2\pi}\!\!\int_{-\infty}^{+\infty}\!\!d\varepsilon\big[f_{L}(\varepsilon)-f_{R}(\varepsilon)\big]\Gamma_{L\pm\pm}\Gamma_{R\pm\pm}\left|G_{\pm\pm}^{R}(\varepsilon)\right|^{2}\!\!\!,
\label{stdylmtIaIb}
\end{equation}
\begin{align}
I_{+-} & =\frac{e}{2\pi}\!\!\int_{-\infty}^{+\infty} \!\! d\varepsilon\left[f_{L}(\varepsilon)-f_{R}(\varepsilon)\right]\!\Big\{\Gamma_{L++}\Gamma_{R--}\left|G_{-+}^{R}\right|^{2}\nonumber \\
 & +\Gamma_{L--}\Gamma_{R++}\left|G_{+-}^{R}\right|^{2}+2\mathrm{Re}\big\{ G_{++}^{A}\Gamma_{R++}G_{+-}^{R}\Gamma_{L-+}\nonumber \\
 & +G_{+-}^{A}\Gamma_{R-+}G_{++}^{R}\Gamma_{L++}+G_{+-}^{A}\Gamma_{R-+}G_{+-}^{R}\Gamma_{L-+}\nonumber \\
 & +G_{++}^{A}\Gamma_{R+-}G_{--}^{R}\Gamma_{L-+}+G_{+-}^{A}\Gamma_{R--}G_{--}^{R}\Gamma_{L-+}\nonumber \\
 & +G_{--}^{A}\Gamma_{R-+}G_{+-}^{R}\Gamma_{L--}\big\}\Big\},
\end{align}
where $\Gamma_{L\pm\pm}\Gamma_{R\pm\pm}\left|G_{\pm\pm}^{R}(\varepsilon)\right|^{2}$ are the effective transmission coefficients of the bonding (antibonding) state channels.
The transport current component $I_{+-}$ is the second order term of $a_{L,R}$, and hence its contribution to the total transport current is ignorable in the weak indirect coupling limit, $I_{+-} \simeq 0$.

In Ref.~\cite{Kubo2006}, it is found that the full destructive interference of the Fano resonance only happens for the strongest indirect coupling, $|a_{L,R}|=1$.
When $|a_{L,R}|$ decreases from 1 to 0, the Fano resonance is gradually suppressed, the remaining result is the Breit-Wigner resonance containing two peaks associated with the bonding and antibonding states.
In the present formalism, $I_{\pm}$ in Eq.~(\ref{stdylmtIaIb}) are the transport currents flowing through the bonding and antibonding state channels, respectively, which gives the two peaks in the electron conductance for Breit-Wigner resonance, as shown in Ref.~\cite{Kubo2006}, and $I_{+-}$ is the transport current due to interference between the bonding and antibonding state channels, which induce the Fano resonance in the electron conductance when $|a_{L,R} | \rightarrow 1$, as shown in Ref.~\cite{Kang2004,Kubo2006}.
The transport currents flowing through the bonding and antibonding state channels was explicitly detected later \cite{Hatano2011}.
The theoretical analysis in Ref.~\cite{Kubo2006}
and the experimental analysis in Ref.~\cite{Hatano2011} inspire an explicit relation between the DQD reduced density matrix elements and the transport currents in the molecular state basis.

\begin{figure}[t]
\centerline{\scalebox{0.5}{\includegraphics{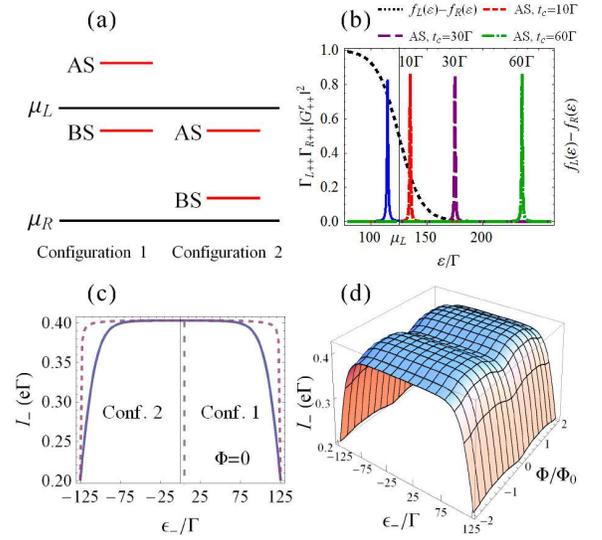}}}
\caption{
(a) The schematic plot of the energy levels of the bonding and antibonding states in configuration 1 and 2 with the chemical potential of the left and right leads, $\mu_{L}$ and $\mu_{R}$. (b) The difference of the left and right lead particle distributions, $f_{L}(\varepsilon)-f_{R}(\varepsilon)$, and the effective transmission coefficients of the bonding and antibonding channels in configuration 1 for different interdot coupling $t_{c}$ are plotted.
In this case, the energy $\epsilon_{-}$ of the bonding state is fixed at $115\Gamma$, and the corresponding transmission is plotted with the blue line.
The transmissions of the antibonding state for $t_{c}=10, 30, 60\Gamma$ are plotted with the red dashed line, purple long dashed line, and green dot-dashed line, respectively.
(c) $I_{-}$ as a function of $\epsilon_{-}$ is plotted.
The blue solid line is for temperature $k_{B}T=10\Gamma$, and the purple dashed line is for zero temperature.
The numbers 1, 2 in the plot denote the corresponding energy configurations 1 and 2 for $\left| t_{c}\right| =60\Gamma$.
(d) $I_{-}$ is plotted as a function of $\epsilon_{-}$ and $\Phi$ \cite{Liu2016}.}
\label{Liu1}
\end{figure}

In the experiment \cite{Hatano2011}, the electron currents are measured under two different energy configurations for the bonding and antibonding state channels with the fixed bias and indirect interdot weak couplings, as shown in Fig.~\ref{Liu1}(a).
Other parameter settings in Ref.~\cite{Hatano2011} are as follow:
the level broadenings of the left lead $\Gamma_{L}=0.3\Gamma$
and the right lead $\Gamma_{R}=0.7\Gamma$ ($\Gamma=\Gamma_{L}+\Gamma_{R}$),
the indirect interdot coupling parameters $a_{L}=-0.1$ for the left lead and $a_{R}=0.15$ for the right lead,
the direct interdot coupling $t_{c}=-60\Gamma$,
the chemical potentials of the left lead $\mu_{L}=125\Gamma$ and the right lead $\mu_{R}=-125\Gamma$,
and the temperature of the reservoirs is set at $k_{B}T=10\Gamma$.
The measured currents are the total electron currents in each configuration.
As shown by Fig.~\ref{Liu1}(a), in configuration 1, only the energy of the bonding state locates within the bias window ($\mu_{L}-\mu_{R}$). In configuration 2, both the energies of the bonding and antibonding states lie in the bias window.
These two energy configurations can be succeeded by tuning gate voltages.

In configuration 1, the current flowing through the bonding state channel, denoted by $I_{1-}$, is dominant such that the total current is almost given by $I\simeq I_{1-}$, where the current $I_{1+}$ flowing through the antibonding state channel in configuration 1 is negligible.
In configuration 2, the total current $I_{2}=I_{2+}+I_{2-}+I_{2+-}$, where $I_{2+}$, $I_{2-}$ are the currents flowing through the antibonding and bonding state channels in configuration 2, respectively, and $I_{2+-}$ is the current due to the interference between the bonding and antibonding state channels.
The latter is negligible in the weak indirect coupling regime \cite{Kubo2006}.
Therefore, the total current in configuration 2 is mainly given by $I_{2}\simeq I_{2+}+I_{2-}$.
With the assumption that currents flowing through the bonding state channel in configuration 1 and 2 are almost the same \cite{Hatano2011}, $I_{1-}\simeq I_{2-}$, one can determine the currents flowing through the bonding and antibonding state channels, respectively by the total currents measured separately in configuration 1 and 2.
This is the method used in Ref.~\cite{Hatano2011} for analysing the currents flowing through the bonding and antibonding state channels.

For the above experimental analysis, one shall ask that whether the current $I_{1+}$ flowing through the antibonding channel in configuration 1 is really negligible; and what are the conditions that should be satisfied such that the assumption $I_{1-}\approx I_{2-}$ is valid.
According to Eq.~(\ref{stdylmtIaIb}), $I_{1+}$ depends on the overlap of the difference of particle number distributions in the two leads, $f_{L}(\varepsilon)-f_{R}(\varepsilon)$, with the effective transmission coefficient of antibonding state channel, $\Gamma_{L++}\Gamma_{R++}\left|G_{++}^{R}(\varepsilon)\right|^{2}$.
In Fig.~\ref{Liu1}(b), the difference $f_{L}(\varepsilon)-f_{R}(\varepsilon)$ is shown by the black dashed line.
The energy of the bonding state is theoretically fixed, $\epsilon_{-}=\epsilon_{0}-|t_{c}|$, and the interdot coupling $t_{c}$ is changed to compare the corresponding antibonding state channel contributions to the current.
In experiments, $\epsilon_{-}$ can be manipulated through tuning the energy of DQDs and the interdot coupling simultaneously.
The effective transmission coefficient $\Gamma_{L--}\Gamma_{R--}\left|G_{--}^{R}(\varepsilon)\right|^{2}$ of the bonding state channel is fixed because of constant $\epsilon_{-}$, which is shown by the blue peak in Fig.~\ref{Liu1}(b).
Other peaks are the corresponding effective transmission coefficient $\Gamma_{L++}\Gamma_{R++}\left|G_{++}^{R}(\varepsilon)\right|^{2}$ of the antibonding state channel for different $t_{c}$.
As shown by Fig.~\ref{Liu1}(b), the larger $t_{c}$ gives the smaller overlap of $f_{L}(\varepsilon)-f_{R}(\varepsilon)$ with $\Gamma_{L++}\Gamma_{R++}\left|G_{++}^{R}(\varepsilon)\right|^{2}$ and hence the smaller current $I_{1+}$ flowing through the antibonding state channel in configuration 1.
So one can conclude that $I_{1+}$ is negligible when $t_{c}$ is properly large enough to make $\Gamma_{L++}\Gamma_{R++}\left|G_{++}^{R}(\varepsilon)\right|^{2}$ lesser overlap with $f_{L}(\varepsilon)-f_{R}(\varepsilon)$ \cite{Liu2016}.

On the other hand, the current $I_{-}$ flowing through the bonding state channel as a function of the energy $\epsilon_{-}$ of the bonding state is shown in Fig.~\ref{Liu1}(c).
Figure \ref{Liu1}(c) shows that the current $I_{-}$ flowing through the bonding state channel becomes maximum when the energy $\epsilon_{-}$ of the bonding state is located in the middle of the bias window.
The current $I_{-}$ symmetrically and dramatically decays when $\epsilon_{-}$ approaches closely to $\mu_{L}$ or $\mu_{R}$.
In Fig.~\ref{Liu1}(c), the blue solid line gives the current $I_{-}$ as a function of $\epsilon_{-}$ for temperature $k_{B}T=10\Gamma$.
It shows that $I_{-}$ is almost a constant within $\left| \epsilon_{-} \right| \lesssim 80\Gamma$.
This indicates that the condition $I_{1-}\simeq I_{2-}$ is well satisfied for $\left| \epsilon_{-} \right| \lesssim 80\Gamma$.
The purple dashed line in Fig.~\ref{Liu1}(c) shows $I_{-}$ at zero temperature .
In this case, the range for $I_{-}$ being almost a constant is wider.
Also, this flat pattern is maintained for arbitrary magnetic flux $\Phi$ (see Fig.\ref{Liu1}(d)).

The experiment of Ref.~\cite{Hatano2011} was performed under wide band limit, weak coupling, and large bias regime, which is a typical regime for transport experiment of DQDs devices.
As shown in Eq.~(\ref{rho(1)}), the steady-state single-particle reduced density matrix in the wide band limit is simply given by $\rho^{(1)}(t\rightarrow\infty)=\boldsymbol{v}(t,t\rightarrow\infty)$.
Because of the indirect interdot weak coupling (small $a_{L,R}$), one can ignore the higher order terms of $a_{L,R}$ \cite{Liu2016}.
The steady-state diagonal elements $v_{\pm \pm}$ then have the simple forms as:
\begin{align}
v_{\pm\pm} & \simeq \int_{-\infty}^{+\infty}\frac{d\varepsilon}{2\pi}\sum_{\alpha=L,R}f_{\alpha}(\varepsilon)\Gamma_{\alpha\pm\pm}\left|G_{\pm\pm}^{R}(\varepsilon)\right|^{2},\nonumber \\
& = v_{L\pm\pm} + v_{R\pm\pm}.
\label{eq v}
\end{align}
The steady-state transport currents through the bonding and antibonding state channels given in Eq.~(\ref{stdylmtIaIb}) can be approximately expressed in terms of $v_{\alpha\pm\pm}$
\begin{equation}
I_{\pm}=e\Gamma_{R\pm\pm}v_{L\pm\pm}-e\Gamma_{L\pm\pm}v_{R\pm\pm}.
\end{equation}

\begin{figure}[t]
\centerline{\scalebox{0.5}{\includegraphics{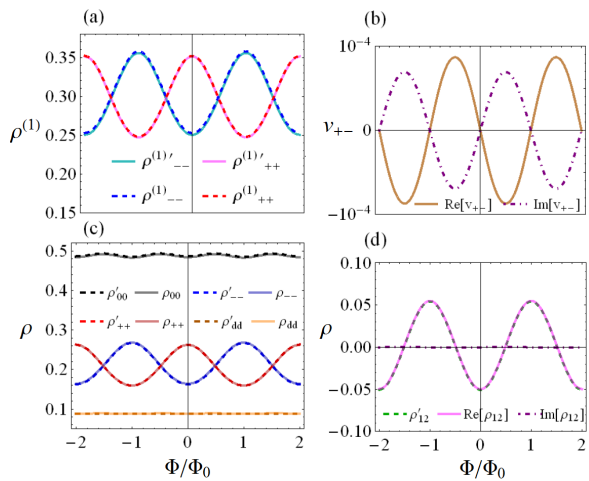}}}
\caption{
(a) The exact and approximate occupation numbers given by Eq.~(\ref{eq rho1}) in the bonding and antibonding states.
(b) The real part and the imaginary part of $v_{+-}$ ($\rho_{+-}$).
(c) The exact and approximate diagonal elements given by Eq.~(\ref{eq rp}) of the reduced density matrix in the molecular basis.
(d) The exact and approximate off-diagonal reduced density matrix elements in the dot basis \cite{Liu2016}.
}
\label{Liu2}
\end{figure}

From the above results, one obtains the relations between occupation numbers of the bonding and antibonding states and the corresponding currents approximately:
\begin{equation}
\rho^{(1)}_{\pm\pm} \simeq \frac{I_{\pm}}{e\Gamma_{R\pm\pm}}.
\label{eq rho1}
\end{equation}
The comparison between this approximated solution with the exact one
given by Eqs.~(\ref{sv}) and (\ref{ssrdm}) at the steady-state limit $t \rightarrow \infty$
are presented in Fig.~\ref{Liu2}(a), where energy configuration $\epsilon_{-}=-40\Gamma$ is chosen as an example.
As one see, the approximation solution is almost the same as the exact one.
Equation (\ref{eq rho1}) implies that the currents flowing through the bonding or antibonding state channels can be used to determine the particle occupations in the corresponding state.
The bonding and antibonding state components of the retarded Green function $|G_{\pm\pm}^{R}(\varepsilon)|^2$ in Eq.~(\ref{eq v}) have sharp peaks located at $\epsilon_{\pm}$, respectively, as the effective transmission shown in Fig. \ref{Liu1}.
When the bias is large ($\epsilon_{\pm} \gg \mu_{R}$), $v_{R\pm\pm}$ are ignorable.
This is because electrons in the right lead hardly tunnel back into DQDs.
The off-diagonal elements $v_{\pm \mp}$ relates to the tunneling probability between the bonding and antibonding states.
Because there is no direct coupling between the bonding and antibonding states, the electrons must hop to the leads, then hop back to the other state.
The weak couplings to the leads suppress the probability, and hence $v_{\pm \mp}$ are ignorable, as shown in Fig.~\ref{Liu2}(b) in which the magnitude of $v_{+-}$ is the order of $10^{-3}$ of the magnitude of the diagonal elements.
Consequently, the reduced density matrix of Eq.~(\ref{rdm0}) in the steady-state limit can be approximately given by the bonding and antibonding currents:
\begin{subequations}
\begin{align}
&\rho_{00}  \simeq \Big(1-\frac{I_{+}}{e\Gamma_{R++}}\Big)\Big(1-\frac{I_{-}}{e\Gamma_{R--}}\Big),\\
&\rho_{--}  \simeq \frac{I_{-}}{e\Gamma_{R--}}\Big(1-\frac{I_{+}}{e\Gamma_{R++}}\Big),\\
&\rho_{++}  \simeq \frac{I_{+}}{e\Gamma_{R++}}\Big(1-\frac{I_{-}}{e\Gamma_{R--}}\Big),\\
&\rho_{dd}  \simeq \frac{I_{+}}{e\Gamma_{R++}}\frac{I_{-}}{e\Gamma_{R--}},\\
&\rho_{+-}  = v_{+-} \simeq 0.
\end{align}
\label{eq rp}
\end{subequations}
The comparison between the above approximated solution with the exact elements of Eq.~(\ref{rdm0}) in the steady-state limit is shown in Fig.~\ref{Liu2}(c), which give almost the same results between the approximated solution and the exact one.

For practical application of DQDs as a promising qubit, one is interested in the quantum coherence between the two physical dots, which is described by the off-diagonal matrix element $\rho_{12}(t)$ in the physical dot basis.
The reduced density matrix elements in the physical dot basis (the charge qubit basis) of the DQDs is given by the following relation \cite{Tu2008,Tu2012a}:
\begin{subequations}
\label{eq r12t}
\begin{align}
\rho_{12}(t) & = \frac{1}{2} \big( \rho_{--}(t) - \rho_{++}(t) \big) + i~\!\textrm{Im} \rho_{+-}(t),\label{eq r12}\\ \nonumber
             & \simeq \frac{1}{2} \big( \rho_{--}(t) - \rho_{++}(t) \big).\\
\rho_{{}^1_2 {}^1_2}(t) & = \frac{1}{2} \big( \rho_{--}(t) + \rho_{++}(t) \big) \pm \textrm{Re} \rho_{+-}(t), \\
                        & \simeq \frac{1}{2} \big( \rho_{--}(t) + \rho_{++}(t) \big). \nonumber
\end{align}
\end{subequations}
The off-diagonal element $\rho_{12}(t)$ is presented in Fig.~\ref{Liu2}(d).
In the charge qubit basis, the probability of the diagonal elements, $\rho_{11}(t)$ and $\rho_{22}(t)$ can also be determined from the diagonal density matrix element, $\rho_{--}$ and $\rho_{++}$ of the bonding and antibonding states, as shown in the above equation.
Thus, the complete information of the reduced density matrix of the DQDs can be obtained experimentally from the measured currents through the relations given by Eqs.~(\ref{eq rp}) and (\ref{eq r12t}).

\section{Conclusion}\label{last_chap}
In summary, we have established a non-equilibrium quantum theory for the transient electron
dynamics of various nanodevices, based on the path integral method in the fermion coherent-state representation.
Our theory builds on the master equation of the reduced density matrix. The non-equilibrium transport current
is directly derived from the reduced density matrix. The master equation for the reduced
density matrix (i.e. equation (\ref{Master Equation}), which provides all the information about the electron quantum
coherence in the device) plus the transient current (i.e. equation (\ref{I_trans}), which determines transient
electron transport phenomena) together provide a unique
procedure to address the quantum decoherence problem in nonequilibrium quantum transport.
The master equation takes a convolutionless form and hence the non-Markovian dynamics are
fully encoded in the time-dependent coefficients. Explicitly, the back-reaction effect of the
gating electrodes on the central system is fully taken into account by these time-dependent
coefficients through the integrodifferential equations of motion (\ref{Dyson_eq_uv}) for
the nonequilibrium Green functions. The non-Markovian
memory structure is non-perturbatively built into the integral kernels in these equations of
motion. All the physical observables can be calculated directly from the master equation. In
particular, the transient transport current (\ref{I_trans}), and the single particle density
matrix, (\ref{rho(1)}), are found directly from the master equation in a rather simple way.
The master equation and the transient transport current are also explicitly related to each other
in terms of the superoperators acting on the reduced density matrix [see equation (\ref{current_Mas})].

This exact non-equilibrium formalism should provide a very intuitive picture showing how the
change in the electron quantum coherence in the devices is intimately related to the electron
tunneling processes through the leads and therefore responds nonlinearly to the corresponding
external bias and gate controls. This theory is applicable to a variety of quantum decoherence and
quantum transport phenomena involving non-Markovian memory effects, in both stationary
and transient scenarios, and at arbitrary initial temperatures of the different contacts.
The examples are given in Sec.~\ref{application}.
As we have also presented, one can simply reproduce the non-equilibrium
transport theory in terms of the non-equilibrium Green function technique
from the master equation formalism. However, we should
point out that the quantum transport theory based on the non-equilibrium Green function
technique does not explicitly give the connection to the reduced density matrix of the device
and thereby lacks a direct description of the quantum decoherence processes of the electrons
and the non-Markovian memory dynamics in nanostructures. Besides, the master equation approach
can be easily extended to incorporate the initial correlations, the formula for the master equation
and the transient current remain unchanged, the only change is given by the system-lead
nonlocal time correlation (\ref{wtg}) and (\ref{corr}) in the
determination of the correlation Green function (\ref{vt1}).

\section*{Acknowledgment}
This research is supported by the Ministry of Science and Technology of ROC under
Contract No. MOST-105-2112-M-006-008-MY3 and MOST-105-2811-M-006-033.
It is also supported in part by the
Headquarters of University Advancement at the National Cheng Kung University,
which is sponsored by the Ministry of Education of ROC.

\end{document}